\documentclass[aps,prd,twocolumn,floatfix,
nofootinbib,10pt]{revtex4-1}
\usepackage{amsmath,amssymb,multirow,bm,longtable,booktabs}
\usepackage{graphics,hyperref,subfigure,siunitx,makecell}
\usepackage{ytableau,graphicx,youngtab}
\hypersetup{pdfnewwindow=true,
colorlinks=true,linkcolor=blue,
citecolor=blue,filecolor=blue,urlcolor=blue
}
\newcommand{\ba}{\begin{eqnarray}}
\newcommand{\ea}{\end{eqnarray}}
\newcommand{\icn}{Instituto de Ciencias Nucleares, Universidad Nacional Aut\'onoma de M\'exico, \\ AP 70-543, Coyoac\'an, 04510 Ciudad de M\'exico, M\'exico}
\newcommand{\ijs}{Jozef Stefan Institute, Jamova 39, 1000 Ljubljana, Slovenia}

\begin{document}

\title{Masses and radiative decay widths of $S$- and $P$-wave singly-, doubly- and triply-heavy charm and bottom baryons}

\author{Emmanuel Ortiz-Pacheco}
\affiliation{\ijs}
\email{emmanuel.ortiz@ijs.si }
\author{Roelof Bijker}
\affiliation{\icn}
\email{bijker@nucleares.unam.mx}

\date{\today}

\begin{abstract}
We present a study of mass spectra and electromagnetic couplings of $S$- and $P$-wave baryons containing one, two or three heavy quarks, either charm ($c$) or bottom ($b$), in the framework of a non-relativistic harmonic oscillator quark model. A simultaneous fit to 41 known masses of heavy baryons (40 singly-heavy and 1 doubly-heavy) shows a r.m.s. deviation of 19 MeV. We present equal-spacing mass rules for excited baryons which may help to assign quantum numbers to experimentally observed heavy baryons. Explicit expressions are derived for electromagnetic couplings, both for spin-flavor matrix elements and radial integrals. 
\end{abstract}

\maketitle

\section{Introduction}
\label{sec:introduction}

The study of baryons with heavy quark content has taken great relevance to understand hadron structure, the dynamics of the strong interaction and heavy quark symmetry since the discovery of a large amount of hadrons with heavy quark flavors, either with charm $c$ or bottom $b$ quarks, by the LHCb, CMS, BelleII, and BESIII collaborations \cite{RPP76,Chen_2023}. The new discoveries also include exotic baryons and mesons with more than the minimal quark content \cite{Esposito:2016noz,Karliner:2017qhf,Olsen:2017bmm,RevModPhys.90.015004,LIU2019237,li2022physics,bulava2022hadron,accardi2023strong}.

In recent years there has been a large amount of new experimental information on heavy baryons, especially on singly-heavy baryons with one charm $c$ or one bottom $b$ quark. In particular, all but one ground state heavy baryons predicted by the quark model have been observed and identified \cite{PhysRevD.89.091102,PhysRevD.84.012003,PhysRevD.72.052006,PhysRevLett.122.012001,Workman:2022ynf}. The only exception is the $\Omega_b^\ast$ baryon with $J^P=3/2^+$.  

The LHCb Collaboration discovered five narrow $\Omega_c^0$ states in the $\Xi_c^+K^-$ decay channel \cite{PhysRevLett.118.182001} which were later confirmed by the Belle Collaboration \cite{PhysRevD.97.051102}. In a previous study it was shown that these states can be interpreted in the quark model as $P$-wave excited states \cite{Santopinto2019}. In addition, predictions made for the existence of excited $\Omega_b^-$ baryons were confirmed by the observation of the four $\Omega_b^-$ resonances in the $\Xi_b^0K^-$ mass spectrum by the LHCb Collaboration \cite{PhysRevLett.124.082002}. More recently, the LHCb Collaboration reported the discovery of two new excited states, $\Omega_c(3185)^0$ and $\Omega_c(3327)^0$ in the $\Xi_c^+ K^-$ channel \cite{lhcbcollaboration2023observation2omegas}, whose interpretation and quantum number assignments are still under discussion. 

The PDG compilation shows that for the strange-charm baryons, $\Xi_{c}$ and $\Xi'_{c}$, all ground state $S$-wave baryons have been observed as well as several candidates for excited $P$-wave baryons \cite{Workman:2022ynf}. 
Recently, the LHCb Collaboration reported the discovery of three new resonances in the $\Lambda_c^+ K^-$ channel, $\Xi_c(2923)^0$, $\Xi_c(2939)^0$ and $\Xi_c(2965)^0$ \cite{PhysRevLett.124.222001}. The Belle Collaboration determined the spin and parity of the charmed-strange baryon, $\Xi_c(2970)^+$ to be $J^P=1/2^+$ \cite{MoonPRD103} and measured the radiative decay widths of the excited charm baryons $\Xi_c(2790)$ and $\Xi_c(2815)$ \cite{PhysRevD.102.071103}. 
In the bottom sector, the LHCb collaboration claimed the observation of $\Xi^-_b(5935)$ and $\Xi^-_b(5955)$ resonances close to the threshold  $\Xi_b\pi$ \cite{PhysRevLett.114.062004}, and the $\Xi_b$ and $\Xi^*_b$ states  
\cite{PhysRevLett.113.242002,PhysRevD.99.052006}. The $\Xi^0_b(5945)$ baryon was observed in the $\Xi_b\pi$ decay \cite{XiJHEP}, and $\Xi^-_b(6227)$ in both the $\Lambda_b K$ and $\Xi_b\pi$ channels \cite{PhysRevLett.121.072002}.

For doubly-heavy baryons, there is experimental evidence from the LHCb collaboration related to the $\Xi^{++}_{cc}(3621)$ signal founded in the $\Lambda^+_c K^-\pi^+\pi^+$ invariant mass \cite{PhysRevLett.119.112001}. The mass reported for this resonance is $3621.40\pm 0.72$ MeV. Earlier evidence reported by the SELEX collaboration \cite{PhysRevLett.89.112001,OCHERASHVILI200518} for the existence of double charm baryons, $\Xi^{+}_{cc}(3519)$, could not be confirmed by the BaBar, Belle and LHCb collaborations \cite{Xicc2019,Workman:2022ynf}. At present, there is no experimental information on triply-heavy baryons, $\Omega_{ccc}$ and $\Omega_{bbb}$. The study of triply-heavy baryons is of special interest to gain a better understanding of baryon structure, heavy quark symmetry and the dynamics of the strong interaction as these systems do not contain any light quarks \cite{Yang_2020,PhysRevD.85.014012,PhysRevD.90.094507,PhysRevD.101.074031}.

The physics of heavy baryons has been studied theoretically in a vast variety of approaches such as non-relativistic quark models (NRQM) \cite{Valcarce2008,PhysRevD.90.094007,PhysRevD.92.114029,PhysRevD.101.074031,doi:10.1142/S0217751X22502256,Chen2017,Santopinto2019,PhysRevD.105.074029,PhysRevD.107.034031,garciatecocoatzi2023decay}, relativistic quark models (RQM) \cite{PhysRevD.66.014008,PhysRevD.84.014025,YU2023116183,Migura2006}, hypercentral quark models (hCQM) \cite{Shah_2016,Shah2016epjc,Shah2017epja,Gandhi2018plus,Gandhi2020,Kakadiya2023,Mutuk_2020}, chiral quark models ($\chi$QM) \cite{PhysRevD.96.094005,WangOme,Wang2017,PhysRevD.103.074025}, Regge phenomenology \cite{PhysRevD.95.116005,Song2023,oudichhya2023}, QCD sum rules (QCDSR) \cite{Aliev2009,Aliev2015,Aliev2019,Aliev2016,PhysRevD.102.114009,PhysRevD.91.054034,PhysRevD.101.114013,PhysRevD.104.034037}, Faddeev methods \cite{Garcilazo_2007,PhysRevD.100.034008}, molecular states \cite{PhysRevD.101.054033,Zhu2020}, large $N_c$ limit \cite{YANG2020135142}, heavy baryon chiral perturbation theory (HB$\chi$PT) \cite{JuanWang2019}, quark-diquark model \cite{Mutuk_2021} and lattice QCD calculations (LQCD) \cite{MeinelPRD85,PhysRevD.90.074504,PhysRevD.87.094512,PhysRevD.90.074501,PhysRevD.91.094502,PhysRevD.92.034504,PhysRevD.96.034511,PhysRevD.90.094507,PhysRevD.86.094504}. Reviews of heavy baryon physics together with a more complete list of references can be found in Refs.~\cite{KORNER1994787,Roberts2008,VIJANDE2013,rev2015,RPP76,Workman:2022ynf,Chen_2023}.

The aim of this work is to present a simultaneous study of mass spectra and electromagnetic couplings of $S$- and $P$-wave baryons with one, two or three heavy quarks, either charm $c$ or bottom $b$. As such, it is an extension of previous studies of $\Omega_Q$ baryons \cite{Ortiz-Pacheco:2020hmj,Santopinto2019}, and $\Xi_Q$ and $\Xi'_Q$ baryons \cite{PhysRevD.105.074029} to include $\Sigma_Q$, $\Lambda_Q$, $\Xi_{QQ}$, $\Omega_{QQ}$ and $\Omega_{QQQ}$ as well. We derive equal-spacing mass rules for excited baryons which may help to assign quantum numbers to experimentally observed heavy baryons. Whenever possible, we compare our results to experimental data or, in its absence, with other theoretical calculations. 

This article is organized as follows. We start by reviewing the classification of the singly-, doubly- and triply-heavy baryons and the structure of the wave functions in the harmonic oscillator quark model in Section~\ref{sec:wf}. 
Next we present the results for the mass spectra of heavy baryons in Section~\ref{sec:masses}. In Section~\ref{sec:em} we derive the spin-flavor and radial matrix elements of the electromagnetic couplings which are used in Section~\ref{sec:widths} to calculate the radiative decay widths. Finally, in Section~\ref{sec:summary} we present a summary and conclusions. 

\section{Heavy baryon wave functions}
\label{sec:wf}

In this section we review the classification of singly-, doubly- and triply-heavy baryon states \cite{Workman:2022ynf,Wang2017,Ortiz_Pacheco_2019P,Bjorken:1964gz}. In the quark model heavy baryons are considered as systems made up of three constituent quarks, where light quarks are denoted by $q=u$, $d$, $s$ and heavy quarks by $Q=c$, $b$. The baryon wave function is obtained by coupling the spatial, spin, flavor and color degrees of freedom. 

\begin{figure}[t]
\centering
\setlength{\unitlength}{0.4pt}
\begin{picture}(550,550)(-50,-225)
\thicklines
\multiput(100,300)(100,0){3}{\circle*{20}}
\multiput(150,250)(100,0){2}{\circle*{20}}
\put(200,200){\circle*{20}}
\put(100,300) {\line(1,0){200}}
\put(150,250) {\line(1,0){100}}
\put(200,200) {\line(1,1){100}}
\put(150,250) {\line(1,1){ 50}}
\put(200,200) {\line(-1,1){100}}
\put(250,250) {\line(-1,1){ 50}}
\put(200,100) {\circle*{20}}
\multiput(150, 50)(100,0){2}{\circle*{20}}
\put(150, 50) {\line(1,0){100}}
\put(150, 50) {\line(1,1){ 50}}
\put(250, 50) {\line(-1,1){ 50}}
\multiput(150,-50)(100,0){2}{\circle*{20}}
\put(200,-100) {\circle*{20}}
\put(150, -50) {\line(1,0){100}}
\put(150, -50) {\line(1,-1){ 50}}
\put(250, -50) {\line(-1,-1){ 50}}
\put(200,-200) {\circle*{20}}
\put(375, 290) {$\Sigma_Q(nnQ)$}
\put(375, 240) {$\Xi'_Q(nsQ)$}
\put(375, 190) {$\Omega_Q(ssQ)$}
\put(375,  90) {$\Lambda_Q(nnQ)$}
\put(375,  40) {$\Xi_Q(nsQ)$}
\put(375, -60) {$\Xi_{QQ}(QQn)$}
\put(375,-110) {$\Omega_{QQ}(QQs)$}
\put(-50, 240) {$[2] \equiv {\bf 6}$}
\put(-50,  65) {$[11] \equiv {\bf \bar{3}}$}
\put(-50, -85) {$[3] \equiv {\bf 3}$}
\put(-50,-210) {$[0] \equiv {\bf 1}$}
\put(375,-210) {$\Omega_{QQQ}(QQQ)$}
\end{picture}
\caption{Singly-heavy flavor sextet ${\bf 6}$ and anti-triplet ${\bf \bar{3}}$, doubly-heavy flavor triplet ${\bf 3}$ and triply-heavy singlet ${\bf 1}$ configurations. Here $n$ denotes the nonstrange light quarks $n=u$, $d$ and $Q$ the heavy quarks $Q=c$, $b$.}
\label{HeavyB}
\end{figure}
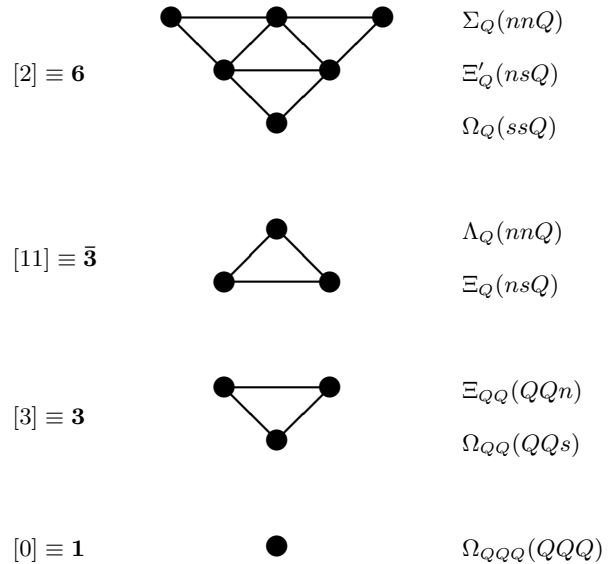

In a four-flavor classification scheme (three light flavors and one heavy) the ground state baryons in the quark model can be separated into two 20-plets: one with spin-parity $J^P=1/2^+$, and the other with $J^P=3/2^+$ \cite{Workman:2022ynf,Ortiz_Pacheco_2019P}. Each one of these multiplets splits into various $SU(3)$ flavor multiplets. The $J^P=3/2^+$ 20-plet consists of the light $uds$ baryon decuplet $\mathbf{10}$, a sextet $\mathbf{6}$ with one heavy quark, a triplet $\mathbf{3}$ with two heavy quarks and a singlet $\mathbf{1}$ with three heavy quarks. The $J^P=1/2^+$ 20-plet consists of the light $uds$ baryon octet $\mathbf{8}$, a sextet $\mathbf{6}$ and an anti-triplet $\mathbf{\bar{3}}$ with one heavy quark and a triplet $\mathbf{3}$ with two heavy quarks. The flavor multiplets are shown in Fig.~\ref{HeavyB}. 

The flavor wave function of singly-heavy baryons is written as $qqQ$, in which we distinguish between the light quarks $q$ and the heavy quark $Q$. The flavor sextet is symmetric under the interchange of the two light quarks, and the flavor anti-triplet antisymmetric. 
The flavor sextet $\mathbf{6}$ consists of three $\Sigma_Q$, two $\Xi'_Q$, and one $\Omega_Q$ states. The flavor wave functions with maximum charge state are given by 
\ba
\Sigma_Q &=& uuQ, \nonumber\\
\Xi'_Q &=& (us+su)Q/\sqrt{2}, \nonumber\\
\Omega_Q &=& ssQ,
\ea
The flavor anti-triplet $\mathbf{\bar{3}}$ consists of one $\Lambda_Q$ and two $\Xi_Q$ charge states. The corresponding flavor wave functions with maximum charge state are given by 
\ba
\Lambda_Q &=& (ud-du)Q/\sqrt{2}, \nonumber\\
\Xi_Q &=& (us-su)Q/\sqrt{2}.
\ea
For doubly-heavy baryons $QQq$ the flavor triplet $\mathbf{3}$ consists of two $\Xi_{QQ}$ and one $\Omega_{QQ}$ charge states
\ba
\Xi_{QQ} &=& QQu, \nonumber\\
\Omega_{QQ} &=& QQs,
\label{}
\ea
while for triply-heavy baryons $QQQ$ the flavor state is a singlet $\mathbf{1}$ 
\ba
\Omega_{QQQ} &=& QQQ .
\ea

The spin wave functions of baryons can be constructed by coupling the spins of the three quarks. The complete states with maximum spin projections are 
\ba
\chi_{E_\rho}&=&(\uparrow\downarrow-\downarrow\uparrow)\uparrow/\sqrt{2} , \nonumber\\
\chi_{E_\lambda}&=&(2\uparrow\uparrow\downarrow-\uparrow\downarrow\uparrow-\downarrow\uparrow\uparrow)/\sqrt{6} , \nonumber\\
\chi_{A_1}&=&\uparrow\uparrow\uparrow , 
\label{sp}
\ea
where the symmetric $A_1$, and mixed symmetric subscripts $E_\rho$ and $E_\lambda$ 
exhibit the symmetry properties under the permutation group $S_3$. $\chi_{E_\rho}$ is antisymmetric under the interchange of the first two quarks, whereas $\chi_{E_\lambda}$ and $\chi_{A_1}$ are symmetric. 

The orbital part is constructed by considering the harmonic oscillator potential to describe the interaction between the three constituyent quarks 
\begin{equation}
H_{ho} =
2m+m'+ \frac{{p_1}^2+{p_2}^2}{2m}+\frac{ {p_3}^2}{2m'}+\frac{1}{2}C \sum_{i<j}^3 |\vec{r_i}-\vec{r_j}|^2. \label{HO} 
\end{equation}
where we have taken $m_1=m_2=m \neq m'=m_3$. 
It is convenient to introduce Jacobi coordinates \cite{Isgurandkarl(1978)} 
\ba
\vec \rho    &=& \frac{1}{\sqrt{2}} (\vec r_1 - \vec r_2) ,
\nonumber\\
\vec \lambda &=& \frac{1}{\sqrt{6}} (\vec r_1 + \vec r_2 - 2 \vec r_3) ,
\nonumber\\
\vec R   &=& \frac{m(\vec r_1 + \vec r_2) + m'\vec r_3}{2m+m'} ,
\label{jacobi}
\ea 
and their conjugate momenta
\ba
\vec{p}_{\rho} &=& \frac{1}{\sqrt{2}} (\vec p_1 - \vec p_2) ,
\nonumber\\
\vec{p}_\lambda &=& \sqrt{\frac{3}{2}} \frac{m'(\vec p_1 + \vec p_2) - 2m \vec p_3}{2m+m'} ,
\nonumber\\
\vec{P}   &=& \vec p_1 + \vec p_2 + \vec p_3 .
\ea 
With this transformation, Eq.~(\ref{HO}) reduces to  
\ba
H_{ho} &=& M + \frac{P^2}{2M} + \frac{ {p_\rho}^2}{2m_\rho} 
+ \frac{ {p_\lambda}^2}{2m_\lambda} + \frac{3}{2}C\rho^2+\frac{3}{2}C\lambda^2. 	
\label{H3}
\ea
with $M=2m+m'$.
The new Hamiltonian now includes two independent harmonic oscillators in the $\rho$- and  $\lambda$-modes, with the same spring constant $C$, but different reduced masses
\begin{equation}
m_\rho  \equiv m, \hspace{0.6cm} m_\lambda \equiv \frac{3mm'}{2m+m'}.
\end{equation}
Consequently, the orbital wave functions are given by 
\ba
\psi_i(\vec{\rho}, \vec{\lambda},\vec R) =\frac{1}{(2\pi)^{3/2}}e^{\vec P \cdot\vec R}\psi_i^{ rel}(\vec{\rho}, \vec{\lambda}), 
\ea
with $i=\{0, \rho, \lambda\}$, and the relative wave functions are expressed by
\ba
\psi_{i}^{ rel}(\vec{\rho}, \vec{\lambda})=\frac{1}{\sqrt{3\sqrt{3}}} \psi_{n_\rho l_\rho m_\rho}(\vec{\rho})
\psi_{n_\lambda l_\lambda m_\lambda}(\vec{\lambda}). 
\ea
For the ground state one has 
\ba
\psi_{000}(\vec{\mu})=\frac{\alpha_\mu^{3/2}}{\pi^{3/4}}e^{-\alpha_\mu^2\mu^2/2},
\ea
with $\mu=\{\rho,\lambda\}$, and for one excited oscillator quantum
\ba
\psi_{11m_{\mu}}(\vec{\mu})=\sqrt{\frac{8\alpha_\mu^5}{3\sqrt{\pi}}}\mu e^{-\alpha_\mu^2\mu^2/2}Y_{1m_\mu}(\hat{\mu}),
\ea
where the corresponding oscillator size parameters and frequencies are given by 
\begin{equation}
\alpha_\mu=(3Cm_\mu)^{1/4}, \hspace{1cm}\omega_\mu=\sqrt{3C/m_\mu}.
\label{size}
\end{equation}
The oscillator size parameters are related by
\ba
\alpha_\lambda=\alpha_\rho\left(\frac{3m'}{2m+m'}\right)^{1/4}.
\label{alpl}
\ea
The ground state baryon wave function $\psi_0$ is symmetric, whereas the baryon wave function with one quantum of orbital excitation, $\psi_\rho$ and $\psi_\lambda$, are symmetric and antisymmetric under the interchange of the first two quarks.

\begin{figure}
\centering
\rotatebox{0}{\scalebox{0.18}[0.18]{\includegraphics{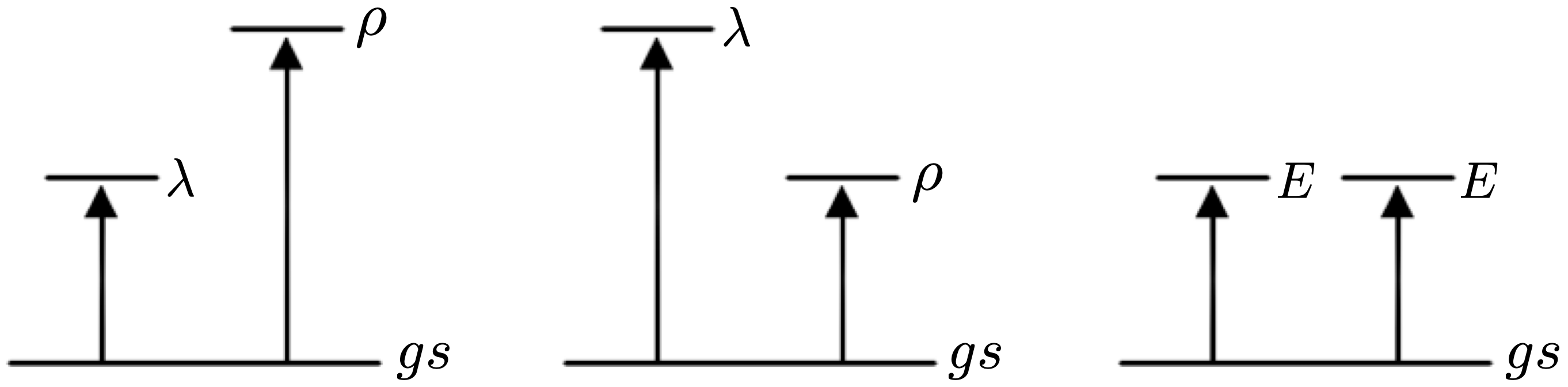}}} 
\caption[]{Orbital excitations in the $\lambda$, $\rho$ and the doubly degenerate $E$ mode, associated to the 
configurations: $qqQ$ with $m<m'$ (left), $QQq$ with $m>m'$ (center), and $QQQ$ with $m=m'$ (right).}
\label{levels}
\end{figure} 

If we consider systems containing two light quarks and one heavy quark $qqQ$ with $m'>m$, then 
$m_\lambda>m_\rho$, and $\omega_\lambda<\omega_\rho$, $i.e.$, the $\lambda$ state is less energetic than $\rho$ state. The last condition obtained suggest that states with one quantum of excitation in $\lambda$ will be very relevant for this configuration. 
By contrast, for the other case where 
$m'<m$ is associated with $QQq$, the conclusion is the opposite, and the first contribution for baryons with two heavy quarks will be the $\rho$ mode. The last case, $QQQ$, when all quarks have the same mass $m'=m$, then $\omega_\lambda=\omega_\rho$, and both modes have the same frequency. In Figure \ref{levels} the corresponding energy levels for these modes are shown. 

The total baryon wave function is given in terms of the coupled product of the spatial, spin, flavor and color parts $\psi = \psi_{\rm o} \chi_{\rm s} \phi_{\rm f}  \psi_{\rm c}$. 
Since all hadrons are colorless, the color wave function is a singlet, and in particular for baryons it corresponds to a completely antisymmetric wave function of three colors. 
Since the total baryon wave function is antisymmetric, this implies that the orbital and spin-flavor parts have the same symmetry for light $qqq$ baryons and triply-heavy $ccc$ and $bbb$ baryons under the interchange of any of the three quarks. For singly-heavy $qqQ$ and doubly-heavy $QQq$ baryons the orbital and spin-flavor parts have the same symmetry under the interchange of the first two quarks. 

With this information, it is now straightforward to write the total wave functions of heavy baryons. For singly-heavy baryons of the flavor sextet $\mathbf{6}$, the flavor part is combined with the symmetric spin-orbit part to obtain
\ba
^2(\Sigma_Q)_J&=&uuQ[\psi_0\times\chi_{E_\lambda}]_{J=1/2} , \nonumber\\
^4(\Sigma_Q)_J&=&uuQ[\psi_0\times\chi_{A_1}]_{J=3/2} , \nonumber\\
^2\rho(\Sigma_Q)_J&=&uuQ[\psi_\rho\times\chi_{E_\rho}]_J , \nonumber\\
^2\lambda(\Sigma_Q)_J&=&uuQ[\psi_\lambda\times\chi_{E_\lambda}]_J , \nonumber\\
^4\lambda(\Sigma_Q)_J&=&uuQ[\psi_\lambda\times\chi_{A_1}]_J,
\label{fq1}
\ea
for the $\Sigma_Q$ hyperons, 
\ba
^2(\Xi'_Q)_J&=&\frac{1}{\sqrt{2}}(us+su)Q[\psi_0\times\chi_{E_\lambda}]_{J=1/2} , \nonumber\\
^4(\Xi'_Q)_J&=&\frac{1}{\sqrt{2}}(us+su)Q[\psi_0\times\chi_{A_1}]_{J=3/2} , \nonumber\\
^2\rho(\Xi'_Q)_J&=&\frac{1}{\sqrt{2}}(us+su)Q[\psi_\rho\times\chi_{E_\rho}]_J , \nonumber\\
^2\lambda(\Xi'_Q)_J&=&\frac{1}{\sqrt{2}}(us+su)Q[\psi_\lambda\times\chi_{E_\lambda}]_J , \nonumber\\
^4\lambda(\Xi'_Q)_J&=&\frac{1}{\sqrt{2}}(us+su)Q[\psi_\lambda\times\chi_{A_1}]_J,
\ea
for the $\Xi'_Q$ hyperons, and
\ba
^2(\Omega_Q)_J&=&ssQ[\psi_0\times\chi_{E_\lambda}]_{J=1/2} , \nonumber\\
^4(\Omega_Q)_J&=&ssQ[\psi_0\times\chi_{A_1}]_{J=3/2} , \nonumber\\
^2\rho(\Omega_Q)_J&=&ssQ[\psi_\rho\times\chi_{E_\rho}]_J , \nonumber\\
^2\lambda(\Omega_Q)_J&=&ssQ[\psi_\lambda\times\chi_{E_\lambda}]_J , \nonumber\\
^4\lambda(\Omega_Q)_J&=&ssQ[\psi_\lambda\times\chi_{A_1}]_J ,
\ea
for the $\Omega_Q$ hyperons.

Similarly, the singly-heavy baryons of the anti-triplet $\mathbf{\bar{3}}$ are combined with 
the antisymmetric spin-orbit part 
\ba
^2(\Lambda_Q)_J&=&\frac{1}{\sqrt{2}}(ud-du)Q[\psi_0\times\chi_{E_\rho}]_{J=1/2} , \nonumber\\
^2\rho(\Lambda_Q)_J&=&\frac{1}{\sqrt{2}}(ud-du)Q[\psi_\rho\times\chi_{E_\lambda}]_J , \nonumber\\
^4\rho(\Lambda_Q)_J&=&\frac{1}{\sqrt{2}}(ud-du)Q[\psi_\rho\times\chi_{A_1}]_J , \nonumber\\
^2\lambda(\Lambda_Q)_J&=&\frac{1}{\sqrt{2}}(ud-du)Q[\psi_\lambda\times\chi_{E_\rho}]_J,
\ea
for the $\Lambda_Q$ hyperons, and
\ba
^2(\Xi_Q)_J&=&\frac{1}{\sqrt{2}}(us-su)Q[\psi_0\times\chi_{E_\rho}]_{J=1/2} , \nonumber\\
^2\rho(\Xi_Q)_J&=&\frac{1}{\sqrt{2}}(us-su)Q[\psi_\rho\times\chi_{E_\lambda}]_J , \nonumber\\
^4\rho(\Xi_Q)_J&=&\frac{1}{\sqrt{2}}(us-su)Q[\psi_\rho\times\chi_{A_1}]_J , \nonumber\\
^2\lambda(\Xi_Q)_J&=&\frac{1}{\sqrt{2}}(us-su)Q[\psi_\lambda\times\chi_{E_\rho}]_J ,
\label{fq2}
\ea
for the $\Xi_Q$ hyperons.

The total wave function for the doubly-heavy baryons of the flavor triplet $\bf 3$ (as maximum charge states) are given by
\ba
^2(\Xi_{QQ})_J&=&QQu[\psi_0\times\chi_{E_\lambda}]_{J=1/2} , \nonumber\\
^4(\Xi_{QQ})_J&=&QQu[\psi_0\times\chi_{A_1}]_{J=3/2} , \nonumber\\
^2\rho(\Xi_{QQ})_J&=&QQu[\psi_\rho\times\chi_{E_\rho}]_J , \nonumber\\
^2\lambda(\Xi_{QQ})_J&=&QQu[\psi_\lambda\times\chi_{E_\lambda}]_J , \nonumber\\
^4\lambda(\Xi_{QQ})_J&=&QQu[\psi_\lambda\times\chi_{A_1}]_J ,
\label{fqq1}
\ea
for the $\Xi_{QQ}$ hyperons, and by
\ba
^2(\Omega_{QQ})_J&=&QQs[\psi_0\times\chi_{E_\lambda}]_{J=1/2} , \nonumber\\
^4(\Omega_{QQ})_J&=&QQs[\psi_0\times\chi_{A_1}]_{J=3/2} , \nonumber\\
^2\rho(\Omega_{QQ})_J&=&QQs[\psi_\rho\times\chi_{E_\rho}]_J , \nonumber\\
^2\lambda(\Omega_{QQ})_J&=&QQs[\psi_\lambda\times\chi_{E_\lambda}]_J , \nonumber\\
^4\lambda(\Omega_{QQ})_J&=&QQs[\psi_\lambda\times\chi_{A_1}]_J ,
\label{fqq2}
\ea
for the $\Omega_{QQ}$ hyperons.

Finally, the total wave function of the $\Omega_{QQQ}$ flavor singlet $\bf 1$ is given as follows
\ba
^4(\Omega_{QQQ})_J&=&QQQ[\psi_0\times\chi_{A_1}]_{J=3/2} , \nonumber\\
^2E(\Omega_{QQQ})_J&=& \frac{1}{\sqrt{2}}QQQ[\psi_\rho\times\chi_{E_\rho}+\psi_\lambda\times\chi_{E_\lambda}]_J .
\label{fqqq}
\ea

\section{Mass spectra}
\label{sec:masses}

The mass spectra of heavy baryons with either charm or bottom quark content are calculated by considering the Hamiltonian (or mass operator)
\ba
H=H_{ho}+A\:\vec{S}^2+B\:\vec{S}\cdot \vec{L}+E\:\vec{I}^2+G\:{C}_{2SU_f(3)}.\:\:\:\: \label{wm}
\ea
The term $H_{ho}$ corresponds to the sum of the constituent quark masses and the harmonic oscillator Hamiltonian, previously defined in Eq.~(\ref{H3}). The remaining terms denote the spin-flavor dependent contributions for which we use a G\"ursey-Radicati form, {\it i.e.} the sum of a spin term, a spin-orbit splitting, an isospin term and a flavor term \cite{Santopinto2019,GR}. Consequently, the mass spectrum of heavy baryons is analyzed with the mass formula
\ba
M&=&2m+m'+\omega_\rho n_\rho+\omega_\lambda n_\lambda+A\:S(S+1)
\nonumber\\
&&+B\:\frac{1}{2}\left[J(J+1)-L(L+1)-S(S+1)\right]\nonumber\\
&&+E\:I(I+1)+G\:\frac{1}{3}\left[p(p+3)+q(q+3)+pq\right].\nonumber\\  
\label{mass_formula}
\ea
Here $n_\rho$ and $n_\lambda$ denote the number of quanta in the $\rho$- and $\lambda$-oscillator, respectively. The ground state $\psi_0$ has $(n_\rho,n_\lambda)=(0,0)$, whereas the states with one quantum of excitation in the $\rho$ mode $\psi_\rho$ or the $\lambda$ mode $\psi_\lambda$ are characterized by $(1,0)$ and $(0,1)$, respectively. The labels ${L}$, ${S}$, ${J}$ and ${I}$ denote the orbital angular momentum, spin, total angular momentum and isospin, respectively. The labels $(p,q)$ represent the $SU(3)$ flavor multiplets: the flavor sextets are labeled by $(2,0) \equiv \mathbf{6}$, the anti-triplets by $(0,1) \equiv \mathbf{\bar{3}}$, the triplets by $(1,0) \equiv \mathbf{3}$, and the singlets by $(0,0) \equiv \mathbf{\bar{1}}$. The isospin term in Eq.~(\ref{mass_formula}) splits the baryons with different values of isospin within the same flavor multiplet, whereas the flavor term can be determined from the mass difference between the anti-triplet $\Xi_Q$ and the sextet $\Xi'_Q$ baryons. 

In this study we adopt a harmonic oscillator quark model to describe the spectroscopy of $1S$- and $1P$-wave heavy baryons associated with the ground state ($n_\rho+n_\lambda=0$) and excited state ($n_\rho+n_\lambda=1$), respectively. It is well known from the study of nonstrange baryons that the harmonic oscillator quark model has difficulties in reproducing the relative mass of the Roper $N(1440)_{1/2^+}$ and the $N(1520)_{3/2^-}$ resonance, since the former is associated with a $2S$-wave and the latter with a $1P$-wave. For this reason, we limit the present study to $1S$- and $1P$-wave heavy baryons whose properties can reasonably well described in an effective way by the mass formula of Eq.~(\ref{mass_formula}) as was shown in previous work on the $\Omega_Q$ baryons \cite{Santopinto2019}  and the $\Xi_Q$ and $\Xi'_Q$ baryons \cite{PhysRevD.105.074029}.

\subsection{Determination of parameters}

There are a total of 14 parameters in the mass formula of Eq.~(\ref{mass_formula}): the 4 constituent quark masses $m_u = m_d$, $m_s$, $m_c$ and $m_b$, and the parameters $C$, $A$, $B$, $E$ and $G$, 5 for the charm and 5 for the bottom baryon sector. 

In order to determine the values of these parameters as well as their uncertainties, we fit the experimental data on the available Breit-Wigner masses of 41 heavy baryon resonances, 25 single charm and 15 single bottom baryons and 1 double charm baryon as listed in Tables~\ref{massQ} and \ref{massQQ}. The recently observed $\Omega_c$ baryons \cite{lhcbcollaboration2023observation2omegas} were not included in the fitting procedure. We use the maximum likelihood estimation technique \cite{Press2007}, where we consider the masses of the experimental data, $M^{exp}$, to be Gaussian distributed around the central value such that their uncertainties represent the widths of the distributions. 

We construct the likelihood function $\mathcal{L}$, by taking the product of our Gaussian distributions for each measured mass $M^{exp}_i$ as follows  
\ba
\mathcal{L}=\prod_{i=1}^{41}\frac{1}{\sqrt{2\pi\sigma_i}}\mbox{exp}\left(-\frac{(M^{th}_i-M_i^{exp})^2}{2\sigma_i^2}\right),
\ea
where $\sigma_i^2$ is the standard deviation of each of the experimental masses $M^{exp}_i$. The interest is on maximizing the likelihood function, with the aim of obtaining the optimal value of the 14 parameters, which is equivalent to minimizing the negative of its logarithm. For this purpose, we use Nelder-Mead method implemented in Python library \verb|scipy.optimize| \cite{scipy-optimize}. Furthermore, we compute the uncertainty of the optimal parameters, by re-sampling the parameters from likelihood using Nested sampled technique \cite{bortolato2022learning, buchner2021nested,hal-00216003}. We obtain the confidence intervals from the  the samples above, and report the uncertainties up to 1 sigma. The resulting parameter values and their uncertainties are shown in Tables~\ref{cqm} and \ref{par}. 

\begin{table}
\centering
\caption{Parameter values for the quark masses.}
\label{cqm}
\begin{ruledtabular}
\begin{tabular}{crl}
\noalign{\smallskip}
$m_u=m_d$ &  291.53 $\pm$ 0.60 & MeV \\
$m_s$     &  461.24 $\pm$ 0.73 & MeV \\
$m_c$     & 1606.80 $\pm$ 1.01 & MeV \\
$m_b$     & 4944.34 $\pm$ 1.62 & MeV \\
\noalign{\smallskip}
\end{tabular}
\end{ruledtabular}
\end{table}

\begin{table}
\centering
\caption{Fitted parameters using Eq.~(\ref{wm}) for heavy baryons with $Q=c$ and $Q=b$.}
\label{par}
\begin{ruledtabular}
\begin{tabular}{cccl}
\noalign{\smallskip}
&$Q=c$& $Q=b$ & \\
\noalign{\smallskip}
\hline 
\noalign{\smallskip}
$C$ & $0.02712 \pm 0.00008$ & $0.02377 \pm 0.00014$ & GeV$^3$ \\
$A$ & $22.08 \pm 0.19$ & $ 7.37 \pm 0.25$ & MeV \\
$B$ & $21.84 \pm 0.28$ & $ 4.33 \pm 0.55$ & MeV \\
$E$ & $30.47 \pm 0.40$ & $35.68 \pm 0.61$ & MeV \\
$G$ & $56.19 \pm 0.34$ & $61.77 \pm 0.28$ & MeV \\
\noalign{\smallskip}
\end{tabular}
\end{ruledtabular}
\end{table}

Next, we use these values to calculate the masses of all heavy baryons listed in Tables~\ref{massQ}-\ref{massQQQ}, and proceed to make the error propagation by extracting the standard deviation of the theoretical masses $M^{th}$. In the tables we show a comparison with the available experimental information. We find a good overall fit for 41 resonances (40 singly-heavy and 1 doubly-heavy) with an r.m.s. (root mean square) deviation of $\delta_{\rm rms}=19$ MeV. In previous studies of light $qqq$ baryons (with $q=u$, $d$, $s$) in the framework of a collective string-like model of baryons the r.m.s. deviation was found to be $39$ MeV for 25 non-strange baryons and $33$ MeV for 48 non-strange and strange baryon resonances \cite{Bijker:1994yr,Bijker:2000gq}. 

\begin{table}[b]
\centering
\caption{Mass differences between ground-state charm and bottom baryons $\Delta M_Q=M(^2(B_Q)_{1/2})-M(^2(\Lambda_Q)_{1/2})$.}
\label{massdiff}
\begin{ruledtabular}
\begin{tabular}{crcr}
\noalign{\smallskip}
$B_c$ & $\Delta M_c$ & $B_b$ & $\Delta M_b$ \\ 
\noalign{\smallskip}
\hline
\noalign{\smallskip}
$\Lambda_c$ &   0.0 & $\Lambda_b$ &   0.0 \\
$\Xi_c$     & 182.6 & $\Xi_b$     & 174.9 \\
$\Sigma_c$  & 167.0 & $\Sigma_b$  & 193.5 \\
$\Xi'_c$    & 292.0 & $\Xi'_b$    & 315.4 \\
$\Omega_c$  & 408.7 & $\Omega_b$  & 425.6 \\
\noalign{\smallskip}
\end{tabular}
\end{ruledtabular}
\end{table}

Table~\ref{massQ} and Figs.~\ref{charm} and \ref{bottom} show that in the charm sector all ground-state $S$-wave baryons have been observed, and in the bottom sector all with the exception of $^4(\Omega_b)_{3/2}$. The mass spectra of the charm and bottom baryons are very similar. For example, the mass differences between the ground-state charm and bottom baryons are almost the same (see Table~\ref{massdiff}). The assignment of quantum numbers is mostly based on energy systematics, such as the observed equal-spacing mass rule for the sextet baryons \cite{PhysRevLett.124.222001,PhysRevD.105.074029}
\ba
&& M(\Omega_c(3050)^0)-M(\Xi_c(2923)^0)
\nonumber\\
&& \qquad \simeq M(\Omega_c(3065)^0)-M(\Xi_c(2939)^0)
\nonumber\\
&& \qquad \simeq M(\Omega_c(3090)^0)-M(\Xi_c(2965)^0)
\nonumber\\
&& \qquad \simeq 125 \mbox{ MeV} ~,
\ea
which in the present approach for $\lambda$-mode excited baryons is equal to a constant
\ba
&& M(^{2S+1}\lambda(\Omega_Q)_J)-M(^{2S+1}\lambda(\Xi'_Q)_J) 
\nonumber\\
&& \qquad = m_s-m_{u/d} + \omega_\lambda(\Omega_Q)-\omega_\lambda(\Xi'_Q)-\frac{3}{4}E 
\nonumber\\
&& \qquad = \left\{ \begin{array}{ccc} 126 \mbox{ MeV} && (Q=c) \\ 120 \mbox{ MeV} && (Q=b) \end{array} \right.
\label{esr1}
\ea
in close agreement with the experimental value for single charm baryons. A similar equal-spacing mass rule holds for the anti-triplet baryons
\ba
&& M(\Xi_c(2790))-M(\Lambda_c(2595))
\nonumber\\
&& \qquad \simeq M(\Xi_c(2815))-M(\Lambda_c(2625))
\nonumber\\
&& \qquad \simeq 195 \mbox{ MeV} ~,
\ea
compared to
\ba
&& M(^{2}\lambda(\Xi_Q)_J)-M(^{2}\lambda(\Lambda_Q)_J) 
\nonumber\\
&& \qquad = m_s-m_{u/d} + \omega_\lambda(\Xi_Q)-\omega_\lambda(\Lambda_Q)+\frac{3}{4}E 
\nonumber\\
&& \qquad = \left\{ \begin{array}{ccc} 161 \mbox{ MeV} && (Q=c) \\ 164 \mbox{ MeV} && (Q=b) \end{array} \right.
\label{esr2}
\ea
For the other single charm or single bottom baryons there is little experimental information available. The mass difference between $\Xi'_Q$ and $\Sigma_Q$ can be calculated as
\ba
&& M(^{2S+1}\lambda(\Xi'_Q)_J)-M(^{2S+1}\lambda(\Sigma_Q)_J) 
\nonumber\\
&& \qquad = m_s-m_{u/d} + \omega_\lambda(\Xi'_Q)-\omega_\lambda(\Sigma_Q)-\frac{5}{4}E 
\nonumber\\
&& \qquad = \left\{ \begin{array}{ccc} 101 \mbox{ MeV} && (Q=c) \\ 93 \mbox{ MeV} && (Q=b) \end{array} \right.
\label{esr3}
\ea
The equal-spacing mass rules for $\rho$-mode excited sextet baryons are given by
\ba
&& M(^{2}\rho(\Omega_Q)_J)-M(^{2}\rho(\Xi'_Q)_J) 
\nonumber\\
&& \qquad = m_s-m_{u/d} + \omega_\rho(\Omega_Q)-\omega_\rho(\Xi'_Q)-\frac{3}{4}E 
\nonumber\\
&& \qquad = \left\{ \begin{array}{ccc} 102 \mbox{ MeV} && (Q=c) \\ 101 \mbox{ MeV} && (Q=b) \end{array} \right.
\nonumber\\
&& M(^{2}\rho(\Xi'_Q)_J)-M(^{2}\rho(\Sigma_Q)_J) 
\nonumber\\
&& \qquad = m_s-m_{u/d} + \omega_\rho(\Xi'_Q)-\omega_\rho(\Sigma_Q)-\frac{5}{4}E 
\nonumber\\
&& \qquad = \left\{ \begin{array}{ccc} 68 \mbox{ MeV} && (Q=c) \\ 66 \mbox{ MeV} && (Q=b) \end{array} \right.
\label{esr4}
\ea
For $\rho$-mode excited anti-triplet baryons the equal-spacing mass rule is given by
\ba
&& M(^{2S+1}\rho(\Xi_Q)_J)-M(^{2S+1}\rho(\Lambda_Q)_J) 
\nonumber\\
&& \qquad = m_s-m_{u/d} + \omega_\rho(\Xi_Q)-\omega_\rho(\Lambda_Q)+\frac{3}{4}E 
\nonumber\\
&& \qquad = \left\{ \begin{array}{ccc} 129 \mbox{ MeV} && (Q=c) \\ 137 \mbox{ MeV} && (Q=b) \end{array} \right.
\label{esr5}
\ea

Finally, the mass splitting between the $S=\frac{3}{2}$ and $\frac{1}{2}$ ground-state baryons of the flavor sextet is $\approx 65-70$ MeV in the charm sector and $\approx 20$ MeV in the bottom sector.

In the following subsections, we discuss the results of the calculated masses of ground-state $S$- and $P$-wave heavy baryons, which could be used as benchmarks for future measurements as well as lattice QCD calculations
\cite{padmanath2019heavy,PhysRevD.101.094503,PhysRevLett.119.042001,PhysRevD.102.054513}.

\begin{table*}
\centering
\caption{Mass spectra of single charm and single bottom $S$- and $P$-wave baryons. The states are labeled according to Eqs.~(\ref{fq1}-\ref{fq2}) and the quantum numbers $(n_\rho,n_\lambda)$, $I$, $L$, $S$ and $J^P$. The parity is given by $P=(-1)^L$. The experimental values are taken from \cite{Workman:2022ynf,PhysRevLett.124.222001,lhcbcollaboration2023observation2omegas}.}
\label{massQ}
\begin{ruledtabular}
\begin{tabular}{ccccccccccccccc}
\noalign{\smallskip}
&&&&&
&\multicolumn{3}{c}{Single-charm baryons \: $Q=c$}  & & \multicolumn{3}{c}{Single-bottom baryons \: $Q=b$} & \\ 
\noalign{\smallskip}
\cline{7-10} \cline{11-14}
\noalign{\smallskip}
State & $(n_\rho,n_\lambda)$ & $I$ & $L$ & $S$ & $J^{P}$ & $M^{th}(\text{MeV})$ 
& $M^{exp}(\text{MeV})$ & Name & $\ast$ & $M^{th}(\text{MeV})$ & $M^{exp}(\text{MeV})$ 
& Name & $\ast$ \\
\noalign{\smallskip}
\hline
\noalign{\smallskip}
$^2(\Sigma_{Q})_{{1/2}}$&(0,0)&1&0&$\frac{1}{2}$&$\frac{1}{2}^+$ & $2455 \pm 2$ & $2453.5\pm0.1$ & $\Sigma_{c}(2455)$ & 4 & $5810 \pm 3$ & $5813.1 \pm 0.2$ & $\Sigma_{b}$ & 3 \\
$^4(\Sigma_{Q})_{{3/2}}$&(0,0)&1&0&$\frac{3}{2}$&$\frac{3}{2}^+$ & $2521 \pm 2$ 
& $2518.1 \pm 0.2$ & $\Sigma_{c}(2520)$&3& $5832 \pm 3$ & $5832.5 \pm 0.2$ & $\Sigma_{b}^{*}$ &3\\
$^2\rho(\Sigma_{Q})_{{1/2}}$&(1,0)&1&1&$\frac{1}{2}$&$\frac{1}{2}^-$ & $2961 \pm 2$ &$\cdots$&$\cdots$&& $6300 \pm 3$ &$\cdots$&$\cdots$&\\
$^2\rho(\Sigma_{Q})_{{3/2}}$&(1,0)&1&1&$\frac{1}{2}$&$\frac{3}{2}^-$ & $2994 \pm 2$ &$\cdots$&$\cdots$&& $6307 \pm 3$ &$\cdots$&$\cdots$&\\
$^2\lambda(\Sigma_{Q})_{{1/2}}$&(0,1)&1&1&$\frac{1}{2}$&$\frac{1}{2}^-$ & $2789 \pm 2$ 
& $2800 \pm 4$ & $\Sigma_{c}(2800)$&3& $6108 \pm 3$ & $6096.9 \pm 1.2$ & $\Sigma_{b}(6097)$&3\\
$^2\lambda(\Sigma_{Q})_{{3/2}}$&(0,1)&1&1&$\frac{1}{2}$&$\frac{3}{2}^-$ & $2822 \pm 2$  &$\cdots$&$\cdots$&& $6114 \pm 3$ &$\cdots$&$\cdots$&\\
$^4\lambda(\Sigma_{Q})_{{1/2}}$&(0,1)&1&1&$\frac{3}{2}$&$\frac{1}{2}^-$ & $2822 \pm 2$ &$\cdots$&$\cdots$&& $6123 \pm 3$ &$\cdots$&$\cdots$&\\
$^4\lambda(\Sigma_{Q})_{{3/2}}$&(0,1)&1&1&$\frac{3}{2}$&$\frac{3}{2}^-$ & $2855 \pm 2$ &$\cdots$&$\cdots$&& $6130 \pm 3$ &$\cdots$&$\cdots$&\\
$^4\lambda(\Sigma_{Q})_{{5/2}}$&(0,1)&1&1&$\frac{3}{2}$&$\frac{5}{2}^-$ & $2910 \pm 2$ &$\cdots$&$\cdots$&& $6141 \pm 3$ &$\cdots$&$\cdots$&\\
\noalign{\smallskip}
\hline
\noalign{\smallskip}
$^2(\Xi'_{Q})_{{1/2}}$&(0,0)&$\frac{1}{2}$&0&$\frac{1}{2}$&$\frac{1}{2}^+$ & $2586 \pm 2$  & $2578.5 \pm 0.4$ & $\Xi'_{c}(2578)$&3& $5935 \pm 2$ & $5935.0 \pm 0.1$ & $\Xi'_{b}(5935)$ &3\\
$^4(\Xi'_{Q})_{{3/2}}$&(0,0)&$\frac{1}{2}$&0&$\frac{3}{2}$&$\frac{3}{2}^+$ & $2653 \pm 2$ & $2645.6 \pm 0.2$ & $\Xi_{c}(2645)$&3& $5957 \pm 2$ & $5953.8 \pm 0.3$ & $\Xi'^*_{b}(5955)$ &3\\
$^2\rho(\Xi'_{Q})_{{1/2}}$&(1,0)&$\frac{1}{2}$&1&$\frac{1}{2}$&$\frac{1}{2}^-$ & $3029 \pm 2$ & $3055.9 \pm 0.4$ & $\Xi_{c}(3055)$&3& $6366 \pm 3$ &$\cdots$&$\cdots$&\\
$^2\rho(\Xi'_{Q})_{{3/2}}$&(1,0)&$\frac{1}{2}$&1&$\frac{1}{2}$&$\frac{3}{2}^-$ & $3062 \pm 2$ & $3078.6 \pm 0.7$ & $\Xi_{c}(3080)$&3& $6373 \pm 3$ &$\cdots$&$\cdots$&\\
$^2\lambda(\Xi'_{Q})_{{1/2}}$&(0,1)&$\frac{1}{2}$&1&$\frac{1}{2}$&$\frac{1}{2}^-$ & $2890 \pm 2$  &$\cdots$&$\cdots$&& $6201 \pm 3$ &$\cdots$&$\cdots$ &\\
$^2\lambda(\Xi'_{Q})_{{3/2}}$&(0,1)&$\frac{1}{2}$&1&$\frac{1}{2}$&$\frac{3}{2}^-$ & $2922 \pm 2$  & $2938.5 \pm 0.3$ &$\Xi_{c}(2939)$&& $6207 \pm 3$ &$\cdots$&$\cdots$&\\
$^4\lambda(\Xi'_{Q})_{{1/2}}$&(0,1)&$\frac{1}{2}$&1&$\frac{3}{2}$&$\frac{1}{2}^-$ & $2923 \pm 2$ 
& $2923.0 \pm 0.3$ & $\Xi_{c}(2923)$&2& $6216 \pm 3$ &$\cdots$&$\cdots$ &\\
$^4\lambda(\Xi'_{Q})_{{3/2}}$ &(0,1)& $\frac{1}{2}$ &1& $\frac{3}{2}$ & $\frac{3}{2}^-$ & $2956 \pm 2$ & $2964.9 \pm 0.3$ & $\Xi_{c}(2965)$&& $6223 \pm 3$ &$\cdots$&$\cdots$ &\\
$^4\lambda(\Xi'_{Q})_{{5/2}}$&(0,1)&$\frac{1}{2}$&1&$\frac{3}{2}$&$\frac{5}{2}^-$ & $3011 \pm 2$ &$\cdots$&$\cdots$&& $6234 \pm 3$ & $6226.3 \pm 0.9$ & $\Xi_{b}(6227)$&3\\
\noalign{\smallskip}
\hline
\noalign{\smallskip}
$^2(\Omega_{Q})_{{1/2}}$&(0,0)&0&0&$\frac{1}{2}$&$\frac{1}{2}^+$ & $2733 \pm 2$ 
& $2695.2 \pm 1.7$ & $\Omega_{c}(2695)$ &3& $6078 \pm 2$ & $6045.2 \pm 1.2$ & $\Omega_{b}$&3\\
$^4(\Omega_{Q})_{{3/2}}$&(0,0)&0&0&$\frac{3}{2}$&$\frac{3}{2}^+$ & $2799 \pm 2$ & $2765.9 \pm 2.0$ & $\Omega_{c}(2770)$ &3& $6100 \pm 3$ &$\cdots$&$\cdots$ &\\
$^2\rho(\Omega_{Q})_{{1/2}}$&(1,0)&0&1&$\frac{1}{2}$&$\frac{1}{2}^-$ & $3131 \pm 2$ &$\cdots$&$\cdots$&& $6467 \pm 3$ &$\cdots$&$\cdots$&\\
$^2\rho(\Omega_{Q})_{{3/2}}$&(1,0)&0&1&$\frac{1}{2}$&$\frac{3}{2}^-$ & $3164 \pm 2$ & $3185.1 \pm 1.7$ & $\Omega_c(3185)$&& $6474 \pm 3$ &$\cdots$&$\cdots$&\\
$^2\lambda(\Omega_{Q})_{{1/2}}$&(0,1)&0&1&$\frac{1}{2}$&$\frac{1}{2}^-$& $3016 \pm 2$ 
& $3000.4 \pm 0.2$ & $\Omega_{c}(3000)$&3& $6321 \pm 2$ & $6315.6 \pm 0.6$ & $\Omega_b(6316)$&1\\
$^2\lambda(\Omega_{Q})_{{3/2}}$&(0,1)&0&1&$\frac{1}{2}$&$\frac{3}{2}^-$ & $3048 \pm 2$ & $3065.5 \pm 0.3$ & $\Omega_{c}(3065)$&3& $6328 \pm 2$ & $6330.3 \pm 0.6$ & $\Omega_b(6330)$ &1\\
$^4\lambda(\Omega_{Q})_{{1/2}}$&(0,1)&0&1&$\frac{3}{2}$&$\frac{1}{2}^-$ & $3049 \pm 2$ & $3050.2 \pm 0.1$ & $\Omega_{c}(3050)$&3& $6337 \pm 3$ & $6339.7 \pm 0.6$ & $\Omega_b(6340)$ &1\\
$^4\lambda(\Omega_{Q})_{{3/2}}$&(0,1)&0&1&$\frac{3}{2}$&$\frac{3}{2}^-$ & $3082 \pm 2$ & $3090.1 \pm 0.5$ &$\Omega_{c}(3090)$&3& $6343 \pm 3$ & $6349.8 \pm 0.6$ & $\Omega_b(6350)$ &1\\
$^4\lambda(\Omega_{Q})_{{5/2}}$&(0,1)&0&1&$\frac{3}{2}$&$\frac{5}{2}^-$ & $3136 \pm 2$ & $3119.1 \pm 1.0$ & $\Omega_{c}(3120)$&3& $6354 \pm 3$ &$\cdots$&$\cdots$&\\
\noalign{\smallskip}
\hline
\noalign{\smallskip}
$^2(\Lambda_{Q})_{{1/2}}$&(0,0)&0&0&$\frac{1}{2}$&$\frac{1}{2}^+$ & $2281 \pm 2$ &$2286.5 \pm 0.1$ & $\Lambda_{c}$ &4& $5615 \pm 2$ & $5619.6 \pm 0.2$ & $\Lambda_{b}$&3\\
$^2\rho(\Lambda_{Q})_{{1/2}}$&(1,0)&0&1&$\frac{1}{2}$&$\frac{1}{2}^-$ & $2788 \pm 2$ & $2766.6 \pm 2.4$ & $\Lambda_{c}(2765)$&1& $6106 \pm 2$ &$\cdots$&$\cdots$&\\
$^2\rho(\Lambda_{Q})_{{3/2}}$&(1,0)&0&1&$\frac{1}{2}$&$\frac{3}{2}^-$ & $2821 \pm 2$ &$\cdots$&$\cdots$&& $6112 \pm 2$ &$\cdots$&$\cdots$&\\
$^4\rho(\Lambda_{Q})_{{1/2}}$&(1,0)&0&1&$\frac{3}{2}$&$\frac{1}{2}^-$ & $2821 \pm 2$ &$\cdots$&$\cdots$&& $6121 \pm 3$ &$\cdots$&$\cdots$&\\
$^4\rho(\Lambda_{Q})_{{3/2}}$&(1,0)&0&1&$\frac{3}{2}$&$\frac{3}{2}^-$ & $2854 \pm 2$ &$\cdots$&$\cdots$&& $6128 \pm 3$ &$\cdots$&$\cdots$&\\
$^4\rho(\Lambda_{Q})_{{5/2}}$&(1,0)&0&1&$\frac{3}{2}$&$\frac{5}{2}^-$ & $2909 \pm 2$ & $2939.6 \pm 1.5$ & $\Lambda_{c}(2940)$&3& $6138 \pm 3$ &$\cdots$&$\cdots$&\\
$^2\lambda(\Lambda_{Q})_{{1/2}}$&(0,1)&0&1&$\frac{1}{2}$&$\frac{1}{2}^-$ & $2616 \pm 2$ &$2592.3\pm0.3$&$\Lambda_{c}(2595)$&3 & $5913 \pm 2$ & $5912.2 \pm 0.2$ & $\Lambda_{b}(5912)$&3\\
$^2\lambda(\Lambda_{Q})_{{3/2}}$&(0,1)&0&1&$\frac{1}{2}$&$\frac{3}{2}^-$ & $2648 \pm 2$ & $2628.1 \pm 0.2$ & $\Lambda_{c}(2625)$&3& $5919 \pm 2$ & $5920.1 \pm 0.2$ & $\Lambda_{b}(5920)$&3\\
\noalign{\smallskip}
\hline
\noalign{\smallskip}
$^2(\Xi_{Q})_{{1/2}}$&(0,0)&$\frac{1}{2}$&0&$\frac{1}{2}$&$\frac{1}{2}^+$ & $2474 \pm 2$ & $2469.1 \pm 0.2$ & $\Xi_{c}(2469)$&4& $5812 \pm 2$ & $5794.5 \pm 0.4$ & $\Xi_{b}(5794)$ &3\\
$^2\rho(\Xi_{Q})_{{1/2}}$&(1,0)&$\frac{1}{2}$&1&$\frac{1}{2}$&$\frac{1}{2}^-$ & $2917 \pm 2$ &$\cdots$&$\cdots$&& $6243 \pm 3$ &$\cdots$&$\cdots$&\\
$^2\rho(\Xi_{Q})_{{3/2}}$&(1,0)&$\frac{1}{2}$&1&$\frac{1}{2}$&$\frac{3}{2}^-$ & $2950 \pm 2$ &$\cdots$&$\cdots$&& $6249 \pm 3$ &$\cdots$&$\cdots$&\\
$^4\rho(\Xi_{Q})_{{1/2}}$&(1,0)&$\frac{1}{2}$&1&$\frac{3}{2}$&$\frac{1}{2}^-$ & $2950 \pm 2$ &$\cdots$&$\cdots$&& $6258 \pm 3$ &$\cdots$&$\cdots$&\\
$^4\rho(\Xi_{Q})_{{3/2}}$&(1,0)&$\frac{1}{2}$&1&$\frac{3}{2}$&$\frac{3}{2}^-$ & $2983 \pm 2$ &$\cdots$&$\cdots$&& $6265 \pm 3$ &$\cdots$&$\cdots$&\\
$^4\rho(\Xi_{Q})_{{5/2}}$&(1,0)&$\frac{1}{2}$&1&$\frac{3}{2}$&$\frac{5}{2}^-$ & $3038 \pm 2$ &$\cdots$&$\cdots$&& $6276 \pm 3$ &$\cdots$&$\cdots$&\\
$^2\lambda(\Xi_{Q})_{{1/2}}$&(0,1)&$\frac{1}{2}$&1&$\frac{1}{2}$&$\frac{1}{2}^-$ & $2777 \pm 2$ & $2792.9 \pm 0.4$ & $\Xi_{c}(2790)$&3& $6077 \pm 3$ &$\cdots$&$\cdots$&\\
$^2\lambda(\Xi_{Q})_{{3/2}}$&(0,1)&$\frac{1}{2}$&1&$\frac{1}{2}$&$\frac{3}{2}^-$ & $2810 \pm 2$ & $2818.2 \pm 0.2$ & $\Xi_{c}(2815)$&3& $6084 \pm 3$ & $6100.3 \pm 0.6$ & $\Xi_{b}(6100)$ & 3 \\
\noalign{\smallskip}
\end{tabular}
\end{ruledtabular}
\end{table*}

\subsection{$\Sigma_Q$ and $\Lambda_Q$ baryons}

\begin{figure*}
\centering
\rotatebox{0}{\scalebox{0.51}[0.51]{\includegraphics{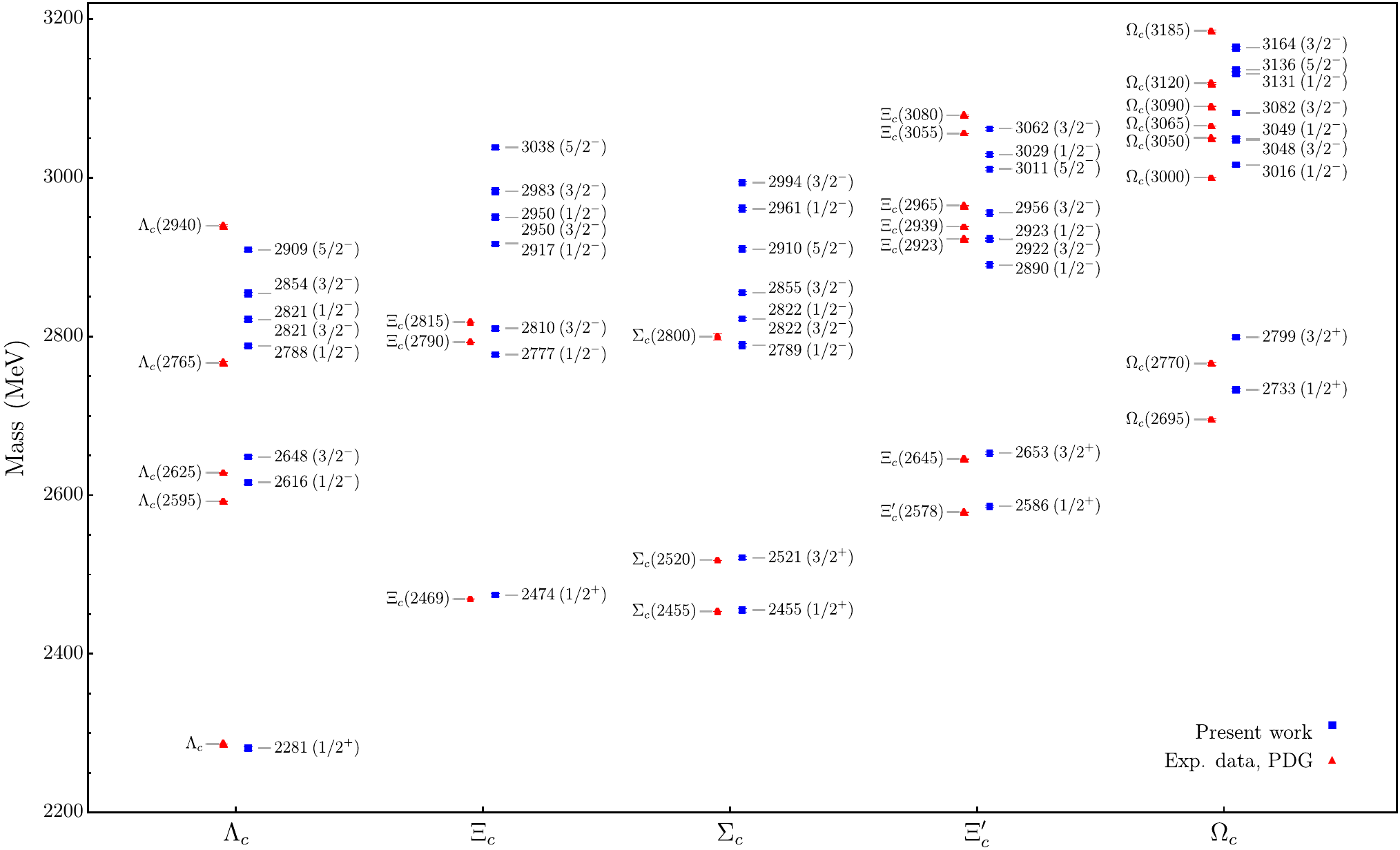}}} 
\caption[]{Comparison between experimental data (red triangles) and our quark model calculations (blue squares) of the mass spectra for the single charm baryons: $\Lambda_c$ and $\Xi_c$ from the baryon anti-triplet $\mathbf{\Bar{3}}$, and $\Sigma_c$,  $\Xi'_c$ and $\Omega_c$ from the baryon sextet $\mathbf{6}$ \cite{Workman:2022ynf}. The suggested assignment of $J^P$ quantum numbers for each state us given in parenthesis.} 
\label{charm}
\end{figure*}

We start by studying singly-heavy $nnQ$ baryons made up of two non-strange quarks with $n=u$, $d$ and one heavy quark with $Q=c$, $b$. The $nnQ$ baryons can be divided into an isospin triplet $\Sigma_Q$ belonging to the flavor sextet $\bf 6$ and an isospin singlet $\Lambda_Q$ of the flavor anti-triplet $\bf \bar{3}$, see Fig.~\ref{HeavyB}. 

The most recent PDG compilation \cite{Workman:2022ynf} shows that there are three known $\Sigma_c$ states as well as three $\Sigma_b$ states. In the charm sector, 
the $\Sigma_{c}(2455)$ and $\Sigma_{c}(2520)$ resonances \cite{YeltonPRD104} correspond to the ground states with $J^P=S^P=1/2^+$ and $3/2^+$, respectively, while $\Sigma_{c}(2800)$ is assigned as a $\lambda-$mode excitation with $J^P=1/2^-$. Similarly, the $\Sigma_b$ and $\Sigma^{*}_b$ resonances observed by LHCb \cite{PhysRevLett.122.012001} are assigned as ground state $S$-wave with $J^P=S^P=1/2^+$ and $3/2^+$, respectively, and $\Sigma_b(6097)$ as a $P$-wave excitation in the $\lambda$-mode with $J^P=1/2^-$.  

For the $\Lambda_Q$ states there is more experimental data available. In Table~\ref{massQ} we have omitted $\Lambda_c(2860)$, $\Lambda_c(2880)$, $\Lambda_b(6070)$ $\Lambda_b(6146)$ and $\Lambda_b(6152)$, since according to PDG~\cite{Workman:2022ynf} the most likely assignment of these resonances is as excited $2S$-wave or $D$-wave states. Here we focus on the five $\Lambda_c$ states and three $\Lambda_b$ states that can be assigned as ground-state $S$-wave or excited $P$-wave states. In the bottom sector, $\Lambda_b$ corresponds to the ground state $S$-wave with $J^P=S^P=1/2^+$, whereas the $\Lambda_b(5912)$ and $\Lambda_b(5920)$ resonances are assigned as a $\lambda$-mode excitation with $S=1/2$, and $J^P=1/2^-$ and $J^P=3/2^-$, respectively \cite{PhysRevLett.109.172003}. The corresponding states in the charm sector are $\Lambda_c$, $\Lambda_c(2595)$ and $\Lambda_c(2625)$. The $\Lambda_c^+$ spin was determined recently by the BESIII Collaboration to be $J^P=1/2^+$ \cite{AblikimPRD103}. In addition, one-star $\Lambda_c(2765)$ and three-star $\Lambda_c(2940)$ resonances are assigned as a $\rho$-mode excitation with $S=1/2$, $J^P=1/2^-$ and $S=3/2$, $J^P=5/2^-$.

The comparison between the calculated (blue squares) and experimental masses (red triangles) in Fig.~\ref{charm} for single charm baryons $\Sigma_c$ and $\Lambda_c$, and in Fig.~\ref{bottom} for single bottom baryons, $\Sigma_b$ and $\Lambda_b$, shows a good overall agreement with the available experimental information. 

\subsection{$\Xi_Q$ and $\Xi'_Q$ baryons}

Next we consider singly-heavy $nsQ$ baryons consisting of a non-strange quark with $n=u$, $d$, a strange quark $s$ and a heavy quark with $Q=c$, $b$. The $nsQ$ baryons are isospin doublets belonging to the flavor sextet $\bf 6$ or the anti-triplet $\bf \bar{3}$, called $\Xi'_Q$ and $\Xi_Q$, respectively. Since for $\Xi_Q$ and $\Xi'_Q$ baryons we distinguish between the mass of the two light quarks, $m_u=m_d \neq m_s$, the reduced masses are taken as $m_{\rho}=(m_{u/d}+m_s)/2$, {\it i.e} the average of the strange and nonstrange quark masses, and $m_{\lambda}=3 m_{\rho}m_Q/(2m_{\rho}+m_Q)$. The approximation to take the average of the strange and nonstrange quark masses is justified in the limit of $m_s-m_{u/d} \ll m_Q$. 

Over the last couple of years there has been a lot of progress in the understanding of charmed-strange baryons, $\Xi_Q$ and $\Xi^{'}_Q$. The LHCb Collaboration reported the discovery of three new resonances in the $\Lambda_c^+ K^-$ channel, $\Xi_c(2923)^0$, $\Xi_c(2939)^0$ and $\Xi_c(2965)^0$ \cite{PhysRevLett.124.222001}. In our previous analysis we established the  following assignment of these states $^4\lambda(\Xi'_{c})_{{1/2}^-}\rightarrow\Xi_c(2923)$, $^2\lambda(\Xi'_{c})_{{3/2}^-}\rightarrow\Xi_c(2939)$ and $^4\lambda(\Xi'_{c})_{{3/2}^-}\rightarrow\Xi_c(2965)$ \cite{PhysRevD.105.074029}.
The Belle Collaboration determined the spin and parity of the charmed-strange baryon, $\Xi_c(2970)^+$ to be $J^P=1/2^+$ \cite{MoonPRD103} and studied the electromagnetic decay of the excited charm baryons $\Xi_c(2790)$ and $\Xi_c(2815)$ \cite{PhysRevD.102.071103}. 

The ground-state $S$-wave $\Xi'_c$ and $\Xi_c$ baryons have all been identified. There are in total 7 $P$-wave states for the flavor sextet and another 7 for the flavor anti-triplet. Table~\ref{massQ} shows that for the flavor sextet 5 candidates have been assigned tentatively, only for the $^2\lambda(\Xi'_c)_{1/2}$ and $^4\lambda(\Xi'_c)_{5/2}$ states there are no obvious candidates. For the anti-triplet, the $\Xi_c(2790)$ and $\Xi_c(2815)$ baryons are assigned as 
the doublet $^2\lambda(\Xi_c)_J$ with $J^P=1/2^-$ and $3/2^-$, respectively. The latter assignment is based both on masses and electromagnetic decay widths \cite{PhysRevD.102.071103,PhysRevD.105.074029}. 

In the bottom sector, there is much less experimental information. The LHCb Collaboration has  observed the $\Xi^-_b(5935)$ and $\Xi^-_b(5954)$ resonances close to the threshold  $\Xi_b\pi$ \cite{PhysRevLett.114.062004}, which are assigned as the ground-state $S$-wave baryons of the flavor sextet with spin and parity $J^P=1/2^+$ and $3/2^+$, respectively. Also the ground-state $S$-wave $\Xi_b$ baryon of the flavor anti-triplet has been identified. The $\Xi_b(6227)$ baryon is assigned as the $^4\lambda(\Xi'_b)_{5/2}$ state of the flavor sextet. The $\Xi_b(6100)$ baryon which was recently observed by the CMS Collaboration in the $\Xi_b^-\pi^+\pi^-$ invariant mass spectrum \cite{PhysRevLett.126.252003}, is assigned as the $^2\lambda(\Xi_b)_{3/2}$ state of the flavor anti-triplet. 

\begin{figure*}
\centering
\rotatebox{0}{\scalebox{0.49}[0.49]{\includegraphics{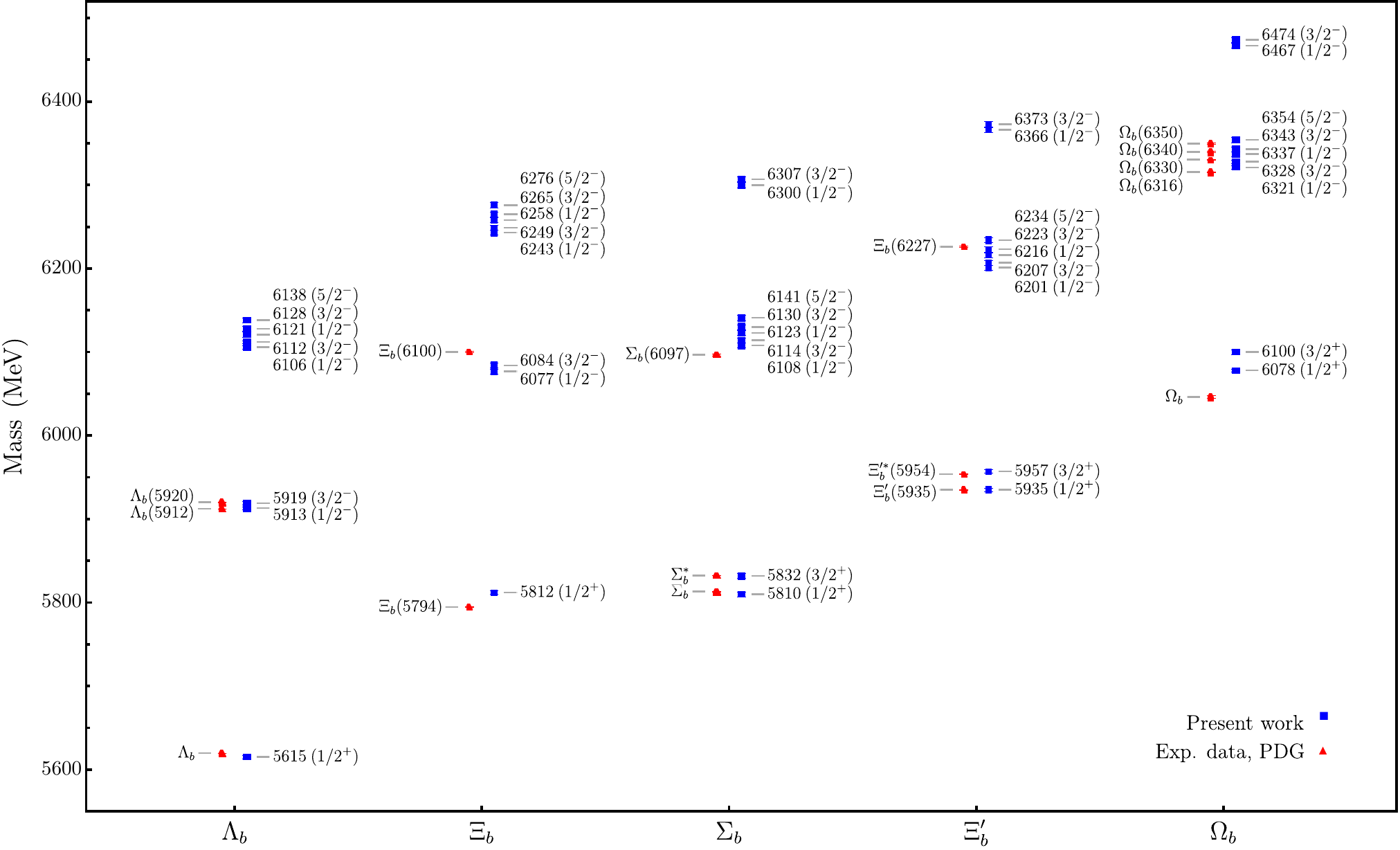}}} 
\caption[]{As Fig.~\ref{charm}, but for single bottom baryons.}
\label{bottom}
\end{figure*}

In Figs.~\ref{charm} and \ref{bottom} we present a comparison between our results for the $\Xi_Q$ and $\Xi'_Q$ mass spectrum and the experimental data. The conclusions of Ref.~\cite{PhysRevD.105.074029} remain valid. 

\subsection{$\Omega_Q$ baryons}

The LHCb Collaboration made the identification of five resonances $\Omega_c(3000)$, $\Omega_c(3050)$, $\Omega_c(3065)$, $\Omega_c(3090)$ and $\Omega_c(3120)$ \cite{PhysRevLett.118.182001,PhysRevD.97.051102}. At the same time, a further resonance was reported $\Omega_c(3188)$, but because the absence of enough statistic, they did not claim it as an authentic resonance. All these signals were discovered in the $\Xi_c^+K^-$ decay channel. One year later, the Belle Collaboration confirmed the observation of the first four resonances together with $\Omega_c(3188)$, however,  they do not observe the $\Omega_c(3119)$. Very recently the LHCb Collaboration observed two new excited states, $\Omega_c(3185)$ and $\Omega_c(3327)$ \cite{lhcbcollaboration2023observation2omegas}, whose interpretation and assignment of quantum numbers is still under discussion. The $\Omega_c(3327)$ resonance is most often interpreted as a $D$-wave state \cite{luo2023newly,feng2023description}, or alternatively as a $2S$ state \cite{karliner2023excited}. Our calculations suggest to assign the $\Omega_c(3185)$ resonance as a $P$-wave $^2\rho$-excited state with $J^P=3/2^-$.

The experimental knowledge on $\Omega_b$ has been expanded recently with the inclusion of more resonances to the spectra since the observation of 4 excited $\Omega_b^-$ states in the $\Xi_b^0K^-$ mass spectrum by the LHCb collaboration: $\Omega_b(6316)$, $\Omega_b(6330)$, $\Omega_b(6340)$ and $\Omega_b(6350)$ \cite{PhysRevLett.124.082002}. The masses and widths of these resonances were predicted in a quark model analysis of $\Omega_Q$ baryons before their experimental discovery \cite{Santopinto2019}. The present study is an extension of this earlier work as well as of a previous analysis of $\Xi_Q$ and $\Xi'_Q$ baryons \cite{PhysRevD.105.074029} in which now all singly-, doubly- and triply-heavy baryons of the type $B_{Q}$, $B_{QQ}$, $B_{QQQ}$ are included.  

In Table~\ref{massQ} we present the assignment of quantum numbers and compare our predictions for the mass spectra with the experimental data. Table~\ref{massQ} and Figs.~\ref{charm} and \ref{bottom} show a good agreement with the available experimental data. 

\begin{table*}
\centering
\caption{Mass spectra of double charm and double bottom $S$- and $P$-wave baryons. The states are labeled according to Eqs.~(\ref{fqq1}-\ref{fqq2}) and the quantum numbers $(n_\rho,n_\lambda)$, $I$, $L$, $S$ and $J^P$. The parity is given by $P=(-1)^L$. The experimental value is taken from \cite{Workman:2022ynf}.}
\label{massQQ}
\begin{ruledtabular}
\begin{tabular}{cccccc@{\hspace*{.86cm}}cccc@{\hspace*{.85cm}}rcc}
\noalign{\smallskip}
&&&&&
&\multicolumn{3}{c}{Double-charm baryons \: $Q=c$}  & & \multicolumn{3}{c}{Double-bottom baryons \: $Q=b$}\\ 
\noalign{\smallskip}
\cline{7-10} \cline{11-13}
\noalign{\smallskip}
State & $(n_\rho,n_\lambda)$&$I$&$L$&$S$&$J^{P}$&$M^{th} (\text{MeV})$&$M^{exp} (\text{MeV})$&Name  &  $\ast$ &$M^{th} (\text{MeV})$&$M^{exp} (\text{MeV})$&Name\\
\noalign{\smallskip}
\hline
\noalign{\smallskip}
$^2(\Xi_{QQ})_{{1/2}}$&(0,0)&$\frac{1}{2}$&0&$\frac{1}{2}$&$\frac{1}{2}^+$ & $3619 \pm 2$  & $3621.6 \pm 0.4$ &$\Xi_{cc}$&3& $10295 \pm 3$ &$\cdots$&$\Xi_{bb}$   \\
$^4(\Xi_{QQ})_{{3/2}}$&(0,0)&$\frac{1}{2}$&0&$\frac{3}{2}$&$\frac{3}{2}^+$ & $3686 \pm 2$ &$\cdots$& $\Xi^*_{cc}$&& $10317 \pm 4$ &$\cdots$& $\Xi^*_{bb}$ \\
$^2\rho(\Xi_{QQ})_{{1/2}}$&(1,0)&$\frac{1}{2}$&1&$\frac{1}{2}$&$\frac{1}{2}^-$ & $3823 \pm 2$ &$\cdots$&$\cdots$&& $10411 \pm 3$ &$\cdots$&$\cdots$\\
$^2\rho(\Xi_{QQ})_{{3/2}}$&(1,0)&$\frac{1}{2}$&1&$\frac{1}{2}$&$\frac{3}{2}^-$ & $3855 \pm 2$  &$\cdots$&$\cdots$&& $10417 \pm 3$ &$\cdots$&$\cdots$\\
$^2\lambda(\Xi_{QQ})_{{1/2}}$&(0,1)&$\frac{1}{2}$&1&$\frac{3}{2}$&$\frac{1}{2}^-$ & $4048 \pm 2$ &$\cdots$&$\cdots$&& $10700 \pm 4$ &$\cdots$&$\cdots$ \\
$^2\lambda(\Xi_{QQ})_{{3/2}}$&(0,1)&$\frac{1}{2}$&1&$\frac{1}{2}$&$\frac{3}{2}^-$ & $4081 \pm 2$  &$\cdots$ &$\cdots$&& $10707 \pm 4$ &$\cdots$&$\cdots$\\
$^4\lambda(\Xi_{QQ})_{{1/2}}$&(0,1)&$\frac{1}{2}$&1&$\frac{3}{2}$&$\frac{1}{2}^-$ & $4082 \pm 2$  &$\cdots$ &$\cdots$&& $10716 \pm 4$ &$\cdots$&$\cdots$ \\
$^4\lambda(\Xi_{QQ})_{{3/2}}$&(0,1)&$\frac{1}{2}$&1&$\frac{3}{2}$&$\frac{3}{2}^-$ & $4114 \pm 2$  &$\cdots$ &$\cdots$&& $10722 \pm 4$ &$\cdots$&$\cdots$ \\
$^4\lambda(\Xi_{QQ})_{{5/2}}$&(0,1)&$\frac{1}{2}$&1&$\frac{3}{2}$&$\frac{5}{2}^-$ & $4169 \pm 2$  &$\cdots$&$\cdots$&& $10733 \pm 4$ &$\cdots$&$\cdots$\\
\noalign{\smallskip}
\hline
\noalign{\smallskip}
$^2(\Omega_{QQ})_{{1/2}}$&(0,0)&0&0&$\frac{1}{2}$&$\frac{1}{2}^+$& $3766 \pm 2$ 
&$\cdots$& $\Omega_{cc}$ && $10438 \pm 3$ &$\cdots$& $\Omega_{bb}$\\
$^4(\Omega_{QQ})_{{3/2}}$&(0,0)&0&0&$\frac{3}{2}$&$\frac{3}{2}^+$ & $3833 \pm 2$ &$\cdots$&$\Omega^*_{cc}$ && $10460 \pm 4$ &$\cdots$&$\Omega^*_{bb}$ \\
$^2\rho(\Omega_{QQ})_{{1/2}}$&(1,0)&0&1&$\frac{1}{2}$&$\frac{1}{2}^-$ & $3970 \pm 2$ &$\cdots$&$\cdots$&& $10554 \pm 3$ &$\cdots$&$\cdots$\\
$^2\rho(\Omega_{QQ})_{{3/2}}$&(1,0)&0&1&$\frac{1}{2}$&$\frac{3}{2}^-$ & $4002 \pm 2$ &$\cdots$&$\cdots$&& $10560 \pm 3$ &$\cdots$&$\cdots$\\
$^2\lambda(\Omega_{QQ})_{{1/2}}$&(0,1)&0&1&$\frac{1}{2}$&$\frac{1}{2}^-$ & $4111 \pm 2$  &$\cdots$&$\cdots$&& $10762 \pm 4$ &$\cdots$&$\cdots$\\
$^2\lambda(\Omega_{QQ})_{{3/2}}$&(0,1)&0&1&$\frac{1}{2}$&$\frac{3}{2}^-$ & $4144 \pm 2$  &$\cdots$&$\cdots$&& $10768 \pm 4$ &$\cdots$&$\cdots$ \\
$^4\lambda(\Omega_{QQ})_{{1/2}}$&(0,1)&0&1&$\frac{3}{2}$&$\frac{1}{2}^-$ & $4145 \pm 2$ 
&$\cdots$ &$\cdots$&& $10778 \pm 4$ &$\cdots$&$\cdots$ \\
$^4\lambda(\Omega_{QQ})_{{3/2}}$&(0,1)&0&1&$\frac{3}{2}$&$\frac{3}{2}^-$ & $4177 \pm 2$ 
&$\cdots$ &$\cdots$&& $10784 \pm 4$ &$\cdots$&$\cdots$ \\
$^4\lambda(\Omega_{QQ})_{{5/2}}$&(0,1)&0&1&$\frac{3}{2}$&$\frac{5}{2}^-$ & $4232 \pm 2$ 
&$\cdots$ &$\cdots$&& $10795 \pm 4$ &$\cdots$&$\cdots$ \\
\noalign{\smallskip}
\end{tabular}
\end{ruledtabular}
\end{table*}

\subsection{$\Xi_{QQ}$ and $\Omega_{QQ}$ baryons}

In this section, we extend the previous discussion of single charm and single bottom baryons to doubly-heavy $\Xi_{QQ}$ and $\Omega_{QQ}$ baryons using the mass formula of Eq.~(\ref{mass_formula}). The only doubly-heavy baryon resonance observed by the LHCb Collaboration \cite{PhysRevLett.119.112001}, $\Xi^{++}_{cc}$, is assigned as the $^2(\Xi_{cc})_{1/2}$ ground state with $J^P=1/2^+$. The calculated mass of $3619 \pm 2$ MeV is in excellent agreement with the experimental value of $3621.6 \pm 0.4$ MeV as is shown in Table~\ref{massQQ}. The analogous state in the bottom sector associated to $^2(\Xi_{bb})_{1/2}$ has a calculated mass of $10295 \pm 3$ MeV. The $J^P=3/2^+$ counterparts of these baryons, $\Xi^*_{cc}$ and $\Xi^*_{bb}$, are assigned as $^4(\Xi_{cc})_{3/2}$ and $^4(\Xi_{bb})_{3/2}$. The corresponding $\Omega_{QQ}$ ground states assignments in our notation are $^2(\Omega_{cc})_{1/2}$, $^2(\Omega_{bb})_{1/2}$, $^4(\Omega_{cc})_{3/2}$ and $^4(\Omega_{bb})_{3/2}$. In addition to the $S$-wave ground state baryons, we also present the results for the $P$-wave baryons with one quantum of excitation in the $\rho$ and $\lambda$ coordinates in Table~\ref{massQQ} and Figs.~\ref{spcc} and \ref{spbb}. 

The masses of the doubly-heavy baryons are comparable to the values reported in a quark model descriptions by Yoshida {\it et al.} \cite{PhysRevD.92.114029} and Karliner and Rosner \cite{PhysRevD.90.094007}, Regge phenomenology 
\cite{PhysRevD.95.116005}, heavy diquark-light quark model \cite{Song2023} and lattice QCD calculations \cite{Workman:2022ynf,PhysRevD.90.094507}. In Table~\ref{heavybaryons} we present a comparison of the present results for the masses of ground state baryons with those obtained in constituent quark models \cite{PhysRevD.92.114029,PhysRevD.90.094007} and lattice QCD calculations \cite{PhysRevD.90.094507}. In general, there is a good agreement between the three different calculations and the available experimental data. For double bottom baryons there is no experimental information. For these baryons our values as well as the CQM results of \cite{PhysRevD.92.114029} are about 150 MeV higher than the lattice QCD calculations \cite{PhysRevD.90.094507}.

Finally, we present the equal-spacing mass rules for the doubly-heavy baryons. For the $\lambda$-mode excited baryons one has 
\ba
&& M(^{2S+1}\lambda(\Omega_{QQ})_J)-M(^{2S+1}\lambda(\Xi_{QQ})_J) 
\nonumber\\
&& \qquad = m_s-m_{u/d} + \omega_\lambda(\Omega_{QQ})-\omega_\lambda(\Xi_{QQ})-\frac{3}{4}E 
\nonumber\\
&& \qquad = \left\{ \begin{array}{ccc} 63 \mbox{ MeV} && (Q=c) \\ 62 \mbox{ MeV} && (Q=b) \end{array} \right.
\label{esr6}
\ea
For the $\rho$-mode excitations the frequencies are the same, $\omega_\rho(\Omega_{QQ})=\omega_\rho(\Xi_{QQ})$, 
and the equal-spacing mass rule reduces to
\ba
&& M(^{2}\rho(\Omega_{QQ})_J)-M(^{2}\rho(\Xi_{QQ})_J) 
\nonumber\\
&& \qquad = m_s-m_{u/d} -\frac{3}{4}E 
\nonumber\\
&& \qquad = \left\{ \begin{array}{ccc} 147 \mbox{ MeV} && (Q=c) \\ 143 \mbox{ MeV} && (Q=b) \end{array} \right.
\label{esr7}
\ea

\subsection{$\Omega_{QQQ}$ baryons}

Even though triply-heavy baryons, $\Omega_{ccc}$ and $\Omega_{bbb}$, have not yet been observed experimentally, there are many ongoing efforts to study these states and understand their implications for hadron structure and the strong interaction. According to the quark model classification the ground state of these baryons has spin and parity $J^P=3/2^+$, and is labeled as $^4(\Omega_{ccc})_{3/2}$ and $^4(\Omega_{bbb})_{3/2}$. Just as for three identical light quarks, for three identical heavy quarks the $\rho$- and $\lambda$-modes become degenerate. The wave function of triply-heavy $P$-wave baryons is given by Eq.~(\ref{fqqq}), a single state $^2E(\Omega_{QQQ})_J$ with $J^p=1/2^-$ and $3/2^-$.

\begin{figure}
\centering
\rotatebox{0}{\scalebox{0.43}[0.43]{\includegraphics{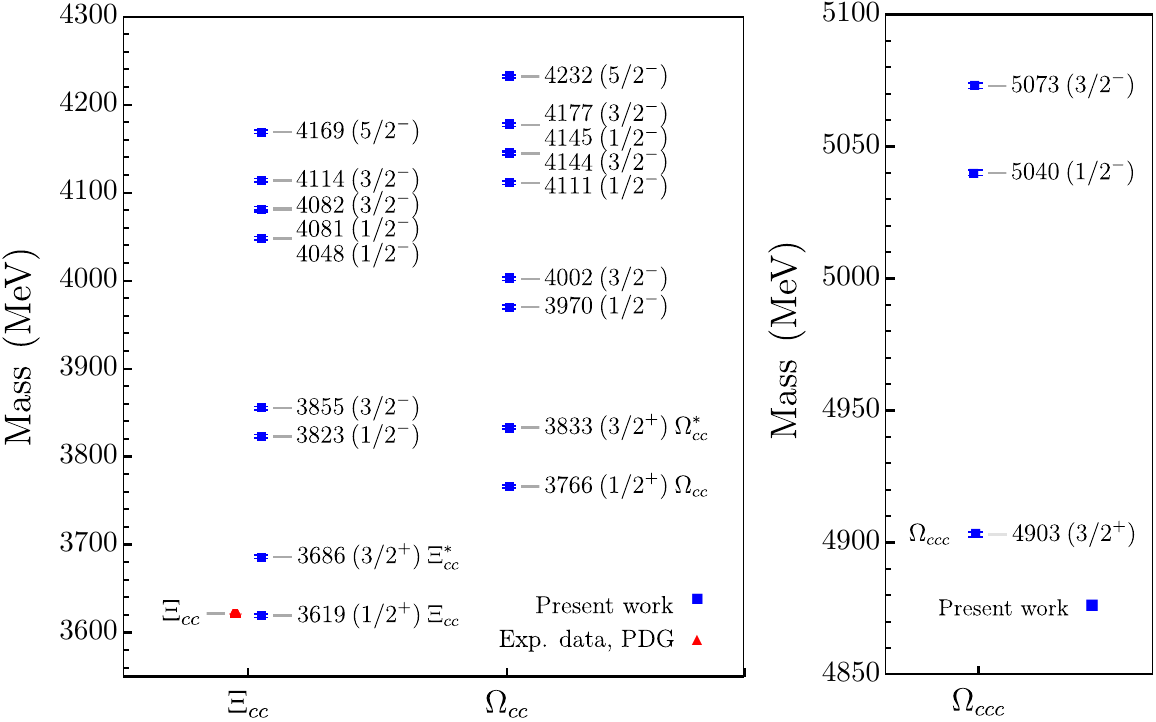}}} 
\caption[]{Comparison between the experimental data (red triangles) and our quark model calculations (blue squares) of the mass spectra for the double charm baryons, $\Xi_{cc}$ and $\Omega_{cc}$, and triple charm baryons, $\Omega_{ccc}$.} 
\label{spcc}
\end{figure}

\begin{figure}
\centering
\rotatebox{0}{\scalebox{0.43}[0.43]{\includegraphics{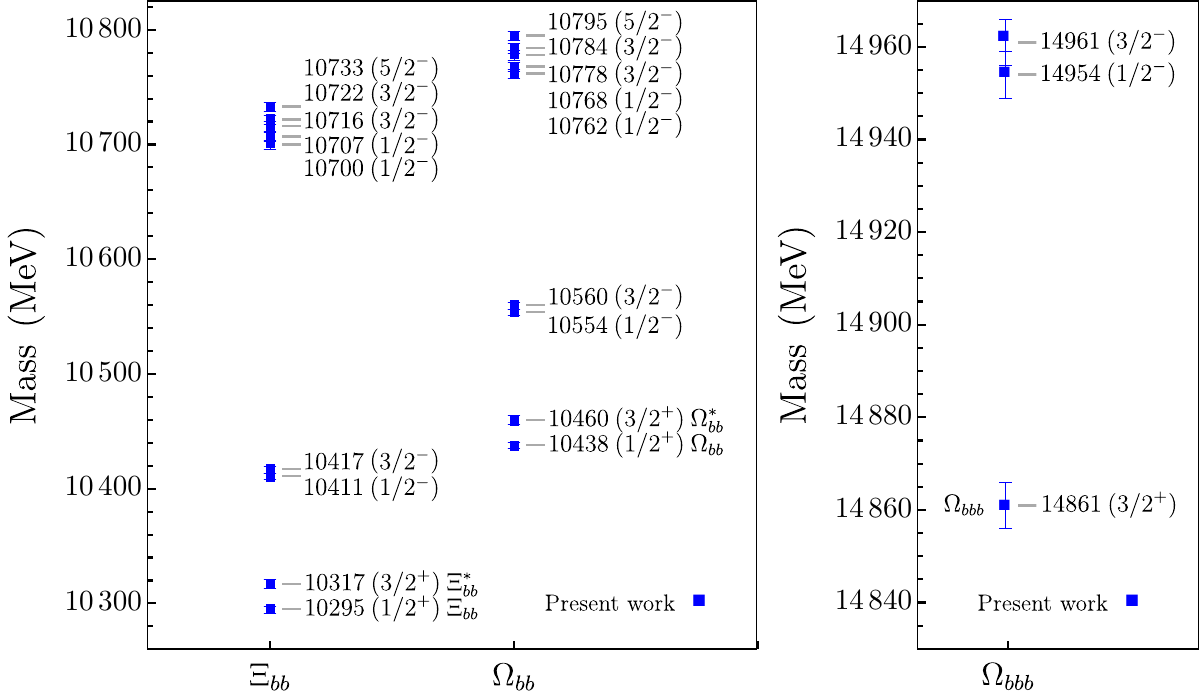}}} 
\caption[]{As Fig.~\ref{spcc}, but for double and triple bottom baryons.}
\label{spbb}
\end{figure}

The mass spectrum of triply-heavy $S$- and $P$-wave baryons consists of three states with angular momentum $J^p=3/2^+$, $1/2^-$ and $3/2^-$. The results are presented in Table~\ref{massQQQ} and the right panel of Figs.~\ref{spcc} and \ref{spbb}. Table~\ref{massQQQ} shows that the splitting between the masses of $S$- and $P$-wave baryons is about 150 MeV for $\Omega_{ccc}$ and 100 MeV for $\Omega_{bbb}$, in comparison with values of more than 300 MeV in a recent calculation in a non-relativistic constituent quark model \cite{PhysRevD.101.074031}.

The PDG review of lattice calculations of the mass of the ground state $\Omega_{ccc}$ baryon shows a range of 4700 - 4800 MeV \cite{Workman:2022ynf}. The same approximate range was found in recent reviews of theoretical values in both model and lattice calculations \cite{PhysRevD.101.074031,doi:10.1142/S0217751X22502256}. Our calculation yields a slightly higher value of 4903 MeV which is about 100 MeV above the result of the lattice calculation of Brown {\it et al.} \cite{PhysRevD.90.094507} (see Table~\ref{heavybaryons}). 

Similarly, recent reviews of the ground state mass of the triple bottom $\Omega_{bbb}$ baryon show a range of 14000 - 15000 MeV \cite{PhysRevD.95.116005,PhysRevD.101.074031,doi:10.1142/S0217751X22502256}. We calculate the ground state mass to be 14861 MeV which is within the this range, and about 500 MeV higher than the lattice calculation of Brown {\it et al.} \cite{PhysRevD.90.094507} (see Table~\ref{heavybaryons}). 

\begin{table*}
\centering
\caption{Mass spectra of triple charm and triple bottom $S$- and $P$-wave baryons. The states are labeled according to Eq.~(\ref{fqqq}) and the quantum numbers $(n_\rho,n_\lambda)$, $I$, $L$, $S$ and $J^P$. 
The parity is given by $P=(-1)^L$.}
\label{massQQQ}
\begin{ruledtabular}
\begin{tabular}{cccccc@{\hspace*{.86cm}}cccc@{\hspace*{.85cm}}rcc}
\noalign{\smallskip}
&&&&&&
\multicolumn{3}{c}{Triple-charm baryons \: $Q=c$} && \multicolumn{3}{c}{Triple-bottom baryons \: $Q=b$} \\ 
\noalign{\smallskip}
\cline{7-9} \cline{11-13}
\noalign{\smallskip}
State & $(n_\rho,n_\lambda)$ & $I$ & $L$ & $S$ & $J^P$ 
& $M^{th}$ (MeV) & NRCQM \cite{PhysRevD.101.074031} & Name &
& $M^{th}$ (MeV) & NRCQM \cite{PhysRevD.101.074031} & Name \\
\noalign{\smallskip}
\hline
\noalign{\smallskip}
 $^4(\Omega_{QQQ})_{{3/2}}$ & (0,0) & 0 & 0 & $\frac{3}{2}$ & $\frac{3}{2}^+$ & $4903 \pm 3$ & 4828 & $\Omega_{ccc}$ && $14861 \pm 5$ & 14432 &$\Omega_{bbb}$\\
$^2E(\Omega_{QQQ})_{{1/2}}$ & (1,0) & 0 & 1 & $\frac{1}{2}$ & $\frac{1}{2}^-$ & $5040 \pm 3$ 
& 5142 & $\cdots$ && $14954 \pm 5$ & 14773 & $\cdots$ \\
$^2E(\Omega_{QQQ})_{{3/2}}$ & (1,0) & 0 & 1 & $\frac{1}{2}$ & $\frac{3}{2}^-$ & $5073 \pm 3$ 
& 5162 & $\cdots$ && $14961 \pm 5$ & 14779 & $\cdots$ \\
\noalign{\smallskip}
\end{tabular}
\end{ruledtabular}
\end{table*}

\begin{table}
\centering
\caption{Mass spectra of single, double and triple charm and bottom ground-state $S$-wave baryons. The first uncertainty in the lattice calculations is statistical and the second systematic.}
\label{heavybaryons}
\begin{ruledtabular}
\begin{tabular}{lccccc}
\noalign{\smallskip}
State & QM \cite{PhysRevD.92.114029} & KR \cite{PhysRevD.90.094007} & LQCD \cite{PhysRevD.90.094507} & Present & Exp. \\
\noalign{\smallskip}
\hline
\noalign{\smallskip}
$\Lambda_c$        & 2285 & 2286.5   & 2254(48)(31) &  2281 & 2286.5(1) \\
$\Sigma_c$         & 2460 & 2444.0   & 2474(41)(25) &  2455 & 2453.5(1) \\
$\Sigma^\ast_c$    & 2523 & 2507.7   & 2551(43)(25) &  2521 & 2518.1(2) \\
$\Xi_c$            &      & 2475.3   & 2433(35)(30) &  2474 & 2469.1(2) \\
$\Xi'_c$           &      & 2565.4   & 2574(37)(23) &  2586 & 2578.5(4) \\
$\Xi^\ast_c$       &      & 2632.6   & 2648(38)(25) &  2653 & 2645.6(2) \\
$\Omega_c$         & 2731 & 2692.1   & 2679(37)(20) &  2733 & 2695.2(17) \\
$\Omega^\ast_c$    & 2779 & 2762.8   & 2755(37)(24) &  2799 & 2765.9(20) \\
$\Xi_{cc}$         & 3685 & 3627(12) & 3610(23)(22) &  3619 & 3621.6(4) \\
$\Xi^\ast_{cc}$    & 3754 & 3690(12) & 3692(28)(21) &  3686 & \\
$\Omega_{cc}$      & 3832 &          & 3738(20)(20) &  3766 & \\
$\Omega^\ast_{cc}$ & 3883 &          & 3822(20)(22) &  3833 & \\
$\Omega_{ccc}$     &      &          & 4796(8)(18)  &  4903 & \\ 
\noalign{\smallskip}
\hline
\noalign{\smallskip}
$\Lambda_b$        &  5618 &  5619.4   &  5626(52)(29) &  5615 & 5619.6(2) \\
$\Sigma_b$         &  5823 &  5805.1   &  5856(56)(27) &  5810 & 5813.1(2) \\
$\Sigma^\ast_b$    &  5845 &  5826.7   &  5877(55)(27) &  5832 & 5832.5(2) \\
$\Xi_b$            &       &  5801.5   &  5771(41)(24) &  5812 & 5794.5(4) \\
$\Xi'_b$           &       &  5921.3   &  5933(47)(24) &  5935 & 5935.0(1) \\
$\Xi^\ast_b$       &       &  5944.1   &  5960(47)(25) &  5957 & 5953.8(3) \\
$\Omega_b$         &  6076 &  6042.8   &  6056(47)(20) &  6078 & 6045.2(12) \\
$\Omega^\ast_b$    &  6094 &  6066.7   &  6085(47)(20) &  6100 & \\
$\Xi_{bb}$         & 10314 & 10162(12) & 10143(30)(23) & 10295 & \\
$\Xi^\ast_{bb}$    & 10339 & 10184(12) & 10178(30)(24) & 10317 & \\
$\Omega_{bb}$      & 10447 &           & 10273(27)(20) & 10438 & \\
$\Omega^\ast_{bb}$ & 10467 &           & 10308(27)(21) & 10460 & \\
$\Omega_{bbb}$     &       &           & 14366( 9)(20) & 14861 & \\
\noalign{\smallskip}
\end{tabular}
\end{ruledtabular}
\end{table}

\section{Electromagnetic couplings}
\label{sec:em}

Electromagnetic couplings provide an important tool to investigate the properties of baryons since they are far more sensitive to wave functions (and models) than masses. Recently, the Belle II Collaboration reported the first measurement of radiative decay widths of charmed baryons. It was found that the widths of the neutral $\Xi_c(2790)^0$ and $\Xi_c(2815)^0$ baryons are large, whereas for the widths of the charged $\Xi_c(2790)^+$ and $\Xi_c(2815)^+$ baryons only an upper limit was established \cite{PhysRevD.102.071103}. In the following we analyze the radiative decay widths of heavy baryons in the harmonic oscillator quark model. 

The Hamiltonian for electromagnetic couplings is given by ~\cite{LeYaouanc:1988fx}
\begin{equation}
H=e\int d^3x \hat J^\mu(\vec x) A_\mu (\vec x) ,
\end{equation}
where $A_\mu(\vec x)$ is the electromagnetic field and $J^\mu(\vec x)$ is the quark current \begin{equation}
\hat J^\mu(\vec x)=\sum_q e_q \bar q(\vec x) \gamma^\mu q(\vec x) ,
\end{equation}
where the sum runs over all quark flavors. For each flavor $q$, there is a quark field $q(\vec x)$ (Dirac spinor field) and $e_q$ is the corresponding electromagnetic charge. 

\begin{figure}[t]
\centering
\rotatebox{0}{\scalebox{0.31}[0.3]{\includegraphics{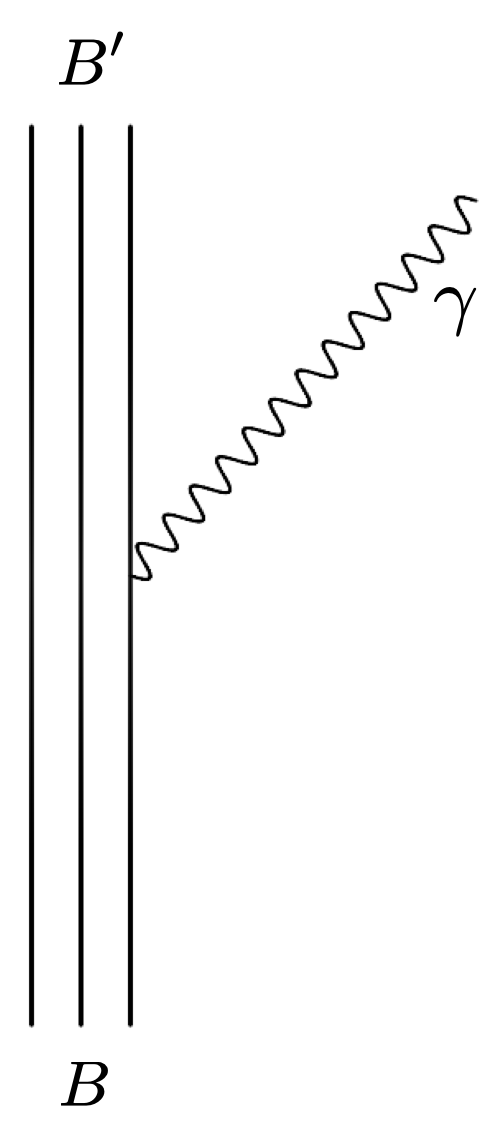}}} 
\caption[]{Photon emission from a heavy baryon in the radiative decay $B \rightarrow B' + \gamma$. } 
\label{BBg}
\end{figure}

The electromagnetic decay of an initial baryon $B$ to a final baryon $B'$ is described by the emission of a left-handed photon, $B \rightarrow B'+ \gamma$ (see Fig.~\ref{BBg}). The final baryon is a ground state baryon without radial excitation. 
The  non-relativistic interaction Hamiltonian for the electromagnetic couplings for the emission of a left-handed photon is given by \cite{Close:1979bt,Bijker:1994yr}
\ba
\mathcal{H}_{\rm em} &=& \sum_{j=1}^{3} \mathcal{H}_{{\rm em},j} = 2\sqrt{\frac{\pi}{k_0}} \sum_{j=1}^{3} \mu_j \left[ks_{j,-} \hat{U}_j 
- \frac{1}{g} \hat{T}_{j,-} \right] ~,
\nonumber\\
\hat{U}_j &=& e^{-i\vec{k}\cdot\vec{r}_j} ~,
\nonumber\\ 
\hat{T}_{j,-} &=& \frac{1}{2} \left( p_{j,-} e^{-i\vec{k}\cdot\vec{r}_j} 
+ e^{-i\vec{k}\cdot\vec{r}_j} p_{j,-} \right) ~, 
\label{HED1}
\ea
where $\vec{r}_j$, $\vec{p}_{j}$,  $\vec{s}_j$ and $\mu_j$ are the coordinate, momentum, spin and magnetic moment of the $j$th constituent quark, respectively. The photon energy is denoted by $k_0$ and $\vec{k}=k\hat z$ is the momentum carried by the emitted photon along the quantization axis, $z$. The coefficient $g$ is related to the quark scale magnetic moment, $g=1$. 

The electromagnetic decay widths of a given baryon is expressed in terms of the helicity amplictudes which in turn correspond to the transition matrix elements of the interaction Hamiltonian of Eq.~(\ref{HED1})
\ba
\mathcal{A}_\nu = \langle \psi_{B'},J',\nu-1 | \mathcal{H}_{\rm em} | \psi_{B},J,\nu \rangle,
\ea
where $\nu$ is the helicity quantum number. For three identical quarks, the contribution of each one of the quarks is the same and one can write
\ba
\mathcal{A}_\nu = \langle \psi_{B'},J',\nu-1 | 3\mathcal{H}_{{\rm em},3} | \psi_{B},J,\nu \rangle,
\ea
whereas for two identical quarks which are different from the third one has
\ba
\mathcal{A}_\nu = \langle \psi_{B'},J',\nu-1 | 2\mathcal{H}_{{\rm em},1}+\mathcal{H}_{{\rm em},3} | \psi_{B},J,\nu \rangle.
\ea
In order to calculate the helicity amplitudes of the baryons, we expand the coupled basis $|(n_\rho l_\rho)(n_\lambda l_\lambda)L,S;JM\rangle$ into the uncoupled basis 
\ba
| (L,S)J,M \rangle = \sum_{M_L M_S}\langle L,M_L,S,M_S |J,M \rangle |LM_L;SM_S \rangle,
\nonumber\\ 
\ea 
and use that the final state is a ground state baryon with $L'=0$ and $J'=S'$. The helicity amplitude can be re-expressed in the following way 
\begin{widetext}
\ba
\mathcal{A}_\nu(k) &=& \sum_{M_L M_S} \langle L,M_L,S,M_S | J,\nu \rangle
\langle \psi_{B'},0,0;S'=J',M_{S'}=\nu-1 | \mathcal{H}_{em} | \psi_{B},L,M_L;S,M_S \rangle 
\nonumber\\
&=& 2\sqrt{\frac{\pi}{k_0}} \sum_{j=1}^3 \left\lbrace k \langle L,0;S,\nu | J,\nu \rangle 
\langle \psi_{B'},0,0;S',\nu-1 | \mu_j s_{j,-} \hat{U}_j | \psi_{B},L,0;S,\nu \rangle \right. 
\nonumber\\ 
&& \left. - \frac{1}{g} \langle L,1;S,\nu-1 | J,\nu \rangle 
\langle \psi_{B'},0,0;S',\nu-1 | \mu_j \hat{T}_{j,-} 
| \psi_{B},L,1;S,\nu-1 \rangle \right\rbrace ~. 
\label{helamp}  
\ea
\end{widetext}

\subsection{Radial integrals}

The helicity amplitudes for the emission of a left-handed photon can be expressed 
in terms of the radial integrals 
\ba
U_{i,j} &=& \left< \psi_{0} \left| \hat{U}_j \right| \psi_{i} \right> ,
\nonumber\\
T_{i,j} &=& \left< \psi_{0} \left| \hat{T}_{j,-} \right| \psi_{i} \right> ,
\ea
with $i=0$ for ground-state $S$-wave baryons and $i=\rho$, $\lambda$ 
for excited-state $P$-wave baryons. The radial integrals can be evaluated by using Jacobi coordinates and their conjugate momenta of Eq.~(\ref{jacobi}). 
For ground state baryons one has  
\ba
U_{0,1} = U_{0,2} &=& \mbox{e}^{-k^2/8\alpha_{\rho}^2} \, 
\mbox{e}^{-(\frac{3m'}{2m+m'} k)^2/24\alpha_{\lambda}^2} ,
\nonumber\\ 
U_{0,3} &=& \mbox{e}^{-(\frac{3m}{2m+m'} k)^2/6\alpha_{\lambda}^2} ,
\nonumber\\
T_{0,1} = T_{0,2} &=& T_{0,3} = 0 .
\ea
For the radially excited states one finds
\ba
U_{\rho,1} &=& - U_{\rho,2} 
\nonumber\\
&=& -i \frac{k}{2\alpha_{\rho}} \, 
\mbox{e}^{-k^2/8\alpha_{\rho}^2} \, 
\mbox{e}^{-(\frac{3m'}{2m+m'} k)^2/24\alpha_{\lambda}^2} ,
\nonumber\\
U_{\rho,3} &=& 0 ,
\nonumber\\
U_{\lambda,1} &=& U_{\lambda,2} 
\nonumber\\
&=& -i \frac{k}{2\alpha_{\lambda}\sqrt{3}} \, \frac{3m'}{2m+m'} 
\, \mbox{e}^{-k^2/8\alpha_{\rho}^2} \, 
\mbox{e}^{-(\frac{3m'}{2m+m'} k)^2/24\alpha_{\lambda}^2} ,
\nonumber\\
U_{\lambda,3} &=& i \frac{k}{\alpha_{\lambda}\sqrt{3}} \, \frac{3m}{2m+m'} \, 
\mbox{e}^{-(\frac{3m}{2m+m'} k)^2/6\alpha_{\lambda}^2} ,
\ea
and
\ba
T_{\rho,1} &=& - T_{\rho,2} 
\nonumber\\
&=& i\frac{mk_0}{\alpha_{\rho}\sqrt{2}} \, 
\mbox{e}^{-k^2/8\alpha_{\rho}^2} \, 
\mbox{e}^{-(\frac{3m'}{2m+m'} k)^2/24\alpha_{\lambda}^2} ,
\nonumber\\
T_{\rho,3} &=& 0 ,
\nonumber\\
T_{\lambda,1} &=& T_{\lambda,2} 
\nonumber\\
&=& i\frac{mk_0}{\alpha_{\lambda}\sqrt{6}} \, \frac{3m'}{2m+m'} \, 
\mbox{e}^{-k^2/8\alpha_{\rho}^2} \, 
\mbox{e}^{-(\frac{3m'}{2m+m'} k)^2/24\alpha_{\lambda}^2} ,
\nonumber\\
T_{\lambda,3} &=& -i\frac{m'k_0\sqrt{2}}{\alpha_{\lambda}\sqrt{3}} \, \frac{3m}{2m+m'} \, 
\mbox{e}^{-(\frac{3m}{2m+m'} k)^2/6\alpha_{\lambda}^2} .
\label{Tj-}
\ea
Here $\alpha_\rho$ and $\alpha_\lambda$ are the harmonic oscillator size parameters of Table~\ref{alphas}. Their values can be calculated by using the quark masses of Table~\ref{cqm} and the string constants $C$ of Table~\ref{par}, and are summarized in Table~\ref{alphas}.

\begin{table}
\centering
\caption{Values of oscillator size parameters $\alpha_\rho$ and $\alpha_\lambda$ in MeV calculated using Eqs.~(\ref{size}) and (\ref{alpl}) for heavy baryons with $Q=c$ and $Q=b$.}
\label{alphas}
\begin{ruledtabular}
\begin{tabular}{ccccc}
\noalign{\smallskip}
& \multicolumn{2}{c}{$Q=c$} & \multicolumn{2}{c}{$Q=b$} \\
\noalign{\smallskip}
\cline{2-3} \cline{4-5}	 
\noalign{\smallskip}
& $\alpha_\rho$ & $\alpha_\lambda$ & $\alpha_\rho$ & $\alpha_\lambda$ \\
\noalign{\smallskip}
\hline
\noalign{\smallskip}
$\Sigma_Q$, $\Lambda_Q$ & 392 & 478 & 380 & 486 \\
$\Xi_Q$, $\Xi'_Q$       & 418 & 500 & 405 & 514 \\
$\Omega_Q$              & 440 & 517 & 426 & 537 \\
$\Xi_{QQ}$              & 601 & 425 & 771 & 417 \\
$\Omega_{QQ}$           & 601 & 471 & 771 & 466 \\
$\Omega_{QQQ}$          & 601 & 601 & 771 & 771 \\
\noalign{\smallskip}
\end{tabular}
\end{ruledtabular}
\end{table}

\subsection{Spin-flavor matrix elements}

The spin-flavor matrix contribution to the helicity amplitudes can be evaluated in a straightforward manner. In Table~\ref{sf66}, we present the results for the spin-flip and nonvanishing orbit-flip amplitudes for the decays $\Sigma_Q(uuQ) \rightarrow {^{2}\Sigma_Q}(uuQ) + \gamma$ and 
${^{4}\Sigma_Q}(uuQ) + \gamma$. 
The contributions 
for the other members of the flavor sextet can be obtained by the replacements
\ba
\begin{array}{ccccc}
\mu_u &\rightarrow& \frac{1}{2}(\mu_u + \mu_d) &:& \Sigma_Q(udQ) \\
\mu_u &\rightarrow& \mu_d &:& \Sigma_Q(ddQ) \\
\mu_u &\rightarrow& \frac{1}{2}(\mu_u + \mu_s) &:& \Xi'_Q(usQ) \\
\mu_u &\rightarrow& \frac{1}{2}(\mu_d + \mu_s) &:& \Xi'_Q(dsQ) \\
\mu_u &\rightarrow& \mu_s &:& \Omega_Q(ssQ)
\end{array} 
\label{ec66}
\ea

\begin{table}
\centering
\caption{Spin-flip and nonvanishing orbit-flip amplitudes for the decays  
$\Sigma_Q(uuQ) \rightarrow {^{2}\Sigma_Q}(uuQ) + \gamma$ and 
${^{4}\Sigma_Q}(uuQ) + \gamma$ with helicities $\nu=1/2$ and $3/2$.}
\label{sf66}
\begin{ruledtabular}
\begin{tabular}{ccccc}
\noalign{\smallskip}
Spin-flip & $\nu$ & $\left< \mu_1 s_{1,-} \right>$ & $\left< \mu_2 s_{2,-} \right>$ 
& $\left< \mu_3 s_{3,-} \right>$ \\
\noalign{\smallskip}
\hline
\noalign{\smallskip}
$^{4}\Sigma_Q \rightarrow {^{2}\Sigma_Q}$ & $\frac{1}{2}$ & $-\frac{1}{3\sqrt{2}} \mu_u$ 
& $-\frac{1}{3\sqrt{2}} \mu_u$ & $\frac{\sqrt{2}}{3} \mu_Q$ \\
\noalign{\smallskip}
& $\frac{3}{2}$ & $-\frac{1}{\sqrt{6}} \mu_u$ 
& $-\frac{1}{\sqrt{6}} \mu_u$ & $\frac{\sqrt{2}}{\sqrt{3}} \mu_Q$ \\
\noalign{\smallskip}
$^{2}\rho_J(\Sigma_Q) \rightarrow {^{2}\Sigma_Q}$ & $\frac{1}{2}$ & $-\frac{1}{\sqrt{3}} \mu_u$ 
& $\frac{1}{\sqrt{3}} \mu_u$ & 0 \\
\noalign{\smallskip}
$^{2}\lambda_J(\Sigma_Q) \rightarrow {^{2}\Sigma_Q}$ & $\frac{1}{2}$ & $\frac{2}{3} \mu_u$ 
& $\frac{2}{3} \mu_u$ & $-\frac{1}{3} \mu_Q$ \\
\noalign{\smallskip}
$^{4}\lambda_J(\Sigma_Q) \rightarrow {^{2}\Sigma_Q}$ & $\frac{1}{2}$ & $-\frac{1}{3\sqrt{2}} \mu_u$ 
& $-\frac{1}{3\sqrt{2}} \mu_u$ & $\frac{\sqrt{2}}{3} \mu_Q$ \\
\noalign{\smallskip}
& $\frac{3}{2}$ & $-\frac{1}{\sqrt{6}} \mu_u$ 
& $-\frac{1}{\sqrt{6}} \mu_u$ & $\frac{\sqrt{2}}{\sqrt{3}} \mu_Q$ \\
\noalign{\smallskip}
$^{2}\rho_J(\Sigma_Q) \rightarrow {^{4}\Sigma_Q}$ & $\frac{1}{2}$ & $\frac{1}{\sqrt{6}} \mu_u$ 
& $-\frac{1}{\sqrt{6}} \mu_u$ & 0 \\
\noalign{\smallskip}
$^{2}\lambda_J(\Sigma_Q) \rightarrow {^{4}\Sigma_Q}$ & $\frac{1}{2}$ & $\frac{1}{3\sqrt{2}} \mu_u$ 
& $\frac{1}{3\sqrt{2}} \mu_u$ & $-\frac{\sqrt{2}}{3} \mu_Q$ \\
\noalign{\smallskip}
$^{4}\lambda_J(\Sigma_Q) \rightarrow {^{4}\Sigma_Q}$ & $\frac{1}{2}$ & $\frac{2}{3} \mu_u$ 
& $\frac{2}{3} \mu_u$ & $\frac{2}{3} \mu_Q$ \\
\noalign{\smallskip}
& $\frac{3}{2}$ & $\frac{1}{\sqrt{3}} \mu_u$ 
& $\frac{1}{\sqrt{3}} \mu_u$ & $\frac{1}{\sqrt{3}} \mu_Q$ \\
\noalign{\smallskip}
\hline
\noalign{\smallskip}
Orbit-flip & $\nu$ & $\left< \mu_1 \right>$ & $\left< \mu_2 \right>$ 
& $\left< \mu_3 \right>$ \\
\noalign{\smallskip}
\hline
\noalign{\smallskip}
$^{2}\lambda_J(\Sigma_Q) \rightarrow {^{2}\Sigma_Q}$ & $\frac{1}{2}$ 
& $\mu_u$ & $\mu_u$ & $\mu_Q$ \\
\noalign{\smallskip}
& $\frac{3}{2}$ & $\mu_u$ & $\mu_u$ & $\mu_Q$ \\
\noalign{\smallskip}
$^{4}\lambda_J(\Sigma_Q) \rightarrow {^{4}\Sigma_Q}$ & $\frac{1}{2}$ 
& $\mu_u$ & $\mu_u$ & $\mu_Q$ \\
\noalign{\smallskip}
& $\frac{3}{2}$ & $\mu_u$ & $\mu_u$ & $\mu_Q$ \\
\noalign{\smallskip}
\end{tabular}
\end{ruledtabular}
\end{table}

In Table~\ref{sf63}, we present the results for the 
spin-flip and nonvanishing orbit-flip amplitudes for the decays  
$\Sigma_Q(udQ) \rightarrow {^{2}\Lambda_Q}(udQ) + \gamma$. 
The spin-flavor contributions for the other couplings for transitions from the 
flavor sextet to the flavor anti-triplet can be obtained by the replacements
\ba
\begin{array}{ccccc}
\mu_u-\mu_d &\rightarrow& \mu_u-\mu_s &:& \Xi'_Q(usQ) \\
\mu_u-\mu_d &\rightarrow& \mu_d-\mu_s &:& \Xi'_Q(dsQ) 
\end{array} 
\label{ec63}
\ea

\begin{table}
\centering
\caption{As Table~\ref{sf66}, but for the decay 
$\Sigma_Q(udQ) \rightarrow {^{2}\Lambda_Q}(udQ) + \gamma$.}
\label{sf63}
\begin{ruledtabular}
\begin{tabular}{ccccc}
\noalign{\smallskip}
Spin-flip & $\nu$ & $\left< \mu_1 s_{1,-} \right>$ & $\left< \mu_2 s_{2,-} \right>$ 
& $\left< \mu_3s_{3,-} \right>$ \\
\noalign{\smallskip}
\hline
\noalign{\smallskip}
$^{2}\Sigma_Q \rightarrow {^{2}\Lambda_Q}$ & $\frac{1}{2}$ 
& $-\frac{1}{2\sqrt{3}} (\mu_u-\mu_d)$ 
& $-\frac{1}{2\sqrt{3}} (\mu_u-\mu_d)$ & 0 \\
\noalign{\smallskip}
$^{4}\Sigma_Q \rightarrow {^{2}\Lambda_Q}$ & $\frac{1}{2}$ 
& $-\frac{1}{2\sqrt{6}} (\mu_u-\mu_d)$ 
& $-\frac{1}{2\sqrt{6}} (\mu_u-\mu_d)$ & 0 \\
\noalign{\smallskip}
& $\frac{3}{2}$ & $-\frac{1}{2\sqrt{2}} (\mu_u-\mu_d)$ 
& $-\frac{1}{2\sqrt{2}} (\mu_u-\mu_d)$ & 0 \\
\noalign{\smallskip}
$^{2}\rho_J(\Sigma_Q) \rightarrow {^{2}\Lambda_Q}$ & $\frac{1}{2}$ & 0 & 0 & 0 \\
\noalign{\smallskip}
$^{2}\lambda_J(\Sigma_Q) \rightarrow {^{2}\Lambda_Q}$ & $\frac{1}{2}$ 
& $-\frac{1}{2\sqrt{3}} (\mu_u-\mu_d)$ 
& $-\frac{1}{2\sqrt{3}} (\mu_u-\mu_d)$ & 0 \\
\noalign{\smallskip}
$^{4}\lambda_J(\Sigma_Q) \rightarrow {^{2}\Lambda_Q}$ & $\frac{1}{2}$ 
& $-\frac{1}{2\sqrt{6}} (\mu_u-\mu_d)$ 
& $-\frac{1}{2\sqrt{6}} (\mu_u-\mu_d)$ & 0 \\
\noalign{\smallskip}
& $\frac{3}{2}$ & $-\frac{1}{2\sqrt{2}} (\mu_u-\mu_d)$ 
& $-\frac{1}{2\sqrt{2}} (\mu_u-\mu_d)$ & 0 \\
\noalign{\smallskip}
\hline
\noalign{\smallskip}
Orbit-flip & $\nu$ & $\left< \mu_1 \right>$ & $\left< \mu_2 \right>$ 
& $\left< \mu_3 \right>$ \\
\noalign{\smallskip}
\hline
\noalign{\smallskip}
$^{2}\rho_J(\Sigma_Q) \rightarrow {^{2}\Lambda_Q}$ & $\frac{1}{2}$ 
& $\frac{1}{2}(\mu_u-\mu_d)$ & $-\frac{1}{2}(\mu_u-\mu_d)$ & $0$ \\
\noalign{\smallskip}
& $\frac{3}{2}$ & $\frac{1}{2}(\mu_u-\mu_d)$ 
& $-\frac{1}{2}(\mu_u-\mu_d)$ & $0$ \\
\noalign{\smallskip}
\end{tabular}
\end{ruledtabular}
\end{table}

In Table~\ref{sf36}, we present the results for the 
spin-flip and nonvanishing orbit-flip amplitudes for the decays  
$\Lambda_Q(udQ) \rightarrow {^{2}\Sigma_Q}(udQ) + \gamma$ and
${^{4}\Sigma_Q}(udQ) + \gamma$. 
The spin-flavor contributions for the other couplings for transitions from the flavor anti-triplet to the flavor sextet can be obtained by the replacements
\ba
\begin{array}{ccccc}
\mu_u-\mu_d &\rightarrow& \mu_u-\mu_s &:& \Xi_Q(usQ) \\
\mu_u-\mu_d &\rightarrow& \mu_d-\mu_s &:& \Xi_Q(dsQ) 
\end{array} 
\label{ec36}
\ea

\begin{table}
\centering
\caption{As Table~\ref{sf66}, but for the decays  
$\Lambda_Q(udQ) \rightarrow {^{2}\Sigma_Q}(udQ) + \gamma$ and
${^{4}\Sigma_Q}(udQ) + \gamma$.}
\label{sf36}
\begin{ruledtabular}
\begin{tabular}{ccccc}
\noalign{\smallskip}
Spin-flip & $\nu$ & $\left< \mu_1 s_{1,-} \right>$ & $\left< \mu_2 s_{2,-} \right>$ 
& $\left< \mu_3 s_{3,-} \right>$ \\
\noalign{\smallskip}
\hline
\noalign{\smallskip}
$^{2}\Lambda_Q \rightarrow {^{2}\Sigma_Q}$ & $\frac{1}{2}$ & $-\frac{1}{2\sqrt{3}}(\mu_u-\mu_d)$ 
& $-\frac{1}{2\sqrt{3}}(\mu_u-\mu_d)$ & $0$ \\
\noalign{\smallskip}
$^{2}\rho_J(\Lambda_Q) \rightarrow {^{2}\Sigma_Q}$ & $\frac{1}{2}$ 
& $ \frac{1}{3}(\mu_u-\mu_d)$ 
& $-\frac{1}{3}(\mu_u-\mu_d)$ & $0$ \\
\noalign{\smallskip}
$^{4}\rho_J(\Lambda_Q) \rightarrow {^{2}\Sigma_Q}$ & $\frac{1}{2}$ 
& $-\frac{1}{6\sqrt{2}}(\mu_u-\mu_d)$ 
& $ \frac{1}{6\sqrt{2}}(\mu_u-\mu_d)$ & $0$ \\
\noalign{\smallskip}
& $\frac{3}{2}$ & $-\frac{1}{2\sqrt{6}}(\mu_u-\mu_d)$ 
& $\frac{1}{2\sqrt{6}}(\mu_u-\mu_d)$ & $0$ \\
\noalign{\smallskip}
$^{2}\lambda_J(\Lambda_Q) \rightarrow {^{2}\Sigma_Q}$ & $\frac{1}{2}$ 
& $-\frac{1}{2\sqrt{3}}(\mu_u-\mu_d)$ 
& $-\frac{1}{2\sqrt{3}}(\mu_u-\mu_d)$ & $0$ \\
\noalign{\smallskip}
$^{2}\Lambda_Q \rightarrow {^{4}\Sigma_Q}$ & $\frac{1}{2}$ 
& $\frac{1}{2\sqrt{6}}(\mu_u-\mu_d)$ 
& $\frac{1}{2\sqrt{6}}(\mu_u-\mu_d)$ & $0$ \\
\noalign{\smallskip}
$^{2}\rho_J(\Lambda_Q) \rightarrow {^{4}\Sigma_Q}$ & $\frac{1}{2}$ 
& $ \frac{1}{6\sqrt{2}}(\mu_u-\mu_d)$ 
& $-\frac{1}{6\sqrt{2}}(\mu_u-\mu_d)$ & $0$ \\
\noalign{\smallskip}
$^{4}\rho_J(\Lambda_Q) \rightarrow {^{4}\Sigma_Q}$ & $\frac{1}{2}$ 
& $ \frac{1}{3}(\mu_u-\mu_d)$ 
& $-\frac{1}{3}(\mu_u-\mu_d)$ & $0$ \\
\noalign{\smallskip}
& $\frac{3}{2}$ & $\frac{1}{2\sqrt{3}}(\mu_u-\mu_d)$ 
& $-\frac{1}{2\sqrt{3}}(\mu_u-\mu_d)$ & $0$ \\
\noalign{\smallskip}
$^{2}\lambda_J(\Lambda_Q) \rightarrow {^{4}\Sigma_Q}$ & $\frac{1}{2}$ 
& $\frac{1}{2\sqrt{6}}(\mu_u-\mu_d)$ 
& $\frac{1}{2\sqrt{6}}(\mu_u-\mu_d)$ & $0$ \\
\noalign{\smallskip}
\hline
\noalign{\smallskip}
Orbit-flip & $\nu$ & $\left< \mu_1 \right>$ & $\left< \mu_2 \right>$ 
& $\left< \mu_3 \right>$ \\
\noalign{\smallskip}
\hline
\noalign{\smallskip}
$^{2}\rho_J(\Lambda_Q) \rightarrow {^{2}\Sigma_Q}$ & $\frac{1}{2}$ 
& $\frac{1}{2}(\mu_u-\mu_d)$ & $-\frac{1}{2}(\mu_u-\mu_d)$ & $0$ \\
\noalign{\smallskip}
& $\frac{3}{2}$ & $\frac{1}{2}(\mu_u-\mu_d)$ 
& $-\frac{1}{2}(\mu_u-\mu_d)$ & $0$ \\
\noalign{\smallskip}
$^{4}\rho_J(\Lambda_Q) \rightarrow {^{4}\Sigma_Q}$ & $\frac{1}{2}$ 
& $ \frac{1}{2}(\mu_u-\mu_d)$ 
& $-\frac{1}{2}(\mu_u-\mu_d)$ & $0$ \\
\noalign{\smallskip}
& $\frac{3}{2}$ & $\frac{1}{2}(\mu_u-\mu_d)$ 
& $-\frac{1}{2}(\mu_u-\mu_d)$ & $0$ \\
\noalign{\smallskip}
\end{tabular}
\end{ruledtabular}
\end{table}

In Table~\ref{sf33}, we present the results for the 
spin-flip and nonvanishing orbit-flip amplitudes for the decays  
$\Lambda_Q(udQ) \rightarrow {^{2}\Lambda_Q}(udQ) + \gamma$. 
The contributions for the other members of the flavor antitriplet can be obtained by the replacements
\ba
\begin{array}{ccccc}
\mu_u+\mu_d &\rightarrow& \mu_u+\mu_s &:& \Xi_Q(usQ) \\ 
\mu_u+\mu_d &\rightarrow& \mu_d+\mu_s &:& \Xi_Q(dsQ) 
\end{array} 
\label{ec33}
\ea

\begin{table}
\centering
\caption{As Table~\ref{sf66}, but for the decays  
$\Lambda_Q(udQ) \rightarrow {^{2}\Lambda_Q}(udQ) + \gamma$.}
\label{sf33}
\begin{ruledtabular}
\begin{tabular}{ccccc}
\noalign{\smallskip}
Spin-flip & $\nu$ & $\left< \mu_1 s_{1,-} \right>$ & $\left< \mu_2 s_{2,-} \right>$ 
& $\left< \mu_3 s_{3,-} \right>$ \\
\noalign{\smallskip}
\hline
\noalign{\smallskip}
$^{2}\rho_J(\Lambda_Q) \rightarrow {^{2}\Lambda_Q}$ & $\frac{1}{2}$ 
& $-\frac{1}{2\sqrt{3}}(\mu_u+\mu_d)$ 
& $ \frac{1}{2\sqrt{3}}(\mu_u+\mu_d)$ & $0$ \\
\noalign{\smallskip}
$^{4}\rho_J(\Lambda_Q) \rightarrow {^{2}\Lambda_Q}$ & $\frac{1}{2}$ 
& $-\frac{1}{2\sqrt{6}}(\mu_u+\mu_d)$ 
& $ \frac{1}{2\sqrt{6}}(\mu_u+\mu_d)$ & $0$ \\
\noalign{\smallskip}
& $\frac{3}{2}$ & $-\frac{1}{2\sqrt{2}}(\mu_u+\mu_d)$ 
& $\frac{1}{2\sqrt{2}}(\mu_u+\mu_d)$ & $0$ \\
\noalign{\smallskip}
$^{2}\lambda_J(\Lambda_Q) \rightarrow {^{2}\Lambda_Q}$ & $\frac{1}{2}$ 
& $0$ & $0$ & $\mu_Q$ \\
\noalign{\smallskip}
\hline
\noalign{\smallskip}
Orbit-flip & $\nu$ & $\left< \mu_1 \right>$ & $\left< \mu_2 \right>$ 
& $\left< \mu_3 \right>$ \\
\noalign{\smallskip}
\hline
\noalign{\smallskip}
$^{2}\lambda_J(\Lambda_Q) \rightarrow {^{2}\Lambda_Q}$ & $\frac{1}{2}$ 
& $\frac{1}{2}(\mu_u+\mu_d)$ 
& $\frac{1}{2}(\mu_u+\mu_d)$ & $\mu_Q$ \\
\noalign{\smallskip}
& $\frac{3}{2}$ & $\frac{1}{2}(\mu_u+\mu_d)$ 
& $\frac{1}{2}(\mu_u+\mu_d)$ & $\mu_Q$ \\
\end{tabular}
\end{ruledtabular}
\end{table}

The contributions of the spin-flavor part for the doubly-heavy baryon decays
can be obtained from those for the decays $\Sigma_Q(uuQ) \to {}^2\Sigma_Q(uuQ)+\gamma$ 
and ${}^4\Sigma_Q(uuQ)+\gamma$ in Table~\ref{sf66} by interchanging the light and 
heavy quarks 
\ba
\begin{array}{ccccc}
\mu_u \to \mu_Q & \& & \mu_Q \to \mu_u &:& \Xi_{QQ}(QQu) \\ 
\mu_u \to \mu_Q & \& & \mu_Q \to \mu_d &:& \Xi_{QQ}(QQd) \\ 
\mu_u \to \mu_Q & \& & \mu_Q \to \mu_s &:& \Omega_{QQ}(QQs) 
\end{array} 
\label{ecQQ}
\ea

Finally, the spin-flavor matrix elements for triply-heavy baryons 
are given in Table~\ref{sfQQQ}.

\begin{table}[h]
\caption{Spin-flip amplitudes for the decay 
$\Omega_{QQQ}(QQQ) \rightarrow {^4\Omega_{QQQ}}(QQQ)+\gamma$. 
Orbit flip amplitudes are zero.}
\label{sfQQQ}
\begin{ruledtabular}
\begin{tabular}{ccccc}
\noalign{\smallskip}
Spin-flip & $\nu$ & $\left< \mu_1 s_{1,-} \right>$ & $\left< \mu_2 s_{2,-} \right>$ 
& $\left< \mu_3 s_{3,-} \right>$ \\
\noalign{\smallskip}
\hline
\noalign{\smallskip}
$^{2}\lambda_J(\Omega_{QQQ}) \rightarrow {^{4}\Omega_{QQQ}}$ & $\frac{1}{2}$ 
& $\frac{1}{3\sqrt{2}} \mu_Q$ 
& $\frac{1}{3\sqrt{2}} \mu_Q$ & $-\frac{\sqrt{2}}{3} \mu_Q$ \\
$^{2}\rho_J(\Omega_{QQQ}) \rightarrow {^{4}\Omega_{QQQ}}$ & $\frac{1}{2}$ 
& $ \frac{1}{\sqrt{6}} \mu_Q$ 
& $-\frac{1}{\sqrt{6}} \mu_Q$ & $0$ \\
\noalign{\smallskip}
\end{tabular}
\end{ruledtabular}
\end{table}

\subsection{Helicity amplitudes}

The helicity amplitudes of Eq.~(\ref{helamp}) can be obtained by combining the results for the radial integrals and the spin-flavor matrix elements. In Tables~\ref{hamp66}-\ref{hampQQQ} we present the results. Obviously, for the helicity amplitudes one has the same symmetry relations of Eqs.~(\ref{ec66})-(\ref{ecQQ}) as for the spin-flavor matrix elements. 

\begin{table*}
\centering
\caption{Helicity amplitudes for the decays $\Sigma_Q(uuQ) \to {}^2\Sigma_Q(uuQ)+\gamma$ (top) and ${}^4\Sigma_Q(uuQ)+\gamma$ (bottom).}
\label{hamp66}
\begin{ruledtabular}
\begin{tabular}{cccc} 
\noalign{\smallskip}
& $J$ & $\mathcal{A}_{1/2}/2\sqrt{\pi/k_0}$  & $\mathcal{A}_{3/2}/2\sqrt{\pi/k_0}$ \\
\noalign{\smallskip}
\hline
\noalign{\smallskip}
${}^4\Sigma_Q$
& $\frac{3}{2}$ & $k\frac{\sqrt{2}}{3}\left(-\mu_uU_{0,1}+\mu_QU_{0,3}\right)$ & $k\sqrt{\frac{2}{3}} \left(-\mu_uU_{0,1}+\mu_QU_{0,3}\right)$ \\  
${}^2\rho(\Sigma_Q)_J$ & $\frac{1}{2}$& $k\frac{2}{3}\mu_uU_{\rho,1} $ & 0 \\  & $\frac{3}{2}$ & $-k\frac{2\sqrt{2}}{3}\mu_uU_{\rho,1} $ & 0 \\  
${}^2\lambda(\Sigma_Q)_J$& $\frac{1}{2}$ & $-k\frac{1}{3\sqrt{3}}\left(4\mu_uU_{\lambda,1}-\mu_QU_{\lambda,3}\right)-\frac{1}{g}\sqrt{\frac{2}{3}}\left(2\mu_uT_{\lambda,1}+\mu_QT_{\lambda,3}\right)  $ & 0 \\ 
& $\frac{3}{2}$ & $k\frac{\sqrt{2}}{3\sqrt{3}} \left(4\mu_uU_{\lambda,1}-\mu_QU_{\lambda,3}\right)-\frac{1}{g}\frac{1}{\sqrt{3}}\left(2\mu_uT_{\lambda,1}+\mu_QT_{\lambda,3}\right)  $ & $-\frac{1}{g}\left(2\mu_uT_{\lambda,1}+\mu_QT_{\lambda,3}\right)$  \\ ${}^4\lambda(\Sigma_Q)_J$& $\frac{1}{2}$& $k\frac{\sqrt{2}}{3\sqrt{3}} \left(\mu_uU_{\lambda,1}-\mu_QU_{\lambda,3}\right)$& 0 \\  
& $\frac{3}{2}$ & $k\frac{\sqrt{2}}{3\sqrt{15}} \left(\mu_uU_{\lambda,1}-\mu_QU_{\lambda,3}\right)$ & $k\sqrt{\frac{2}{5}} \left(\mu_uU_{\lambda,1}-\mu_QU_{\lambda,3}\right)$ \\ 
& $\frac{5}{2}$ & $-k\sqrt{\frac{2}{15}}\left(\mu_uU_{\lambda,1}-\mu_QU_{\lambda,3}\right)$ & $-k\frac{2}{\sqrt{15}}\left(\mu_uU_{\lambda,1}-\mu_QU_{\lambda,3}\right)$  \\
\noalign{\smallskip}
\hline
\noalign{\smallskip}
${}^2\rho(\Sigma_Q)_J$ & $\frac{1}{2}$ & $-k\frac{\sqrt{2}}{3}\mu_uU_{\rho,1} $ & 0  \\ 
& $\frac{3}{2}$ & $k\frac{2}{3}\mu_uU_{\rho,1} $ & 0  \\ 
${}^2\lambda(\Sigma_Q)_J$& $\frac{1}{2}$ & $-k\frac{\sqrt{2}}{3\sqrt{3}}\left(\mu_uU_{\lambda,1}-\mu_QU_{\lambda,3}\right)$ & 0 \\ 
& $\frac{3}{2}$ &$k\frac{2}{3\sqrt{3}}\left(\mu_uU_{\lambda,1}-\mu_QU_{\lambda,3}\right)$ &0 \\ 
${}^4\lambda(\Sigma_Q)_J$& $\frac{1}{2}$ & $-k\frac{2}{3\sqrt{3}}\left(2\mu_uU_{\lambda,1}+\mu_QU_{\lambda,3}\right)-\frac{1}{g}\frac{1}{\sqrt{6}}\left(2\mu_uT_{\lambda,1}+\mu_QT_{\lambda,3}\right)  $& 0 \\  
& $\frac{3}{2}$ & $-k\frac{2}{3\sqrt{15}}\left(2\mu_uU_{\lambda,1}+\mu_QU_{\lambda,3}\right)-\frac{1}{g}\sqrt{\frac{8}{15}}\left(2\mu_uT_{\lambda,1}+\mu_QT_{\lambda,3}\right)  $ &$-k\frac{1}{\sqrt{5}}\left(2\mu_uU_{\lambda,1}+\mu_QU_{\lambda,3}\right)-\frac{1}{g}\sqrt{\frac{2}{5}}\left(2\mu_uT_{\lambda,1}+\mu_QT_{\lambda,3}\right)  $ \\
& $\frac{5}{2}$ &$k\frac{2}{\sqrt{15}}\left(2\mu_uU_{\lambda,1}+\mu_QU_{\lambda,3}\right)-\frac{1}{g}\sqrt{\frac{3}{10}}\left(2\mu_uT_{\lambda,1}+\mu_QT_{\lambda,3}\right)  $ &$k\sqrt{\frac{2}{15}}\left(2\mu_uU_{\lambda,1}+\mu_QU_{\lambda,3}\right)-\frac{1}{g}\sqrt{\frac{3}{5}}\left(2\mu_uT_{\lambda,1}+\mu_QT_{\lambda,3}\right)  $ \\
\noalign{\smallskip}
\end{tabular}
\end{ruledtabular}
\end{table*}

\begin{table}
\centering
\caption{As Table~\ref{hamp66}, but for the decays $\Sigma_Q(udQ) \to  {}^2\Lambda_Q(udQ)+\gamma$.}
\label{hamp63}
\begin{ruledtabular}
\begin{tabular}{cccc} 
\noalign{\smallskip}
& $J$ & $\mathcal{A}_{1/2}/2\sqrt{\pi/k_0}$  & $\mathcal{A}_{3/2}/2\sqrt{\pi/k_0}$ \\
\noalign{\smallskip}
\hline
\noalign{\smallskip}
${}^2\Sigma_Q$&
$\frac{1}{2}$&$-k\frac{1}{\sqrt{3}}(\mu_u-\mu_d)U_{0,1}$&0 \\ 
${}^4\Sigma_Q$
& $\frac{3}{2}$ & $-k\frac{1}{\sqrt{6}}(\mu_u-\mu_d)U_{0,1}$ &$-k\frac{1}{\sqrt{2}}(\mu_u-\mu_d)U_{0,1}$ \\  
${}^2\rho(\Sigma_Q)_J$ & $\frac{1}{2}$& $-\frac{1}{g}\sqrt{\frac{2}{3}}(\mu_u-\mu_d)T_{\rho,1} $ & 0 \\  
& $\frac{3}{2}$ & $-\frac{1}{g}\frac{1}{\sqrt{3}}(\mu_u-\mu_d)T_{\rho,1} $ & $-\frac{1}{g}(\mu_u-\mu_d)T_{\rho,1} $ \\  
${}^2\lambda(\Sigma_Q)_J$& $\frac{1}{2}$ & $k\frac{1}{3}(\mu_u-\mu_d)U_{\lambda,1}$ & 0 \\ 
& $\frac{3}{2}$ & $-k\frac{\sqrt{2}}{3}(\mu_u-\mu_d)U_{\lambda,1} $ & 0  \\
${}^4\lambda(\Sigma_Q)_J$& $\frac{1}{2}$& $k\frac{1}{3\sqrt{2}}(\mu_u-\mu_d)U_{\lambda,1}$& 0 \\  
& $\frac{3}{2}$ & $k\frac{1}{3\sqrt{10}}(\mu_u-\mu_d)U_{\lambda,1}$ & $k\sqrt{\frac{3}{10}}(\mu_u-\mu_d)U_{\lambda,1}$ \\  
& $\frac{5}{2}$ & $ -k\frac{1}{\sqrt{10}}(\mu_u-\mu_d)U_{\lambda,1}$ & $-k\frac{1}{\sqrt{5}}(\mu_u-\mu_d)U_{\lambda,1}$ \\
\noalign{\smallskip}
\end{tabular}
\end{ruledtabular}
\end{table}
 
\begin{table*}	
\centering
\caption{As Table~\ref{hamp66}, but for the decays $\Lambda_Q(udQ) \to  {}^2\Sigma_Q(udQ)+\gamma$ (top) and  ${}^4\Sigma_Q(udQ)+\gamma$ (bottom). }
\label{hamp36}
\begin{ruledtabular}
\begin{tabular}{cccc} 
\noalign{\smallskip}
& $J$ & $\mathcal{A}_{1/2}/2\sqrt{\pi/k_0}$  & $\mathcal{A}_{3/2}/2\sqrt{\pi/k_0}$ \\
\noalign{\smallskip}
\hline
\noalign{\smallskip}
${}^2\rho(\Lambda_Q)_J$ & $\frac{1}{2}$& $-k\frac{2}{3\sqrt{3}}(\mu_u-\mu_d)U_{\rho,1}-\frac{1}{g}\sqrt{\frac{2}{3}}(\mu_u-\mu_d)T_{\rho,1} $ & 0 \\ 
& $\frac{3}{2}$ & $k\frac{2\sqrt{2}}{3\sqrt{3}}(\mu_u-\mu_d)U_{\rho,1}-\frac{1}{g}\frac{1}{\sqrt{3}}(\mu_u-\mu_d)T_{\rho,1} $ & $-\frac{1}{g}(\mu_u-\mu_d)T_{\rho,1} $ \\ 
${}^4\rho(\Lambda_Q)_J$& $\frac{1}{2}$& $k\frac{1}{3\sqrt{6}}(\mu_u-\mu_d)U_{\rho,1}$& 0 \\ 
& $\frac{3}{2}$ & $k\frac{1}{3\sqrt{30}}(\mu_u-\mu_d)U_{\rho,1}$ & $k\frac{1}{\sqrt{10}} (\mu_u-\mu_d)U_{\rho,1}$ \\ 
& $\frac{5}{2}$ & $ -k\frac{1}{\sqrt{30}}(\mu_u-\mu_d)U_{\rho,1}$ & $-k\frac{1}{\sqrt{15}} (\mu_u-\mu_d)U_{\rho,1}$ \\ 
${}^2\lambda(\Lambda_Q)_J$& $\frac{1}{2}$ & $k\frac{1}{3}(\mu_u-\mu_d)U_{\lambda,1} $ & 0 \\ 
& $\frac{3}{2}$ & $-k\frac{\sqrt{2}}{3}(\mu_u-\mu_d)U_{\lambda,1} $ & 0 \\  
\noalign{\smallskip}
\hline
\noalign{\smallskip}
${}^2\rho(\Lambda_Q)_J$ & $\frac{1}{2}$& $-k\frac{1}{3\sqrt{6}}(\mu_u-\mu_d)U_{\rho,1}$ & 0 \\ 
& $\frac{3}{2}$ & $k\frac{1}{3\sqrt{3}}(\mu_u-\mu_d)U_{\rho,1}$ & 0 \\ 
${}^4\rho(\Lambda_Q)_J$& $\frac{1}{2}$& $-k\frac{2}{3\sqrt{3}}(\mu_u-\mu_d)U_{\rho,1}-\frac{1}{g}\frac{1}{\sqrt{6}}(\mu_u-\mu_d)T_{\rho,1}$& 0 \\ 
& $\frac{3}{2}$ & $-k\frac{2}{3\sqrt{15}}(\mu_u-\mu_d)U_{\rho,1}-\frac{1}{g} \sqrt{\frac{8}{15}}(\mu_u-\mu_d)T_{\rho,1}$ & $-k\frac{1}{\sqrt{5}}(\mu_u-\mu_d)U_{\rho,1}-\frac{1}{g}\sqrt{\frac{2}{5}}(\mu_u-\mu_d)T_{\rho,1}$  \\ 
& $\frac{5}{2}$ &  $k\frac{2}{\sqrt{15}}(\mu_u-\mu_d)U_{\rho,1}-\frac{1}{g}\sqrt{\frac{3}{10}}(\mu_u-\mu_d)T_{\rho,1}$ & $k \sqrt{\frac{2}{15}} (\mu_u-\mu_d)U_{\rho,1}-\frac{1}{g}\sqrt{\frac{3}{5}}(\mu_u-\mu_d)T_{\rho,1}$  \\ 
${}^2\lambda(\Lambda_Q)_J$& $\frac{1}{2}$ & $-k\frac{1}{3\sqrt{2}}(\mu_u-\mu_d)U_{\lambda,1} $ & 0 \\ 
& $\frac{3}{2}$ & $k\frac{1}{3}(\mu_u-\mu_d)U_{\lambda,1} $ & 0 \\  
\noalign{\smallskip}
\end{tabular}
\end{ruledtabular}
\end{table*}

\begin{table*}	
\centering
\caption{As Table~\ref{hamp66}, but for the decays $\Lambda_Q(udQ) \to  {}^2\Lambda_Q(udQ)+\gamma$. }
\label{hamp33}
\begin{ruledtabular}
\begin{tabular}{cccc} 
& $J$ & $\mathcal{A}_{1/2}/2\sqrt{\pi/k_0}$  & $\mathcal{A}_{3/2}/2\sqrt{\pi/k_0}$ \\
\noalign{\smallskip}
\hline
\noalign{\smallskip}
${}^2\rho(\Lambda_Q)_J$ & $\frac{1}{2}$& $k\frac{1}{3}(\mu_u+\mu_d)U_{\rho,1} $ & 0 \\  
& $\frac{3}{2}$ & $-k\frac{\sqrt{2}}{3}(\mu_u+\mu_d)U_{\rho,1} $ & 0 \\  
${}^4\rho(\Lambda_Q)_J$& $\frac{1}{2}$& $k\frac{1}{3\sqrt{2}}(\mu_u+\mu_d)U_{\rho,1}$& 0 \\ 
& $\frac{3}{2}$ & $k\frac{1}{3\sqrt{10}}(\mu_u+\mu_d)U_{\rho,1}$ & $k\sqrt{\frac{3}{10}}(\mu_u+\mu_d)U_{\rho,1}$ \\ 
& $\frac{5}{2}$ & $- k\frac{1}{\sqrt{10}}(\mu_u+\mu_d)U_{\rho,1}$ & $-k\frac{1}{\sqrt{5}}(\mu_u+\mu_d)U_{\rho,1}$ \\ 
${}^2\lambda(\Lambda_Q)_J$& $\frac{1}{2}$ & $-k\frac{1}{\sqrt{3}} \mu_Q U_{\lambda,3}-\frac{1}{g} \sqrt{\frac{2}{3}} \left[(\mu_u+\mu_d)T_{\lambda,1}+\mu_QT_{\lambda,3}\right]  $ & 0 \\ 
& $\frac{3}{2}$ & $k\sqrt{\frac{2}{3}}\mu_QU_{\lambda,3}-\frac{1}{g}\frac{1}{\sqrt{3}} \left[(\mu_u+\mu_d)T_{\lambda,1}+\mu_QT_{\lambda,3}\right]$ & $-\frac{1}{g} \left[ (\mu_u+\mu_d)T_{\lambda,1}+\mu_QT_{\lambda,3} \right]$ \\  
\noalign{\smallskip}
\end{tabular}
\end{ruledtabular}
\end{table*}

For triply-heavy baryons $QQQ$ the helicity amplitude is given by 
\ba
A_{1/2} &=& -2\sqrt{\frac{\pi}{k_0}} \left< 1,0,\tfrac{1}{2},\tfrac{1}{2} | J,\tfrac{1}{2} \right> U_{\lambda,3}
\nonumber\\
&=& -2\sqrt{\frac{\pi}{k_0}} \left< 1,0,\tfrac{1}{2},\tfrac{1}{2} | J,\tfrac{1}{2} \right> \frac{ik}{\alpha \sqrt{3}} \, \mbox{e}^{-k^2/6\alpha^2}
\nonumber\\
A_{3/2} &=& 0
\ea
with $m=m'=m_Q$ and $\alpha_\rho=\alpha_\lambda=\alpha$.
The explicit expressions are given in Table~\ref{hampQQQ}.

\begin{table}[h]
\centering
\caption{As Table~\ref{hamp66}, but for the decays $\Omega_{QQQ} \to  {}^4\Omega_{QQQ}+\gamma$.}
\label{hampQQQ}
\begin{ruledtabular}
\begin{tabular}{cccc} 
\noalign{\smallskip}
& $J$ & $\mathcal{A}_{1/2}/2\sqrt{\pi/k_0}$  & $\mathcal{A}_{3/2}/2\sqrt{\pi/k_0}$ \\
\noalign{\smallskip}
\hline
\noalign{\smallskip}
${}^2E(\Sigma_{QQQ})_J$ & $\frac{1}{2}$ & $k\frac{1}{\sqrt{3}} \mu_Q U_{\lambda,3}$ & 0 \\ 
& $\frac{3}{2}$ & $-k\sqrt{\frac{2}{3}} \mu_Q U_{\lambda,3}$ & 0 \\
\noalign{\smallskip}
\end{tabular}
\end{ruledtabular}
\end{table}

\section{Radiative decay widths}
\label{sec:widths}

The radiative decay width is related to the helicity amplitudes 
\ba
\Gamma(B\rightarrow B'+\gamma)=2\pi \rho \frac{1}{(2\pi)^3}\frac{2}{2J+1}\sum_{\nu>0}|\mathcal{A}_\nu(k)|^2 . 
\ea
The widths are calculated in the rest frame of the initial baryon $B$. In this reference frame, the momentum of the emitted photon is given by
\ba
k=\frac{m_{B}^2-m_{B'}^2}{2m_{B}},
\ea
and the space phase factor has the form
\ba
\rho=4\pi \frac{E_{B'} k^2}{m_{B}}.
\ea
The energy of the final baryon is  $E_{B'}=\sqrt{m_{B'}^2+k^2}$. The value of $g$ is taken as $g=1$ for all electromagnetic decays.   

The quark magnetic moments of the light quarks are determined by the magnetic moments of the proton, neutron and $\Lambda$ hyperon as $\mu_u=1.852$ $\mu_N$, $\mu_d=-0.972$ $\mu_N$ and $\mu_s=-0.613$ $\mu_N$. In the absence of experimental information on the magnetic moments of heavy baryons, the magnetic moments of the heavy quarks are calculated from the heavy quark masses as $\mu_Q=e_Q m_N/m_Q$ $\mu_N$ to give $\mu_c=0.390$ $\mu_N$ and $\mu_b=-0.063$ $\mu_N$. 

\subsection{Singly-heavy baryons}
\label{shb}

For electromagnetic decays of singly-heavy baryons the final baryon $B'_Q$ is a ground state $S$-wave baryon, {\it i.e.} $^2(B'_Q)_{1/2}$ or $^4(B'_Q)_{3/2}$ of the flavor sextet $\bf 6$ or $^2(B'_Q)_{1/2}$ of the flavor anti-triplet $\bf \bar{3}$. 

First, in Table~\ref{gsgs} we present the radiative decay widths for ground-state $S$-wave baryons. In the absence of experimental data, we make a comparison with the predictions of other approaches: light cone QCD sum rules ($\text{LCQSR}$), bag model (BM), vector-meson dominance (VMD), chiral quark model ($\chi$QM), non-relativistic quark model (NRQM), heavy baryon chiral perturbation theory (HB$\chi$PT), relativistic quark model (RQM), and hypercentral quark model (hCQM). The different theoretical calculations are in qualitative agreement with each other: the transitions $^2\Sigma_Q \rightarrow {}^2\Lambda_Q + \gamma$ and $^4\Sigma_Q \rightarrow {}^2\Lambda_Q + \gamma$ are calculated to have the largest widths, the transitions $^{2,4}\Xi^{'+}_c \rightarrow {}^2\Xi^+_c + \gamma$ and $^{2,4}\Xi^{'0}_b \rightarrow {}^2\Xi^0_b + \gamma$ have somewhat smaller widths, and the remaining transitions have very small widths. 

\begin{table*}
\caption{Radiative decay widths of single charm and bottom $S$-wave baryons in keV.} 
\label{gsgs}
\centering
\begin{ruledtabular}
\begin{tabular}{lrcccccccccccc}
\noalign{\smallskip}
& & LCQSR & BM & VMD & $\chi$QM & NRQM & HB$\chi$PT & RQM & \multicolumn{3}{c}{hCQM} \\
\cline{10-12}
$B_Q\rightarrow B'_Q\gamma$ & Present & \cite{Aliev2009,Aliev2015,Aliev2016} & \cite{Bernotas2013} & \cite{Aliev2012} & \cite{Wang2017} & \cite{Majethiya2009} & \cite{JuanWang2019} &  \cite{Ivanov1999} & \cite{Shah} & \cite{Shah2018} & \cite{Shah_2016} \\
\noalign{\smallskip}
\hline
\noalign{\smallskip}
$^4\Sigma_{c}^{++}\rightarrow{}^2\Sigma_{c}^{++}\gamma$  &2.1&$2.65\pm1.60$&0.826&3.567&3.94&1.15&1.20&&1.32&0.85& \\
$^4\Sigma^{+}_{c}\rightarrow{}^2\Sigma_{c}^{+}\gamma$  &0.0&$0.40\pm0.22$ &0.004&0.187&0.004&$<10^{-4}$&0.04&$0.14\pm0.004$&$1\times10^{-4}$& $9\times10^{-5}$ \\
$^4\Sigma^{0}_{c}\rightarrow{}^2\Sigma_{c}^{0}\gamma$  &1.8&$0.08\pm0.042$&1.08&1.049&3.43&1.12&0.49&&1.072&1.20& 1.55 \\
$^2\Sigma^{+}_{c}\rightarrow{}^2\Lambda_{c}^{+}\gamma$&87.2&$50.0\pm17.0$&46.1&&80.60&60.55&65.6&$60.7\pm1.5$&71.20 & 58.13\\
$^4\Sigma^{+}_{c}\rightarrow{}^2\Lambda_{c}^{+}\gamma$&199.4&$130\pm35$&126&409.3&373&154.48&161.8&$151\pm4$&171.9&143.97&213.3 \\
$^4\Xi^{'+}_{c}\rightarrow{}^2\Xi_{c}^{'+}\gamma$&0.1&0.274&0.011 &0.485&0.004&&0.07\\
$^4\Xi^{'0}_{c}\rightarrow{}^2\Xi_{c}^{'0}\gamma$&1.4&2.142 &1.03&1.317&3.03&&0.42 \\
$^2\Xi^{'+}_{c}\rightarrow{}^2\Xi^{+}_{c}\gamma$&20.6&$8.5\pm2.5$&10.2&&42.3&&5.43&$12.7\pm1.5$\\
$^2\Xi^{'0}_{c}\rightarrow{}^2\Xi^{0}_{c}\gamma$&0.4&$0.27\pm0.06$&0.0015&&0.00&&0.46&$0.17\pm0.002$\\
$^4\Xi^{'+}_{c}\rightarrow{}^2\Xi^{+}_{c}\gamma$&74.2&$52\pm32$&44.3&152.4&139&63.32&21.6&$54\pm3$&&17.48 \\
$^4\Xi^{'0}_{c}\rightarrow{}^2\Xi^{0}_{c}\gamma$&1.6&$0.66\pm0.41$&0.908&1.318&0.00&0.30&1.84&$0.68\pm0.04$&&0.45 & 0.91 \\	
$^4\Omega^0_{c}\rightarrow{}^2\Omega_{c}^{0}\gamma$&1.0&0.932&1.07&1.439&0.89&2.02&0.32&&0.34&& 1.44  \\
\noalign{\smallskip}
\hline
\noalign{\smallskip}
$^4\Sigma^{+}_{b}\rightarrow{}^2\Sigma_{b}^{+}\gamma$&0.1&$0.46\pm0.28$&0.054&0.137&0.25&0.08&0.05 \\
$^4\Sigma^{0}_{b}\rightarrow{}^2\Sigma_{b}^{0}\gamma$&0.0&$0.028\pm0.02$ &0.005& 0.006&0.02&$<10^{-3}$&$3\times10^{-3}$\\
$^4\Sigma^{-}_{b}\rightarrow{}^2\Sigma_{b}^{-}\gamma$&0.0&$0.11\pm0.076$&0.01&0.040&0.06&0.01&0.013 \\
$^2\Sigma^{0}_{b}\rightarrow{}^2\Lambda_{b}^{0}\gamma$&128.1&$152.0\pm60.0$&58.9&&130&94.79&108.0\\
$^4\Sigma^{0}_{b}\rightarrow{}^2\Lambda_{b}^{0}\gamma$&168.8&$114\pm62$&81.1&221.5&335&128.62&142.1\\
$^4\Xi^{'0}_{b}\rightarrow{}^2\Xi_{b}^{'0}\gamma$&0.0&0.131&0.004&0.281&5.19&&$1.5\times10^{-3}$\\
$^4\Xi^{'-}_{b}\rightarrow{}^2\Xi_{b}^{'-}\gamma$&0.0&0.303&0.005&0.702&15.0&&$8.2\times10^{-3}$  \\
$^2\Xi^{'0}_{b}\rightarrow{}^2\Xi^{0}_{b}\gamma$&28.4&$47.0\pm21.0$&14.7&&84.6&&13.0\\
$^2\Xi^{'-}_{b}\rightarrow{}^2\Xi^{-}_{b}\gamma$&0.6&$3.3\pm1.3$&0.118&&0.00&&1.0\\
$^4\Xi^{'0}_{b}\rightarrow{}^2\Xi^{0}_{b}\gamma$&45.2&$135\pm85$&24.7&270.8&104&18.79&17.2 \\
$^4\Xi^{'-}_{b}\rightarrow{}^2\Xi^{-}_{b}\gamma$&1.0&$1.50\pm0.095$&0.278&2.246&0.00&0.09&1.4  \\
$^4\Omega^{-}_{b}\rightarrow{}^2\Omega_{b}^{-}\gamma$&0.0&0.092&0.006&2.873&0.1&0.03&0.031\\
\noalign{\smallskip}
\end{tabular}
\end{ruledtabular}
\end{table*}

Next, the electromagnetic decay widths for singly-heavy $P$-wave baryons are shown in Tables~\ref{6c} and \ref{3barc} for single charm baryons and in Tables~\ref{6b} and \ref{3barb} for single bottom baryons. We compare our results with those obtained in the chiral quark model ($\chi$QM) of Wang {\it et al.} \cite{Wang2017}. In Ref.~\cite{Wang2017}, the electromagnetic decay widths are calculated by combining the masses obtained by Ebert {\it et al.} in a relativistic heavy-quark-light-diquark picture \cite{PhysRevD.84.014025} with non-relativistic harmonic oscillator wave functions. Moreover, the $\chi$QM results for the decays of the $P$-wave $\rho$-mode states of the anti-triplet baryons in Tables~\ref{3barc} and \ref{3barb} were obtained assuming the same masses as for the $\lambda$-mode excitations. In the present calculation, we have calculated the mass spectra and the radiative widths in a consistent manner in the framework of a non-relativistic harmonic oscillator quark model. The largest widths are obtained for the decays of the $\rho$-mode excitations, in particular $^2\rho(\Sigma^{+}_c)_J \rightarrow {}^2\Lambda_c^+ + \gamma$, $^2\rho(\Xi^{'+}_c)_J \rightarrow {}^2\Lambda_c^+ + \gamma$, and the corresponding decays in the bottom sector $^2\rho(\Sigma^{0}_b)_J \rightarrow {}^2\Lambda_b^0 + \gamma$, $^2\rho(\Xi^{'0}_b)_J \rightarrow {}^2\Lambda_b^0 + \gamma$. 

\begin{table}
\caption{Radiative decay widths of sextet singly-heavy charm 
$P$-wave baryons in keV.}
\label{6c}
\centering
\begin{ruledtabular}
\begin{tabular}{lrrrrrr}
\noalign{\smallskip}
& \multicolumn{3}{c}{Present} 
& \multicolumn{3}{c}{$\chi$QM \cite{Wang2017,WangOme}} \\
& $^2\Lambda_{c}$ & $^2\Sigma_{c}$ & $^4\Sigma_{c}$ 
& $^2\Lambda_{c}$ & $^2\Sigma_{c}$ & $^4\Sigma_{c}$ \\
\noalign{\smallskip}
\hline 
\noalign{\smallskip}
$^2\lambda(\Sigma^{++}_{c})_{1/2}$ & &  67.1 &   3.4 & &  283 & 3.04 \\
$^2\lambda(\Sigma^{++}_{c})_{3/2}$ & & 551.8 &   5.6 & &  210 & 14.7 \\
$^4\lambda(\Sigma^{++}_{c})_{1/2}$ & &  12.5 &   1.7 & & 8.54 &  387 \\
$^4\lambda(\Sigma^{++}_{c})_{3/2}$ & &  48.6 &  68.8 & & 17.5 &  181 \\
$^4\lambda(\Sigma^{++}_{c})_{5/2}$ & &  49.7 & 450.3 & & 13.6 &  168 \\
$^2\rho(\Sigma^{++}_{c})_{1/2}$    & & 178.1 &  56.6 & & & \\
$^2\rho(\Sigma^{++}_{c})_{3/2}$    & & 215.5 &  72.2 & & & \\
\noalign{\smallskip}
\hline 
\noalign{\smallskip}
$^2\lambda(\Sigma^{+}_{c})_{1/2}$ & 110.0 &  4.6 & 0.3 & 48.3 & 1.60 & 0.31\\
$^2\lambda(\Sigma^{+}_{c})_{3/2}$ & 132.9 &  4.9 & 0.5 & 87.3 & 4.64 & 1.55 \\
$^4\lambda(\Sigma^{+}_{c})_{1/2}$ &  66.7 &  1.1 & 1.5 & 52.1 & 0.92 & 1.75 \\
$^4\lambda(\Sigma^{+}_{c})_{3/2}$ & 220.9 &  4.5 & 2.0 &  105 & 1.86 & 0.68 \\
$^4\lambda(\Sigma^{+}_{c})_{5/2}$ & 179.8 &  4.7 & 2.3 & 59.4 & 1.46 & 0.89 \\
$^2\rho(\Sigma^{+}_{c})_{1/2}$    & 618.2 & 10.1 & 3.2 & & & \\
$^2\rho(\Sigma^{+}_{c})_{3/2}$    & 624.6 & 12.2 & 4.1 & & & \\
\noalign{\smallskip}
\hline 
\noalign{\smallskip}
$^2\lambda(\Sigma^{0}_{c})_{1/2}$ & & 156.1 &   0.6 & &  205 & 0.39 \\
$^2\lambda(\Sigma^{0}_{c})_{3/2}$ & & 432.2 &   0.9 & &  245 & 1.82 \\
$^4\lambda(\Sigma^{0}_{c})_{1/2}$ & &   2.0 &  13.9 & & 1.02 &  289 \\
$^4\lambda(\Sigma^{0}_{c})_{3/2}$ & &   7.5 & 109.7 & & 2.12 &  159 \\
$^4\lambda(\Sigma^{0}_{c})_{5/2}$ & &   7.4 & 335.1 & & 1.64 &  160 \\
$^2\rho(\Sigma^{0}_{c})_{1/2}$    & &  49.1 &  15.6 & & & \\
$^2\rho(\Sigma^{0}_{c})_{3/2}$    & &  59.4 &  19.9 & & & \\
\noalign{\smallskip}
\hline
\noalign{\smallskip}
& $^2\Xi_{c}$ & $^2\Xi'_{c}$ & $^4\Xi'_{c}$ 
& $^2\Xi_{c}$ & $^2\Xi'_{c}$ & $^4\Xi'_{c}$ \\
\noalign{\smallskip}
\hline 
\noalign{\smallskip}
$^2\lambda(\Xi'^{+}_{c})_{1/2}$ &  37.9 &  0.0 &  0.3 & 46.4 & 0.03 & 1.61 \\
$^2\lambda(\Xi'^{+}_{c})_{3/2}$ &  50.2 & 18.7 &  0.5 & 46.1 & 12.1 & 1.59 \\
$^4\lambda(\Xi'^{+}_{c})_{1/2}$ &  25.2 &  1.3 &  0.1 & 14.5 & 0.33 & 0.16 \\
$^4\lambda(\Xi'^{+}_{c})_{3/2}$ &  90.6 &  5.2 &  0.9 & 54.6 & 2.06 & 1.64 \\
$^4\lambda(\Xi'^{+}_{c})_{5/2}$ &  83.4 &  5.8 & 14.1 & 32.0 & 1.63 & 2.35 \\
$^2\rho(\Xi'^{+}_{c})_{1/2}$    & 709.5 & 12.7 &  3.5 & & & \\
$^2\rho(\Xi'^{+}_{c})_{3/2}$    & 760.5 & 16.4 &  4.8 & & & \\
\noalign{\smallskip}
\hline
\noalign{\smallskip}
$^2\lambda(\Xi'^{0}_{c})_{1/2}$ &  0.8 & 158.3 &   0.1 & 0.0 &  472 & 1.00 \\
$^2\lambda(\Xi'^{0}_{c})_{3/2}$ &  1.1 & 339.3 &   0.3 & 0.0 &  302 & 1.05 \\
$^4\lambda(\Xi'^{0}_{c})_{1/2}$ &  0.5 &   0.6 &  18.5 & 0.0 & 0.20 &  125 \\
$^4\lambda(\Xi'^{0}_{c})_{3/2}$ &  1.9 &   2.5 & 108.0 & 0.0 & 1.21 &  187 \\
$^4\lambda(\Xi'^{0}_{c})_{5/2}$ &  1.8 &   2.7 & 248.2 & 0.0 & 0.93 &  192 \\
$^2\rho(\Xi'^{0}_{c})_{1/2}$    & 15.0 &  20.7 &   5.7 & & & \\
$^2\rho(\Xi'^{0}_{c})_{3/2}$    & 16.1 &  26.8 &   7.8 & & & \\
\noalign{\smallskip}
\hline
\noalign{\smallskip}
& & $^2\Omega_{c}$ & $^4\Omega_{c}$ 
& & $^2\Omega_{c}$ & $^4\Omega_{c}$ \\
\noalign{\smallskip}
\hline 
\noalign{\smallskip}
$^2\lambda(\Omega^{0}_{c})_{1/2}$ & & 135.7 &   0.0 & & 0.36 & 0.02 \\
$^2\lambda(\Omega^{0}_{c})_{3/2}$ & & 251.6 &   0.1 & & 0.35 & $<$0.01 \\
$^4\lambda(\Omega^{0}_{c})_{1/2}$ & &   0.2 &  17.5 & & 0.20 & 0.08 \\
$^4\lambda(\Omega^{0}_{c})_{3/2}$ & &   0.7 &  90.9 & & $<$0.01 & 0.33 \\
$^4\lambda(\Omega^{0}_{c})_{5/2}$ & &   0.7 & 177.0 & & $<$0.01 & 0.18 \\
$^2\rho(\Omega^{0}_{c})_{1/2}$    & &   8.1 &   1.9 & & & \\
$^2\rho(\Omega^{0}_{c})_{3/2}$    & &  11.0 &   2.9 & & & \\
\noalign{\smallskip}
\end{tabular} 
\end{ruledtabular}
\end{table}

\begin{table}
\caption{Radiative decay widths of anti-triplet single charm 
$P$-wave baryons in keV.}
\label{3barc}
\centering
\begin{ruledtabular}
\begin{tabular}{lrrrrrr}
\noalign{\smallskip}
& \multicolumn{3}{c}{Present} 
& \multicolumn{3}{c}{$\chi$QM \cite{Wang2017}} \\
& $^2\Lambda_{c}$ & $^2\Sigma_{c}$ & $^4\Sigma_{c}$ 
& $^2\Lambda_{c}$ & $^2\Sigma_{c}$ & $^4\Sigma_{c}$ \\
\noalign{\smallskip}
\hline 
\noalign{\smallskip}
$^2\lambda(\Lambda^{+}_{c})_{1/2}$ &  0.1 &   1.0 &   0.0 & 0.26 & 0.45 &  0.05 \\
$^2\lambda(\Lambda^{+}_{c})_{3/2}$ &  0.7 &   2.5 &   0.2 & 0.30 & 1.17 &  0.26 \\
$^2\rho(\Lambda^{+}_{c})_{1/2}$    &  9.6 &  97.3 &   1.5 & 1.59 & 41.6 &  0.02 \\
$^2\rho(\Lambda^{+}_{c})_{3/2}$    & 11.8 & 447.0 &   2.5 & 2.35 & 48.0 &  0.09 \\
$^4\rho(\Lambda^{+}_{c})_{1/2}$    &  5.8 &   5.3 &   6.0 & 0.80 & 0.08 &  6.81 \\
$^4\rho(\Lambda^{+}_{c})_{3/2}$    & 19.4 &  21.1 &  79.1 & 3.29 & 0.55 &  17.4 \\
$^4\rho(\Lambda^{+}_{c})_{5/2}$    & 16.1 &  22.2 & 362.8 & & & \\
\noalign{\smallskip}
\hline
\noalign{\smallskip}
& $^2\Xi_{c}$ & $^2\Xi'_{c}$ & $^4\Xi'_{c}$ 
& $^2\Xi_{c}$ & $^2\Xi'_{c}$ & $^4\Xi'_{c}$ \\
\noalign{\smallskip}
\hline 
\noalign{\smallskip}
$^2\lambda(\Xi^{+}_{c})_{1/2}$ &  7.4 &   1.3 &   0.1 & 4.65 & 1.43 & 0.44 \\
$^2\lambda(\Xi^{+}_{c})_{3/2}$ &  4.8 &   2.9 &   0.3 & 2.80 & 2.32 & 0.99 \\
$^2\rho(\Xi^{+}_{c})_{1/2}$    & 12.3 & 145.4 &   1.0 & 1.39 &  128 & 0.25 \\
$^2\rho(\Xi^{+}_{c})_{3/2}$    & 16.2 & 481.6 &   1.7 & 1.88 &  110 & 0.52 \\
$^4\rho(\Xi^{+}_{c})_{1/2}$    &  7.8 &   3.7 &  13.7 & 0.75 & 0.41 & 43.4 \\
$^4\rho(\Xi^{+}_{c})_{3/2}$    & 28.0 &  15.2 & 111.5 & 2.81 & 1.85 & 58.1 \\
$^4\rho(\Xi^{+}_{c})_{5/2}$    & 25.7 &  16.5 & 367.7 & & & \\
\noalign{\smallskip}
\hline
\noalign{\smallskip}
$^2\lambda(\Xi^{0}_{c})_{1/2}$ & 202.5 &  0.0 & 0.0 &  263 & 0.0 & 0.0 \\
$^2\lambda(\Xi^{0}_{c})_{3/2}$ & 292.6 &  0.1 & 0.0 &  292 & 0.0 & 0.0 \\
$^2\rho(\Xi^{0}_{c})_{1/2}$    &  20.1 &  3.1 & 0.0 & 5.57 & 0.0 & 0.0 \\
$^2\rho(\Xi^{0}_{c})_{3/2}$    &  26.5 & 10.2 & 0.0 & 7.50 & 0.0 & 0.0 \\
$^4\rho(\Xi^{0}_{c})_{1/2}$    &  12.8 &  0.1 & 0.3 & 3.00 & 0.0 & 0.0 \\
$^4\rho(\Xi^{0}_{c})_{3/2}$    &  45.9 &  0.3 & 2.4 & 11.2 & 0.0 & 0.0 \\
$^4\rho(\Xi^{0}_{c})_{5/2}$    &  42.1 &  0.4 & 7.8 & & & \\
\noalign{\smallskip}
\end{tabular} 
\end{ruledtabular}
\end{table}

\begin{table}
\caption{Radiative decay widths of sextet singly-heavy bottom 
$P$-wave baryons in keV.}
\label{6b}
\centering
\begin{ruledtabular}
\begin{tabular}{lrrrrrr}
\noalign{\smallskip}
& \multicolumn{3}{c}{Present} 
& \multicolumn{3}{c}{$\chi$QM \cite{Wang2017}} \\
& $^2\Lambda_{b}$ & $^2\Sigma_{b}$ & $^4\Sigma_{b}$ 
& $^2\Lambda_{b}$ & $^2\Sigma_{b}$ & $^4\Sigma_{b}$ \\
\noalign{\smallskip}
\hline 
\noalign{\smallskip}
$^2\lambda(\Sigma^{+}_{b})_{1/2}$ & & 267.7 &   5.3 & & 1016 & 16.9 \\
$^2\lambda(\Sigma^{+}_{b})_{3/2}$ & & 864.7 &   5.8 & &  483 & 15.6 \\
$^4\lambda(\Sigma^{+}_{b})_{1/2}$ & &   8.9 &  19.2 & & 5.31 &  867 \\
$^4\lambda(\Sigma^{+}_{b})_{3/2}$ & &  27.2 & 209.6 & & 13.1 &  527 \\
$^4\lambda(\Sigma^{+}_{b})_{5/2}$ & &  20.0 & 589.4 & & 8.07 &  426 \\
$^2\rho(\Sigma^{+}_{b})_{1/2}$    & & 182.1 &  79.3 & & & \\
$^2\rho(\Sigma^{+}_{b})_{3/2}$    & & 189.2 &  82.8 & & & \\
\noalign{\smallskip}
\hline 
\noalign{\smallskip}
$^2\lambda(\Sigma^{0}_{b})_{1/2}$ & 156.2 & 21.3 &  0.3 &  133 & 74.9 & 1.03 \\
$^2\lambda(\Sigma^{0}_{b})_{3/2}$ & 162.2 & 59.3 &  0.3 &  129 & 37.9 & 0.95 \\
$^4\lambda(\Sigma^{0}_{b})_{1/2}$ &  85.3 &  0.5 &  1.9 & 63.6 & 0.32 & 63.6 \\
$^4\lambda(\Sigma^{0}_{b})_{3/2}$ & 247.4 &  1.5 & 16.2 &  170 & 0.80 & 39.8 \\
$^4\lambda(\Sigma^{0}_{b})_{5/2}$ & 168.2 &  1.1 & 39.5 & 83.3 & 0.49 & 32.6 \\
$^2\rho(\Sigma^{0}_{b})_{1/2}$    & 526.5 & 10.3 &  4.5 & & & \\
$^2\rho(\Sigma^{0}_{b})_{3/2}$    & 523.1 & 10.7 &  4.7 & & & \\
\noalign{\smallskip}
\hline 
\noalign{\smallskip}
$^2\lambda(\Sigma^{-}_{b})_{1/2}$ & &  50.9 &   1.5 & &  212 & 4.36 \\
$^2\lambda(\Sigma^{-}_{b})_{3/2}$ & & 196.2 &   1.6 & &   94 & 4.02 \\
$^4\lambda(\Sigma^{-}_{b})_{1/2}$ & &   2.5 &   2.7 & & 1.37 &  182 \\
$^4\lambda(\Sigma^{-}_{b})_{3/2}$ & &   7.7 &  41.6 & & 3.39 &  107 \\
$^4\lambda(\Sigma^{-}_{b})_{5/2}$ & &   5.7 & 137.0 & & 2.08 & 85.3 \\
$^2\rho(\Sigma^{-}_{b})_{1/2}$    & &  50.2 &  21.9 & & & \\
$^2\rho(\Sigma^{-}_{b})_{3/2}$    & &  52.1 &  22.8 & & & \\
\noalign{\smallskip}
\hline
\noalign{\smallskip}
& $^2\Xi_{b}$ & $^2\Xi'_{b}$ & $^4\Xi'_{b}$ 
& $^2\Xi_{b}$ & $^2\Xi'_{b}$ & $^4\Xi'_{b}$ \\
\noalign{\smallskip}
\hline 
\noalign{\smallskip}
$^2\lambda(\Xi'^{0}_{b})_{1/2}$ &  50.7 &  52.9 &  0.3 & 72.2 & 76.3 & 0.89 \\
$^2\lambda(\Xi'^{0}_{b})_{3/2}$ &  53.9 & 111.5 &  0.3 & 72.8 & 43.9 & 0.90 \\
$^4\lambda(\Xi'^{0}_{b})_{1/2}$ &  29.3 &   0.6 &  6.9 & 34.0 & 0.25 & 69.5 \\
$^4\lambda(\Xi'^{0}_{b})_{3/2}$ &  87.1 &   1.7 & 38.8 & 94.0 & 0.67 & 47.5 \\
$^4\lambda(\Xi'^{0}_{b})_{5/2}$ &  61.6 &   1.3 & 69.4 & 47.7 & 0.44 & 41.5 \\
$^2\rho(\Xi'^{0}_{b})_{1/2}$    & 708.1 &  13.4 &  5.5 & & & \\
$^2\rho(\Xi'^{0}_{b})_{3/2}$    & 714.7 &  14.1 &  5.9 & & & \\
\noalign{\smallskip}
\hline
\noalign{\smallskip}
$^2\lambda(\Xi'^{-}_{b})_{1/2}$ &  1.1 &  50.8 &  0.5 & 0.0 &  190 & 3.54 \\
$^2\lambda(\Xi'^{-}_{b})_{3/2}$ &  1.1 & 128.1 &  0.6 & 0.0 & 92.3 & 3.60 \\
$^4\lambda(\Xi'^{-}_{b})_{1/2}$ &  0.6 &   1.0 &  5.6 & 0.0 & 1.48 &  164 \\
$^4\lambda(\Xi'^{-}_{b})_{3/2}$ &  1.8 &   3.0 & 38.5 & 0.0 & 2.94 &  104 \\
$^4\lambda(\Xi'^{-}_{b})_{5/2}$ &  1.3 &   2.2 & 82.7 & 0.0 & 1.88 & 88.2 \\
$^2\rho(\Xi'^{-}_{b})_{1/2}$    & 15.0 &  21.9 &  9.0 & & & \\
$^2\rho(\Xi'^{-}_{b})_{3/2}$    & 15.2 &  23.0 &  9.6 & & & \\
\noalign{\smallskip}
\hline
\noalign{\smallskip}
& & $^2\Omega_{b}$ & $^4\Omega_{b}$ 
& & $^2\Omega_{b}$ & $^4\Omega_{b}$ \\
\noalign{\smallskip}
\hline 
\noalign{\smallskip}
$^2\lambda(\Omega^{-}_{b})_{1/2}$ & & 36.0 &  0.2 & &  154 &  1.49 \\
$^2\lambda(\Omega^{-}_{b})_{3/2}$ & & 73.7 &  0.2 & & 83.4 &  1.51 \\
$^4\lambda(\Omega^{-}_{b})_{1/2}$ & &  0.4 &  5.1 & & 0.64 & 99.23 \\
$^4\lambda(\Omega^{-}_{b})_{3/2}$ & &  1.1 & 26.7 & & 1.81 & 70.68 \\
$^4\lambda(\Omega^{-}_{b})_{5/2}$ & &  0.9 & 45.1 & & 1.21 & 63.26 \\
$^2\rho(\Omega^{-}_{b})_{1/2}$    & &  8.7 &  3.4 & & & \\
$^2\rho(\Omega^{-}_{b})_{3/2}$    & &  9.2 &  3.7 & & & \\
\noalign{\smallskip}
\end{tabular} 
\end{ruledtabular}
\end{table}


\begin{table}
\caption{Radiative decay widths of anti-triplet singly-heavy bottom 
$P$-wave baryons in keV.}
\label{3barb}
\centering
\begin{ruledtabular}
\begin{tabular}{lrrrrrr}
\noalign{\smallskip}
& \multicolumn{3}{c}{Present} 
& \multicolumn{3}{c}{$\chi$QM \cite{Wang2017}} \\
& $^2\Lambda_{b}$ & $^2\Sigma_{b}$ & $^4\Sigma_{b}$ 
& $^2\Lambda_{b}$ & $^2\Sigma_{b}$ & $^4\Sigma_{b}$ \\
\noalign{\smallskip}
\hline 
\noalign{\smallskip}
$^2\lambda(\Lambda^{0}_{b})_{1/2}$ & 40.7 &   0.2 &   0.0 & 50.2 & 0.14 & 0.09 \\
$^2\lambda(\Lambda^{0}_{b})_{3/2}$ & 43.4 &   0.3 &   0.0 & 52.8 & 0.21 & 0.15 \\
$^2\rho(\Lambda^{0}_{b})_{1/2}$    & 10.2 &  93.3 &   2.0 & 1.62 & 16.2 & 0.02 \\
$^2\rho(\Lambda^{0}_{b})_{3/2}$    & 10.6 & 315.6 &   2.2 & 1.81 & 15.1 & 0.03 \\
$^4\rho(\Lambda^{0}_{b})_{1/2}$    &  5.6 &   3.5 &   6.2 & 0.81 & 0.02 & 8.25 \\
$^4\rho(\Lambda^{0}_{b})_{3/2}$    & 16.2 &  10.6 &  73.9 & 2.54 & 0.07 & 9.90 \\
$^4\rho(\Lambda^{0}_{b})_{5/2}$    & 11.0 &   7.8 & 216.5 & & & \\
\noalign{\smallskip}
\hline
\noalign{\smallskip}
& $^2\Xi_{b}$ & $^2\Xi'_{b}$ & $^4\Xi'_{b}$ 
& $^2\Xi_{b}$ & $^2\Xi'_{b}$ & $^4\Xi'_{b}$ \\
\noalign{\smallskip}
\hline 
\noalign{\smallskip}
$^2\lambda(\Xi^{0}_{b})_{1/2}$ & 83.1 &   0.6 &   0.1 & 63.6 & 1.32 & 2.04 \\
$^2\lambda(\Xi^{0}_{b})_{3/2}$ & 88.9 &   0.7 &   0.2 & 68.3 & 1.68 & 2.64 \\
$^2\rho(\Xi^{0}_{b})_{1/2}$    & 13.3 & 144.5 &   1.7 & 1.86 & 94.3 & 0.62 \\
$^2\rho(\Xi^{0}_{b})_{3/2}$    & 14.0 & 377.2 &   1.9 & 2.10 & 69.4 & 0.80 \\
$^4\rho(\Xi^{0}_{b})_{1/2}$    &  7.5 &   2.9 &  14.7 & 0.93 & 0.16 & 80.0 \\
$^4\rho(\Xi^{0}_{b})_{3/2}$    & 22.2 &   8.8 & 109.7 & 2.94 & 0.80 & 78.0 \\
$^4\rho(\Xi^{0}_{b})_{5/2}$    & 15.4 &   6.5 & 245.2 & & & \\
\noalign{\smallskip}
\hline
\noalign{\smallskip}
$^2\lambda(\Xi^{-}_{b})_{1/2}$ & 91.5 & 0.0 & 0.0 &  135 & 0.0 & 0.0 \\
$^2\lambda(\Xi^{-}_{b})_{3/2}$ & 96.1 & 0.0 & 0.0 &  147 & 0.0 & 0.0 \\
$^2\rho(\Xi^{-}_{b})_{1/2}$    & 21.8 & 3.1 & 0.0 & 7.19 & 0.0 & 0.0 \\
$^2\rho(\Xi^{-}_{b})_{3/2}$    & 23.0 & 8.0 & 0.0 & 8.13 & 0.0 & 0.0 \\
$^4\rho(\Xi^{-}_{b})_{1/2}$    & 12.3 & 0.1 & 0.3 & 3.59 & 0.0 & 0.0 \\
$^4\rho(\Xi^{-}_{b})_{3/2}$    & 36.3 & 0.2 & 2.3 & 11.4 & 0.0 & 0.0 \\
$^4\rho(\Xi^{-}_{b})_{5/2}$    & 25.3 & 0.1 & 5.2 & & & \\
\noalign{\smallskip}
\end{tabular} 
\end{ruledtabular}
\end{table}

The only available experimental information is the measurement of the radiative decay widths of the excited $\Xi_c(2790)$ and $\Xi_c(2815)$ baryons by the Belle II Collaboration \cite{PhysRevD.102.071103}. It was found that the electromagnetic decay widths for neutral baryons are large (albeit with a large uncertainty), $\Gamma(\Xi^0_c(2815)\rightarrow\Xi_c^0+\gamma) = 320 \pm 45$ keV  and $\Gamma(\Xi^0_c(2790)\rightarrow\Xi_c^0+\gamma) = 800 \pm 320$ keV, while for charged baryons only an upper limit was obtained $\Gamma(\Xi^+_c(2815)\rightarrow\Xi_c^++\gamma)< 80$ keV  and $\Gamma(\Xi^+_c(2790)\rightarrow\Xi_c^++\gamma) < 350$ keV. Inspection of Tables~\ref{6c} and \ref{3barc} shows that this behavior is in agreement with the assignment of the $\Xi_c(2790)$ and $\Xi_c(2815)$ baryons as $^2\lambda(\Xi_c)_J$ states with $J^P=1/2^-$ and $3/2^-$, respectively, which confirms the prediction made in Ref.~\cite{Wang2017}. In Table~\ref{p1} we show a comparison with other theoretical calculations. We find a reasonable agreement between our calculation and the experimental data, as well as with the $\chi$QM results. The decay widths of $\Xi_c(2790)$ in Ref.~\cite{Gamermann2011} in which charmed baryons are interpreted as meson-baryon molecular states are calculated to be $\sim$ 250 keV for the charged baryon and $\sim$ 120 keV for the neutral baryon. Even though these results are in qualitative agreement with the data (which have large error bars) their behavior is very different from the present results and the $\chi$QM. In the light-cone QCD sum rule approach of \cite{Aliev2019} the radiative decay width of the charged $\Xi^+_c(2790)$ baryon is calculated to be much larger than that of the neutral $\Xi^0_c(2790)$ baryon, in contradiction with the experimental data. Finally, a calculation in the relativistic quark model shows a larger radiative width for the neutral $\Xi_{c}(2815)^0$ than for the charge $\Xi_{c}(2815)^+$ state \cite{Ivanov1999}, although the calculated value of $\Gamma(\Xi_{c}(2815)^+\rightarrow\Xi_c^+ +\gamma)$ is much larger than the upper limit from experiment.

\begin{table*}
\caption{Radiative decay widths in keV of the $\Xi_c(2790)$ and $\Xi_c(2815)$ baryons.}
\label{p1}
\centering
\begin{ruledtabular}
\begin{tabular}{lrrccccc}
\noalign{\smallskip}
& Present & $\chi$QM \cite{Wang2017} & MB \cite{Gamermann2011} & LCQSR \cite{Aliev2019} & RQM \cite{Ivanov1999} & Exp \cite{PhysRevD.102.071103} \\
\noalign{\smallskip}
\hline 
\noalign{\smallskip}
$\Xi_{c}(2790)^{+}\rightarrow{}^2\Xi_{c}^{+}\gamma$ &   7.4 &   4.6 
& $249.6 \pm 41.9$ & $265 \pm 106$ && $<350$ \\
$\Xi_{c}(2790)^{0}\rightarrow{}^2\Xi_{c}^{0}\gamma$ & 202.5 & 263.0 
& $119.3 \pm 21.7$ & $2.7 \pm 0.8$ && $800\pm320$ \\
$\Xi_{c}(2815)^{+}\rightarrow{}^2\Xi_{c}^{+}\gamma$ &   4.8 &   2.8 
&&& $190 \pm 5$ & $<80$\\
$\Xi_{c}(2815)^{0}\rightarrow{}^2\Xi_{c}^{0}\gamma$ & 292.6 & 292.0 
&&& $497 \pm 14$ & $320 \pm 45^{+45}_{-80}$ \\
\noalign{\smallskip}
\end{tabular} 
\end{ruledtabular}
\end{table*}

\subsection{Doubly- and triply-heavy baryons}
\label{dthb}

For electromagnetic decays of doubly-heavy baryons the final baryon $B'_{QQ}$ is either $^2(B'_{QQ})_{1/2}$ or $^4(B'_{QQ})_{3/2}$ of the flavor triplet $\bf 3$. The electromagnetic decay widths for doubly-heavy $S$- and $P$-wave baryons are shown in Tables~\ref{qcc} and \ref{qbb}.
We compare our results with those obtained in the chiral quark model ($\chi$QM) of Xiao {\it et al.} \cite{Wang2017} in which the electromagnetic decay widths are calculated by combining the masses obtained by Ebert {\it et al.} in a relativistic light-quark-heavy-diquark picture \cite{PhysRevD.66.014008} with non-relativistic harmonic oscillator wave functions. Once again, we note that in the present calculation mass spectra and radiative widths in a consistent manner in the framework of a non-relativistic harmonic oscillator quark model. 

Tables~\ref{qcc} and \ref{qbb} show that the radiative decay widths of $^2\rho(B_{QQ})_J$ baryons are very small. This can be understood as follows. Just as for the case of singly-heavy baryons, for doubly-heavy baryons the main contribution to the radiative decay widths comes from the light quarks. Inspection of Table~\ref{sf66} and Eq.~(\ref{ecQQ}) shows that the light quarks do not contribution to the spin-flip amplitudes for the decays $^2\rho(\Xi_{QQ})_J \rightarrow ^{2,4}\Xi_{QQ}$ and $^2\rho(\Omega_{QQ})_J \rightarrow ^{2,4}\Omega_{QQ}$. In addition, for these decays the orbit-flip amplitude vanishes identically. As a consequence, the electromagnetic decay widths of $^2\rho(B_{QQ})_J$ baryons are very small. 

For electromagnetic decays of triply-heavy baryons the final baryon is the ground-state $S$-wave $^4(\Omega_{QQQ})_{3/2}$ state of the flavor singlet $\bf 1$. The electromagnetic decay widths of triple heavy baryons $\Omega_{QQQ}$ are shown in Table~\ref{widthQQQ}. Since in this case there are only heavy quarks the radiative decays are highly suppressed. 

\begin{table}
\caption{Radiative decay widths of triplet double charm $S$- and 
$P$-wave baryons in keV.}
\label{qcc}
\centering
\begin{ruledtabular}
\begin{tabular}{rrrrr}
\noalign{\smallskip}
& \multicolumn{2}{c}{Present} 
& \multicolumn{2}{c}{$\chi$QM \cite{PhysRevD.96.094005}} \\
& $^2\Xi_{cc}$ & $^4\Xi_{cc}$ & $^2\Xi_{cc}$ & $^4\Xi_{cc}$ \\
\noalign{\smallskip}
\hline 
\noalign{\smallskip}
$^4(\Xi_{cc}^{++})_{3/2}$        &   2.1 &       & 16.7 & \\
\noalign{\smallskip}
$^2\lambda(\Xi^{++}_{cc})_{1/2}$ & 373.1 &  23.0 & 105    &  117 \\
$^2\lambda(\Xi^{++}_{cc})_{3/2}$ & 172.7 &  32.1 & 495    &  196 \\
$^4\lambda(\Xi^{++}_{cc})_{1/2}$ &  55.9 &   0.5 & 35.7   &  287 \\
$^4\lambda(\Xi^{++}_{cc})_{3/2}$ & 196.5 &  41.0 & 147    &  212 \\
$^4\lambda(\Xi^{++}_{cc})_{5/2}$ & 173.9 & 403.1 & 136    &  181 \\
$^2\rho(\Xi^{++}_{cc})_{1/2}$    &   0.1 &   0.0 & $<$0.5 & $<$0.5 \\
$^2\rho(\Xi^{++}_{cc})_{3/2}$    &   0.2 &   0.0 & $<$2.0 & $<$2.0 \\
\noalign{\smallskip}
\hline
\noalign{\smallskip}
$^4(\Xi^{+}_{cc})_{3/2}$        &   1.9 &       & 14.6  & \\
\noalign{\smallskip}
$^2\lambda(\Xi^{+}_{cc})_{1/2}$ & 300.7 &   5.6 & 250    &  24.6 \\
$^2\lambda(\Xi^{+}_{cc})_{3/2}$ & 241.2 &   7.7 & 442    &  40.7 \\
$^4\lambda(\Xi^{+}_{cc})_{1/2}$ &  13.3 &   6.9 & 7.47   &  208 \\
$^4\lambda(\Xi^{+}_{cc})_{3/2}$ &  46.5 &  83.2 & 30.5   &  189 \\
$^4\lambda(\Xi^{+}_{cc})_{5/2}$ &  40.8 & 303.4 & 28.0   &  198 \\
$^2\rho(\Xi^{+}_{cc})_{1/2}$    &   0.1 &   0.0 & $<$0.5 & $<$0.5 \\
$^2\rho(\Xi^{+}_{cc})_{3/2}$    &   0.2 &   0.0 & $<$2.0 & $<$2.0 \\
\noalign{\smallskip}
\hline
\noalign{\smallskip}
& $^2\Omega_{cc}$ & $^4\Omega_{cc}$ & $^2\Omega_{cc}$ & $^4\Omega_{cc}$ \\
\noalign{\smallskip}
\hline 
\noalign{\smallskip}
$^4(\Omega^+_{cc})_{3/2}$          &   1.1 &       & 6.93   & \\
\noalign{\smallskip}
$^2\lambda(\Omega^{+}_{cc})_{1/2}$ & 177.3 &   0.5 & 294    & 9.61 \\
$^2\lambda(\Omega^{+}_{cc})_{3/2}$ & 197.9 &   0.8 & 430    & 157 \\
$^4\lambda(\Omega^{+}_{cc})_{1/2}$ &   1.9 &  15.2 & 3.19   & 209 \\
$^4\lambda(\Omega^{+}_{cc})_{3/2}$ &   7.5 &  82.6 & 12.90  & 202 \\
$^4\lambda(\Omega^{+}_{cc})_{5/2}$ &   8.0 & 171.0 & 12.0   & 225 \\
$^2\rho(\Omega^{+}_{cc})_{1/2}$    &   0.1 &   0.0 & $<$0.4 & $<$0.4 \\
$^2\rho(\Omega^{+}_{cc})_{3/2}$    &   0.2 &   0.0 & $<$2.0 & $<$2.0 \\
\noalign{\smallskip}
\end{tabular} 
\end{ruledtabular}
\end{table}

\begin{table}
\caption{Radiative decay widths of triplet doubly-heavy bottom 
$P$-wave baryons in keV.}
\label{qbb}
\centering
\begin{ruledtabular}
\begin{tabular}{rrrrr}
\noalign{\smallskip}
& \multicolumn{2}{c}{Present} 
& \multicolumn{2}{c}{$\chi$QM \cite{PhysRevD.96.094005}} \\
& $^2\Xi_{bb}$ & $^4\Xi_{bb}$ & $^2\Xi_{bb}$ & $^4\Xi_{bb}$ \\
\noalign{\smallskip}
\hline 
\noalign{\smallskip}
$^4(\Xi^{0}_{bb})_{3/2}$        &   0.1 &       & 1.19 & \\
\noalign{\smallskip}
$^2\lambda(\Xi^{0}_{bb})_{1/2}$ & 674.1 &  35.0 &  455 &  235 \\
$^2\lambda(\Xi^{0}_{bb})_{3/2}$ & 387.8 &  37.2 &  984 &  265 \\
$^4\lambda(\Xi^{0}_{bb})_{1/2}$ &  49.2 &   2.2 &  555 & 1330 \\
$^4\lambda(\Xi^{0}_{bb})_{3/2}$ & 145.2 & 127.8 &  172 &  773 \\
$^4\lambda(\Xi^{0}_{bb})_{5/2}$ & 101.7 & 555.6 &  121 &  569 \\
$^2\rho(\Xi^{0}_{bb})_{1/2}$    &   0.0 &   0.0 & 1.15 & 0.72 \\
$^2\rho(\Xi^{0}_{bb})_{3/2}$    &   0.0 &   0.0 & 3.30 & 2.66 \\
\noalign{\smallskip}
\hline
\noalign{\smallskip}
$^4(\Xi^{-}_{bb})_{3/2}$        &   0.0 &       & 0.24 & \\
\noalign{\smallskip}
$^2\lambda(\Xi^{-}_{bb})_{1/2}$ & 150.0 &   9.7 & 71.1 & 59.3 \\
$^2\lambda(\Xi^{-}_{bb})_{3/2}$ &  79.4 &  10.3 &  182 & 67.1 \\
$^4\lambda(\Xi^{-}_{bb})_{1/2}$ &  13.7 &   0.0 & 14.0 &  271 \\
$^4\lambda(\Xi^{-}_{bb})_{3/2}$ &  40.3 &  23.6 & 43.5 &  149 \\
$^4\lambda(\Xi^{-}_{bb})_{5/2}$ &  28.2 & 128.9 & 30.5 &  104 \\
$^2\rho(\Xi^{-}_{bb})_{1/2}$    &   0.0 &   0.0 & 1.15 & 0.72 \\
$^2\rho(\Xi^{-}_{bb})_{3/2}$    &   0.0 &   0.0 & 3.30 & 2.66 \\
\noalign{\smallskip}
\hline
\noalign{\smallskip}
& $^2\Omega_{bb}$ & $^4\Omega_{bb}$ 
&  $^2\Omega_{bb}$ & $^4\Omega_{bb}$ \\
\noalign{\smallskip}
\hline 
\noalign{\smallskip}
$^4(\Omega^{-}_{bb})_{3/2}$        &  0.0 &      & 0.08 & \\
\noalign{\smallskip}
$^2\lambda(\Omega^{-}_{bb})_{1/2}$ & 63.6 &  1.3 & 76.9 & 26.2 \\
$^2\lambda(\Omega^{-}_{bb})_{3/2}$ & 45.6 &  1.4 &  151 & 30.0 \\
$^4\lambda(\Omega^{-}_{bb})_{1/2}$ &  2.1 &  2.9 & 6.38 &  188 \\
$^4\lambda(\Omega^{-}_{bb})_{3/2}$ &  6.3 & 20.6 & 20.0 &  117 \\
$^4\lambda(\Omega^{-}_{bb})_{5/2}$ &  4.6 & 44.5 & 14.2 & 90.9 \\
$^2\rho(\Omega^{-}_{bb})_{1/2}$    &  0.0 &  0.0 & 1.41 & 1.11 \\
$^2\rho(\Omega^{-}_{bb})_{3/2}$    &  0.0 &  0.0 & 3.38 & 3.16 \\
\noalign{\smallskip}
\end{tabular} 
\end{ruledtabular}
\end{table}

\begin{table}
\centering
\caption{Radiative decay widths of singlet triply-heavy $P$-wave baryons in keV.}
\label{widthQQQ}
\begin{ruledtabular}
\begin{tabular}{cc}
\noalign{\smallskip}
& Present \\
\noalign{\smallskip}
\hline
\noalign{\smallskip}
$^2E(\Omega^+_{ccc})_{1/2}\rightarrow{}^4\Omega_{ccc}^{+}\gamma$ & 0.01 \\
$^2E(\Omega^+_{ccc})_{3/2}\rightarrow{}^4\Omega_{ccc}^{+}\gamma$ & 0.04 \\
$^2E(\Omega^0_{bbb})_{1/2}\rightarrow{}^4\Omega_{bbb}^{0}\gamma$ & 0.00 \\
$^2E(\Omega^0_{bbb})_{3/2}\rightarrow{}^4\Omega_{bbb}^{0}\gamma$ & 0.00 \\
\noalign{\smallskip}
\end{tabular} 
\end{ruledtabular}
\end{table}

\section{Summary and conclusions}
\label{sec:summary}

In this article we presented a study of masses and electromagnetic couplings of singly-, doubly- and triply-heavy baryons where the heavy quark can either be charm or bottom. We adopted a non-relativistic harmonic oscillator quark model in which we use a G\"ursey-Radicati form for the spin-flavor dependence. The parameters are determined in a simultaneous fit to 41 heavy baryon masses, 25 single charm, 15 single bottom and 1 double charm, with an r.m.s. deviation of 19 MeV. 

In the harmonic oscillator quark model there are two radial excitations for $qqQ$ and $QQq$ baryons: a $\rho$-mode for the relative motion between the two identical quarks and a $\lambda$-mode for the relative motion between the pair of identical quarks and the third quark. For $qqQ$ baryons the $\lambda$-mode has a lower frequency than the $\rho$-mode, while for $QQq$ baryons the situation is reversed. In this study we focused on ground-state $S$-wave baryons and excited $P$-wave baryons. The assignment of quantum numbers to the experimentally observed heavy baryons is based mostly on energy systematics. 

All ground-state $S$-wave single charm baryons have been observed, and all but one for the case of single bottom baryons. In addition, there are 7 $P$-wave excited baryons, 5 of which correspond to the $\lambda$-mode and 2 to the $\rho$-mode for the sextet baryons, $\Sigma_Q$, $\Xi'_Q$ and $\Omega_Q$. For the anti-triplet baryons, $\Lambda_Q$ and $\Xi_Q$, there are 2 $P$-wave excitations of the $\lambda$-mode and 5 of the $\rho$-mode. For $\Omega_c$ baryons we have made (tentative) assignments of 6 (out of 7) $P$-wave excitations (5$\lambda$ and 1$\rho$), compared to 5 for $\Xi'_c$ baryons (3$\lambda$ and 2$\rho$), 1 for $\Sigma_c$ baryons (1$\lambda$), 4 for $\Lambda_c$ baryons (2$\lambda$ and 2$\rho$) and 2 for $\Xi_Q$ baryons (2$\lambda$). For single bottom baryons there is much less experimental information available. We have been able to make the following identifications: $\Omega_b$ (4$\lambda$), $\Xi'_b$ (1$\lambda$), $\Sigma_b$ (1$\lambda$), $\Lambda_b$ (2$\lambda$) and $\Xi_b$ (1$\lambda$).

In Section~\ref{sec:masses} we discussed equal-spacing mass rules for $\lambda$- and $\rho$-mode excited 
baryons. For the case of $\lambda$-mode $\Omega_c$ and $\Xi'_c$ baryons the equal-spacing mass rule of Eq.~(\ref{esr1}) is satisfied to a good precision. The other equal-spacing mass rules, Eqs.~(\ref{esr2})-(\ref{esr5}) for singly-heavy baryons and Eqs.~(\ref{esr6})-(\ref{esr7}) for doubly-heavy baryons, may help to identify the missing heavy baryons in the charm and bottom sectors.

The calculated masses in the present harmonic oscillator quark model are in good agreement with the available experimental data. For ground state baryons our results are comparable to those obtained in a quark model description by Karliner and Rosner \cite{PhysRevD.90.094007} and lattice QCD calculations \cite{PhysRevD.90.094507}. 

In the second part of the article we presented the results for electromagnetic couplings. We presented explicit results for all spin-flavor matrix elements and the radial integrals in the harmonic oscillator quark model. The radiative widths are dominated by the contribution of the light quarks. As a consequence the radiative decay widths of triply-heavy baryons $\Omega_{QQQ}$ are very small. The experimental information on radiative widths is rather scarce. A comparison with the observed widths for the charged and neutral $\Xi_c(2790)$ and $\Xi_c(2815)$ baryons confirms the assignment of these resonances as $^2\lambda(\Xi_c)_J$ states with $J^P=1/2^-$ and $3/2^-$, respectively. Our results are in reasonable agreement with those obtained in the $\chi$QM of Wang {\it et al.} \cite{Wang2017}. 

The study of heavy baryons is important to gain a better understanding of hadron structure and the strong interaction. In this paper we focused on mass spectra and radiative decay widths. The study of strong decay widths will be published separately \cite{strongc}, as well as the extension to charm-bottom heavy baryons, $\Omega_{cb}$, $\Xi_{cb}$, $\Omega_{ccb}$ and $\Omega_{cbb}$.

\begin{acknowledgments}
It is a pleasure to thank Mikhail Mikhasenko for interesting discussions. 
This work was supported in part by PAPIIT-DGAPA (UNAM, Mexico) grant No.~IG101423, and by CONACYT (Mexico) grant No.~251817. EOP acknowledges the projects (P1-0035) and (J1-3034) were financially supported by the Slovenian Research Agency. Special thanks to Blaž Bortolato for his useful and interesting discussion during the fitting of the parameters.
\end{acknowledgments}

\

\noindent
{\it Note added:} After this work had been completed, new results were announced by the LHCb collaboration \cite{lhcbcollaboration2023observation} on the confirmation of the $\Xi_b(6100)^-$ baryon and the discovery of $\Xi_b(6087)^0$ and $\Xi_b(6095)^0$. We interpreted the $\Xi_b(6100)^-$ state as a member of the isospin doublet $^2\lambda(\Xi_b)_{J=3/2}$. The recently discovered $\Xi_b(6095)^0$ baryon could be its isospin partner, whereas the the $\Xi_b(6087)^0$ baryon could then be assigned as a member of the isospin doublet $^2\lambda(\Xi_b)_{J=1/2}$. The corresponding theoretical masses are 6084 and 6077 MeV, respectively, in good agreement with the observed values.

\clearpage

\bibliographystyle{apsrev4-1}
%

\end{document}